\documentclass[fleqn,usenatbib]{mnras}

\usepackage{newtxtext,newtxmath}
\usepackage{xspace}
\usepackage[T1]{fontenc}
\usepackage{ae,aecompl}
\usepackage{url}

\usepackage{graphicx}	% Including figure files
\usepackage{amsmath}	% Advanced maths commands
\usepackage{float}

\usepackage[utf8]{inputenc}

\title[VHE emission by 3FHL blazars]{Predictions of TeV emission for a set of hard BL Lac objects\\
}
\author[S. Paiano et al.]{Simona Paiano$^{1,2,3}$\thanks{E-mail:
simona.paiano@inaf.it},
Aldo Treves$^{4,5}$,
Alberto Franceschini$^{6}$,
Renato Falomo$^{7}$ \\
$^{1}$INAF - Osservatorio Astronomico di Roma, via Frascati 33, I-00040, Monteporzio Catone, Italy \\
$^{2}$INAF - IASF Palermo, via Ugo La Malfa, 153, I-90146, Palermo, Italy \\ 
$^{3}$INAF - IASF Milano, via Corti 12, I-20133, Milano, Italy \\
$^{4}$Universita' dell'Insubria, via Valeggio, 22100, Como, Italy\\
$^{5}$INAF - Osservatorio Astronomico di Brera, via Bianchi 46, I-23807, Merate (Lecco), Italy\\
$^{6}$Dipartimento di Fisica e Astronomia, Universita' degli Studi di Padova, vicolo dell'Osservatorio 3, I-35122, Padova, Italy\\
$^{7}$INAF - Osservatorio Astronomico di Padova, vicolo dell'Osservatorio 5, I-35122, Padova, Italy\\
}

\date{Received:~\today; Accepted:~ }

\begin{document}
\label{firstpage}
\pagerange{\pageref{firstpage}--\pageref{lastpage}}
\maketitle

\begin{abstract}
We focus our analysis on 55 BL Lac objects with a hard \textit{Fermi} gamma-ray spectrum, and for which a redshift or a lower limit to it has been determined by a previous study of ours. 
We extrapolate the spectral fits given by the 4FGL catalogue to the VHE band ($>$~0.1 TeV), which can be explored by imaging atmospheric Cherenkov telescopes. 
Furthermore, we take into account the absorption due to the extragalactic background light, strongly depending on the redshift.
Finally, we compare our results with publicly available sensitivity curves for a selection of imaging
atmospheric Cherenkov telescope arrays currently operating or under construction. From our extrapolations and simulations we find a large number of promising candidates for observation with the forthcoming Cherenkov Telescope Array observatory.
\end{abstract}

\begin{keywords}
galaxies: active and redshifts
--- BL Lacertae objects: general 
--- gamma-rays: galaxies
\end{keywords}

\section{Introduction}  
\label{sec:intro}

The current exploration of the gamma-ray sky in the High Energy (HE) band (50~MeV~-~100~GeV) derives from space borne instrumentation, particularly the Large Area Telescope on board the \textit{Fermi}-LAT observatory \citep{atwood2009}. 
Above 100 GeV (i.e. in the so-called Very High Energy regime, VHE) the main instruments are the Imaging Atmospheric Cherenkov Telescopes (IACTs) from ground, allow us observations up to many tens of TeV. 

Beyond these limits, year-long integrations by the water Cherenkov observatories, particularly HAWC  \citep[e.g.][]{2012NIMPA.692...72D} and LHAASO \citep[][]{2016NPPP..279..166D}, can detect steady emission from both galactic and extragalactic sources.
Some indirect information also follows from neutrino detections in the PeV region, a field currently led by the Ice-Cube experiment \citep{icecube17}.
Thanks to all such facilities, high-energy astrophysics in the HE and VHE has experienced enormous progress in the last ten years.

Eventually, the Cherenkov Telescope Array (CTA, particularly the large southern array including tens of small telescopes) will provide us with great sensitivity, quick response, high angular and spectral resolution to cover a wide VHE spectral range up to 100 TeV.

The extragalactic gamma-ray sky at all photon energies is dominated by blazars. 
In these objects a relativistic jet, producing a highly anisotropic emission, is pointed close to  the observer's direction. 
There are two main classes of blazars, the Flat Spectrum Radio Quasars (FSRQs) and the BL Lac objects (BLLs). 
While strong quasar-like emission lines are present in the optical spectra of the former category, very weak or no lines are apparent in the optical spectra of the latter class. 
The ratio of the  space densities of BLLs to FSRQs is poorly known, and might vary with the photon energy of the selection and possibly with redshift. 
There are some indications that BLLs prevail over FSRQs, at least at low to moderate redshifts \citep{ghisellini2017}.

Above a few GeV, for objects at large distances ($z\gtrsim0.1$), the Universe starts becoming opaque, because of the interaction of gamma-rays with photons of the extragalactic background light (EBL), with the consequent production of electron/positron pairs.
Suppose that the HE spectrum is known, as measured by the \textit{Fermi}-LAT observations, then in order to estimate the flux at VHE one should extrapolate to higher energies the observed spectrum, and finally correct it for the EBL absorption.
Since such an absorption depends strongly on the redshift, the estimate of the VHE fluxes, based on the HE ones, can reliably be performed only for sources of known redshift. On the other hand for BLLs, because of the weakness of the lines, the determination of the redshift is difficult and often even impossible.

Known HE BLLs largely outnumber VHE ones, because of the intrinsic spectral shape, and because HE BLL samples derive from all-sky monitoring, which \textit{Fermi}-LAT performs every 3 hours, and have been accumulated for more than a decade. 
By contrast, IACT detections require pointed integrations, of the duration of several hours, and in many cases can be guided by previous lower-energy observations. 
At present $\sim$2000 BLL are detected in HE, but only $\sim$100 at VHE.
\citet{costamante2002}, before the launch of \textit{Fermi}, reported predictions of VHE emission of blazars, based on the Synchrotron-Self-Compton (SSC) emission model, 
and using the available broad-band SEDs. 
The interest in the VHE emission by blazars has increased with the evolution of the IACTs. 
For recent contributions on the prediction of the VHE detection and for an overview of the relevant literature, we refer to \citet{massaro2013, foffano2019, balmaverde2020, costamante2020}.

In  a previous paper \citet{paiano2020} (Paper~I), we highlighted our results of an optical spectroscopic survey for a well defined sample of 55 sources from the third catalog of hard \textit{Fermi}-LAT sources \citep[3FHL,][]{ajello2017} lacking the redshift. 
This homogeneous study exploited the spectroscopic power of the OSIRIS instrument \citep{cepa2003} on the Gran Telescopio de Canarias (GTC).
For 25 objects we determined the redshift, for 5 more we set a robust spectroscopic lower limit to \textit{z} by the detection of intervening absorption systems, and for the remainders we obtained lower limits of the redshift based on some hypotheses on the BLL host galaxies \citep[see details in ][]{paiano2017tev}. 

Based on this information, the present paper provides predictions of the expected TeV fluxes and spectra to direct and optimize observational campaigns with forthcoming IACT arrays. Our estimates follow by extrapolating to the VHE band the spectral shapes in the HE interval, as proposed in the Fermi Large Area Telescope Fourth Source Catalog \citep[4FGL; ][]{4fglcatalog}.
We specifically take into account details of the absorption due to the interaction of the gamma-ray photons with lower energy photons of the EBL to recognize TeV candidates among BLLs of our sample and to evaluate the level of the TeV emission and their VHE spectrum.

Our analysis adopts as a reference for present, forthcoming and future instrumentation at VHE energies the following three IACT arrays: MAGIC \citep{2016APh....72...76A,2016APh....72...61A}, ASTRI-MiniArray \citep{2016SPIE.9906E..5TP}, and CTA North and South \citep{2019scta.book.....C}, respectively. 
In our choice, MAGIC is a dual-telescope array operative since more than 10 years in the northern site of the La Palma Canary Island (our source sample is in the north, following selection for GTC observations), with performances very similar to the other north IACT VERITAS \citep[][]{2006APh....25..391H} and HESS in the south \citep[][]{2004NewAR..48..331H}. 
ASTRI-MiniArray, also located in the Tenerife Canary Island, is  considered because, in a few years time, it will consist in a large array of 9 four-meter class telescopes with quite improved sensitivities above 3 TeV compared to existing instrumentation.
For the complex and ambitious CTA project both the Northern (La Palma Canary Island) and Southern (Atacama, Chile) configurations are considered.

Section 2 of the paper is dedicated to a description of our adopted model for the EBL photon density, and the procedure for evaluating the attenuation of gamma-rays due to pair production. 
The results are reported in Section 3 where we calculate the expected extrapolation of the flux for the 55 BLLs in the VHE domain. 

Implications of these results are briefly discussed in Section 4.
In the present paper we adopt the following cosmology: $H_0=70\ km/s/Mpc$ and $\Omega_\Lambda=0.7$ $\Omega_m=0.3$.

\section{The EBL photon density, pair production, and gamma-ray absorption}
A fundamental limitation of VHE observations is caused by the large volume density of low-energy photons in the Universe and their interaction with those emitted by blazars.
This photon-photon interaction - with a maximal cross-section occurring when the product of photon energies is  equal to that of the electron and positron rest-energies - and the consequent destruction of the gamma-ray, are an inevitable consequence of quantum mechanics \citep[see e.g. ][]{heitler1954}. 
At the same time, however, this phenomenon offers also interesting opportunities to constrain the EBL intensity where not directly measurable (i.e. at almost all UV-to-IR wavelengths), as well as testing fundamental physics in energy regimes not attainable in a laboratory (see Sect. \ref{discussion}).

The EBL spectral intensity contributed by cosmic sources and its time evolution have been modelled by several authors \citep[][among various others]{1998ApJ...494L.159S,2002A&A...386....1K,franceschini2008,2010ApJ...712..238F,dominguez2011,gilmore2012,2018MNRAS.474..898A}.

We use here for our simulations the EBL photon density in the wide waveband interval from the far-UV to the sub-millimeter ($0.1\ \mu m<\lambda < 1000\ \mu m$) from the model by \citet{franceschini2017,2018A&A...614C...1F}.
Their approach was to adopt a {\it backward evolution} model, starting from the detailed knowledge of the local luminosity functions of galaxies and active nuclei all-over the wavelength interval, and the knowledge on how such functions evolve back in cosmic time. 
Thanks to the large variety of deep multi-wavelength surveys, the local EBL intensity and its time evolution appear to be well established. With this information we can precisely calculate photon-photon absorption for both local and high redshift gamma-ray emitters. 
We preferred this empirical approach to the alternative {\it forward evolution} model \citep[][]{gilmore2012}, based on theoretical prescriptions about birth and evolution of galaxies and AGNs, as more reliably grounded on observational data.

Once the redshift-dependent luminosity functions are determined, an integral in luminosity gives the source emissivity as a function of wavelength and redshift. A second integral in redshift of the photon number density, calculated from z~=~0 to that of gamma-ray source $z_{source}$ and properly weighted by the gamma-gamma cross-section, gives us the optical depth $\tau(\epsilon, z_{source})$ as a function of the gamma-ray energy $\epsilon$ (in our local frame). This is the probability that such a VHE photon is  absorbed  during the path along the line-of-sight.
All the details of the calculation can be found in \citet{franceschini2008}, while tables of the photon density and optical depths $\tau(\epsilon, z_{source})$ are reported in \citet{franceschini2017}. The observed spectrum is then calculated as the unabsorbed spectrum multiplied by the absorption factor $e^{-\tau(\epsilon, z_{source})}$.

It should be noticed that, particularly starting from 2008, the most referred to EBL models have converged to a standard pattern, at least below $z\sim 1$ that is of interest for us here. With the implication that differences of only 10\% at most would be found by varying the adopted EBL model from one to the other, well below the uncertainties inherent in the source spectral extrapolations.

\section{Results}

 In order to identify the best candidate BLLs for observation at TeV energies we compare the HE fluxes, extrapolated into the VHE band and absorbed by the EBL, against the sensitivity of current and future IACTs.

\subsection{VHE flux estimates}

For all of the 3FHL targets of our sample (see Section 1 and Paper~I), \textit{Fermi} fluxes at energy 50~MeV~-~1~TeV range are available in the fourth full catalog of LAT sources 4FGL catalog\footnote{In this work we use the version 22 of the 4FGL catalog. Note that from May 2020 a new version of 4FGL-DR2 is available}. 
This catalog is based on 8 years of survey data  compared with the the Third Catalog of Hard \textit{Fermi}-LAT Sources (3FHL) which contains  objects detected above $>$~10~GeV over 7 years of \textit{Fermi} operations and covers the energy range between 10~GeV and 2~TeV.

The \textit{Fermi} fluxes are reported in Figs.~1-2-3 as black points. 
For the VHE extrapolation, we use the 4FGL spectral fits proposed by the collaboration (black solid curve in Figs.1-2-3) which are mainly a power law or, in a few cases, a LogParabola \citep[see details in Section 3.3 of ][]{4fglcatalog}. 
We are making the simplifying and rough assumption that these extrapolations coincide with the spectral shapes emitted by the sources. 
Because all the sources considered in this paper have redshift $\lesssim$~1, we suppose that the fits of the 4FGL fluxes are little affected by the EBL absorption. 
The EBL corrections depend strongly on the redshift. 
We noted that for z$\gtrsim$0.4 the EBL effect is already expected in the \textit{Fermi} data of highest energies (see for instance 3FHLJ0045.3+2127, 3FHLJ0423.8+4149, 3FHLJ0433.6+2905, 3FHLJ0612.8+4122, 3FHLJ1253.1+5300, 3FHLJ1447.9+3608 and 3FHLJ1800.5+7827). 

The extrapolations of the 4FGL best-fitting curves and their correction with the EBL absorption recipes, as outlined in the previous section, yield our estimates of the fluxes in the VHE band. 
For the sources with well-known redshift (see Table 1), our extrapolations are shown in Fig.~1 as red solid curves and we note that for most of the sources they are consistent (within ~2 sigma) with the 3FHL data.

 In the 5 cases (see Table 2) where we have a robust spectroscopic lower limit on \textit{z}  derived by intervening absorbers, we show our extrapolations in Fig. 2 as a red filled band between the measured lower-limit redshift and one corresponding to a distance at a factor 2 larger.
The same is done in Fig. 3 for the cases where the lower limit on the redshift derives from the absence of detected lines from the BLL host galaxy (Table 3).

\subsection{Comparison with IACT sensitivity curves}

For the northern MAGIC array we adopted sensitivity curves taken from \citet[][]{ahnen2015,2016APh....72...61A}, while for the CTA northern and southern sites we referred to the CTA webpage\footnote{https://www.cta-observatory.org/science/cta-performance/}. For the ASTRI-Mini Array\footnote{we assume that the sensitivity curve of the mini array located in the northern hemisphere will not be significantly different from that computed for the southern site } we use the sensitivity curve derived by \citet[][]{lombardi2018, pintore2020}.  
The sensitivity curves have been taken for a zenith angle ZA~$<$~20 deg and $<$~35 deg for CTA and MAGIC, respectively. They are calculated at 5~$\sigma$ for 50~hr exposure and are reported for each source in Figs.~1-2-3. 

In Fig.~1 we also show spectral data-points (red points) derived from simulations for CTA observations of each blazar with well-known redshift, and in Section 3.3 we provide details of how these simulations were performed. Each spectral flux point is given at 3~$\sigma$ significance and the y-axis error bars show the 
$1\sigma$ flux uncertainties.

Based on our EBL-corrected extrapolations and flux simulations, Tables 1, 2 and 3 provide a summary of our perspective on the detection of the BLLs in the TeV band. Each table highlights the BLLs in the respective redshift categories that we considered: the 25 sources with a known redshift, the objects with spectroscopic lower limit, and those where the redshift lower limit is derived by the non-detection of lines from the host galaxy.
Remember that in the cases where only lower limits are available, we calculated bands corresponding to redshift intervals as described in Section 3.1.
Considering the sources of known redshift, from our simulations, we find that all sources should be observable with the CTA arrays with a  detection significance greater than 5$\sigma$ (see Table 1).  
Of these, it is feasible that 5 sources should be detectable with MAGIC-like telescopes.
For the remaining objects (with only a redshift lower limit) we checked if the VHE extrapolated fluxes defining the red band (see  Fig. 2 and 3) are above the CTA sensitivity curve in four energies (0.1~TeV, 0.5~TeV, 1~TeV and 5~TeV)\footnote{we calculated the ratio between the two VHE extrapolated spectra defining the band and the CTA sensitivity curve at four energies (0.1~TeV, 0.5~TeV, 1~TeV and 5~TeV) and consider them above the CTA curve if the ratio is greater than 1.}. We consider as good candidates 18 sources with the VHE fluxes above the CTA curve in at least two energy bands and as acceptable candidates 10 sources with the  VHE flux above the CTA threshold in only one energy bin. 
Based on the current simulations, only two sources (see Table 3) appear as not detectable by CTA within 50 hours.

As it is well known, BLLs may be highly variable. Their gamma-ray variability has been carefully examined by the \textit{Fermi} collaboration. In tables 1-3 we report the
variability index and the fractional variability\footnote{The variability index is derived by the sum of the log(likelihood) difference between the flux fitted in each time interval and the average flux over the full catalog interval; a value greater than 18.48 over 12 intervals indicates $<$1\% chance of being a steady source. The fractional variability of the sources is defined from
the excess variance on top of the statistical and systematic
fluctuations. See \citet{4fglcatalog} for details.} taken from the 4FGL catalog.
According to the variability index,  $\sim$70\% of sources  have a high probability to be variable. For half of them  the fractional variability is less than 0.5. 
When programming very long exposure in the TeV band for the most variable sources (with fractional variability $>$50\%), a preliminary monitoring at lower energies is obviously necessary in order to assess the predicted flux.

\subsection{Simulations of CTA observations}

We performed CTA simulations and analyses of the data by adopting the software {\it Gammapy} v0.18.2 \citep[e.g.][]{2017ICRC...35..766D}. 
We followed the {Gammapy} {\sc MapDatasetEventSampler} class prescriptions\footnote{\url{https://docs.gammapy.org/0.18.2/tutorials/event\_sampling.html}} to generate source and background photons in the energy range 0.05~-~50 TeV. 
The core of the simulator is based on an Inverse Cumulative distribution function \citep[e.g.][]{Kendall&Stuart}.
 
The simulator takes as input a {\sc Dataset} object, which contains the main information about the observation (i.e. IRF properties, background, livetime, pointing position etc...) and the spectral, spatial and temporal properties of the source of interest. From this {\sc Dataset}, the simulator evaluates the predicted source and background event map, and samples it to generate the simulated events. 

The public Instrumental Response Functions (IRFs) of the {\it prod3b-v2} production for CTA north and CTA south configurations, full array and for zenith angle of 0--20 deg, were used. We included energy dispersion in each simulation and as background we only considered the contribution from the IRF. 

We simulated only the blazars of our sample of known redshift (see Table~1), assuming all of them as point-like sources  (PointSpatialModel code in Gammapy) and locating them at the center of a square map of 1 deg x 1 deg, binned with a pixel size of 0.02 deg (i.e. for a total of 2500 pixels), and pointed at the source coordinates. We adopted  a source spectral model given by the extrapolation of the corresponding \textit{Fermi} best-fits taking into account the EBL absorption (see details in Section 3.1). The model was given to the simulator as a template spectral model (i.e. a model having information on the energy and differential flux), by using the {\sc TemplateSpectralModel} code. Source and background events were simulated in the energy range 0.05 TeV - 50 TeV. An exposure of 50 hr was adopted for all simulations.

The simulator produces in output source + background event lists that we analyzed again with Gammapy v.0.18.2. We analyzed these data using a sky map having the same size and properties of the one used for the simulations, hence we did not apply any spatial cut. We also adopted the same spectral source model {\sc TemplateSpectralModel} used for the simulation convolved with a {\sc PowerLawNormSpectralModel} code (in Gammapy). The latter mimics a power-law with a normalized amplitude parameter. We set the model photon index and reference energy at 0 and 1 TeV, respectively, and we let free to vary its normalization. In other words, this allows us to multiply by a constant the {\sc TemplateSpectralModel} model used for the simulation, therefore quantifying the difference in flux with the simulated model. The morphology of each source was assumed to be point-like.

We then fitted the spectra of each simulated source, following a maximum likelihood analysis (with the {\sc Fit} function in {\it Gammapy}). The {\sc Fit} function takes as input a {\sc Dataset} object and a sky model. The {\sc Dataset} object was spatially binned as for the simulation. The background was also modelled with a {\sc PowerLawNormSpectralModel} and we let free to vary only its normalization to take into account the deviation from the IRF background model. The {\sc PowerLawNormSpectralModel} normalizations for both sources and background were always consistent with unity, within the 1$\sigma$ uncertainty, meaning that the source spectra are well reconstructed.

For each source, we calculated the source spectral points that are estimated in the energy range 0.05--50 TeV (in reconstructed energy), considering 10 logarithmic bins. We defined the center of each bin as the geometrical mean between the two energy edges. The flux was estimated independently in each energy bin using the global best-fit model. For every energy bin, the Test-Statistic \citep[TS,][]{1979ApJ...228..939C} was also estimated through a Li \& Ma \citep{1983ApJ...272..317L}
significance estimation and we considered significant only the spectral points for which the TS value was higher than 9 (i.e. corresponding to $\sim3\sigma$ significance). For lower TS values, we estimated a 2$\sigma$ (95\%) upper limit (UL). All results are shown as red points in Fig.~\ref{fig:fig1}.

\section{Conclusions}
\label{discussion}
    
We have made predictions about the detectability for the expected very high energy spectrum of 55 hard \textit{Fermi} BLLs with known redshift or for which confidence limits are available. 
We compared their HE data points and the extrapolation in the VHE band with the sensitivity curves of IACTs, and we also performed detailed simulations of CTA observations.
The result of this comparison shows that the two approaches are completely consistent. 

From our current analysis we found that a large fraction of these sources will be observable with CTA and partly also with existing facilities (like MAGIC, VERITAS, HESS).
 It is also worth noting that most of the sources in our sample are located in the northern hemisphere. As a consequence a good visibility 
(assuming a maximum zenith angle of 60 deg) for 70\% of the targets is found  for CTA-North  while for CTA-South only 20\% of them have a good visibility. 
Since the median value of the redshift of these objects is z$\sim$0.3 \citep[see paper~I and ][]{landoni2020}, we expect that in the future releases by the \textit{Fermi} observatory weaker sources will appear, which will dominate the next catalogues. 
As a consequence higher redshift objects will be present, and because of the strong dependence on \textit{z} of the EBL absorption, the perspective of VHE observability will rely on the determination of the redshift, which will be the bottleneck for the selection of good VHE candidates.
         
The investigation at VHE of extragalactic sources is of paramount importance for the understanding of a number of key phenomena of the Universe.
For instance, VHE photon absorption by EBL offers a direct bridge between VHE and blazar astrophysics and the cosmological production of the EBL by galaxies and active galactic nuclei during the last several Gyrs of the cosmic history. 

Also the observations of the interactions at such energies allow us to probe some fundamental aspects of physics, such as for example the Lorentz invariance, that according to quantum-gravity theories could be violated at the Planck scale \citep{tavecchio2016}.
Another interesting application concerns the physics of photon propagation predicted by various theories attempting to consistently include gravity into the quantum-mechanical framework, like super-symmetric models and super-string theories.
In such case, modifications to the photon-photon opacity effect would be possibly ascribed to the existence of light bosons, the Axion Like Particles (ALP), allowed to interact with two photons, or a photon and a magnetic field, originating a photon-ALP oscillation \citep[][ and references therein]{galanti2020}.

Finally, a reason of interest for VHE blazar observations is the possibility to test alternatives to the standard leptonic models for gamma-ray production, like emissions by hadronic beams \citep{bottcher2013, aharonian2000}, that were invoked to explain VHE neutrinos \citep{mannheim1993, aartsen2017, icecube18}.

\section*{Acknowledgments}
This research has made use of the CTA instrument response functions provided by the CTA Consortium and Observatory, see https://www.ctaobservatory.org/science/cta-performance/ (version prod3b-v2) for more details.
An anonymous referee helped in significantly improving the paper. 
 
\section*{Data availability}
 All data are incorporated into the article and its online supplementary material.

\setcounter{table}{0}
\begin{table*}
\begin{center}
\caption{3FHL objects of known redshift from emission/absorption lines in their optical spectra.}\label{tab:results}
\begin{tabular}{llccccccc} 
\hline
OBJECT           & z & Variability & Fractional   & TS$_{CTA}$ & Flux$_{CTA}$ & CTA & MAGIC & ASTRI \\
                 &   & Index       & variability  &            &   ($\times$10$^{-12}$~ Erg/cm$^2$/s) &    &   &       \\
\hline                                                  
3FHLJ0015.7+5551 & 0.2168  & 10  & 0.2$\pm$0.3 & 343  & 2.1   & X   & - & -  \\ 
3FHLJ0045.3+2127 & 0.4253  & 222 & 0.8$\pm$0.2 &  4914 & 12.2 & X & X & -  \\ 
3FHLJ0045.7+1217 & 0.2549  & 38  & 0.4$\pm$0.1 &  739  & 3.8  & X & - & -  \\ 
3FHLJ0137.9+5815 & 0.2745  & 37  & 0.3$\pm$0.1 &  1890 & 6.3  & X & - & -  \\  
3FHLJ0148.2+5201 & 0.437   & 14  & 0.2$\pm$0.1 &  440  & 3.2  & X & - & -  \\ 
3FHLJ0241.3+6543 & 0.1211  & 18  & 0.2$\pm$0.1 &  1142 & 4.0  & X & - & -  \\ 
3FHLJ0250.5+1712 & 0.2435  & 22  & 0.4$\pm$0.2 &  967  & 3.9  & X & - & -  \\ 
3FHLJ0423.8+4149 & 0.3977  & 9   & 0.03$\pm$0.1 &  3806 & 11.0 & X & - & -  \\ 
3FHLJ0433.6+2905 & 0.91    & 178 & 0.5$\pm$0.1  &  367  & 3.9 & X  & - & -  \\ 
3FHLJ0500.3+5238 & 0.1229  & 10  & 0.1$\pm$0.2  &  2281 & 5.0 & X & - & -  \\ 
3FHLJ0506.0+6113 & 0.538   & 18  & 0.5$\pm$0.1  &  112  & 1.6 & X & - & -  \\ 
3FHLJ0600.3+1245 & 0.0835  & 10  & 0.2$\pm$0.3  &  10852 & 12.6 & X & X & -  \\ 
3FHLJ0601.0+3837 & 0.662   & 9   & 0.1$\pm$0.3  &  143 & 2.2 & X & - & -  \\ 
3FHLJ0602.0+5316 & 0.0522  & 512 & 0.5$\pm$0.1  &  7253 & 8.8 & X & X & -  \\ 
3FHLJ0620.6+2645 & 0.1329  & 14  & 0.4$\pm$0.4  &  6969 & 8.2 & X & X & -  \\ 
3FHLJ0640.0-1254 & 0.1365  & 5   & 0. $\pm$10   &  15477 & 10.5 & X & X & -  \\ 
3FHLJ0708.9+2240 & 0.2966  & 92  & 0.6$\pm$0.2  &  430 & 2.9 & X & - & -  \\ 
3FHLJ0709.1-1525 & 0.1420  & 19  & 0.5$\pm$0.2  &  4739 & 5.5 & X & - & -  \\ 
3FHLJ0723.0-0732 & 0.3285  & 10  & 0.2$\pm$0.2  &  372 & 1.3 & X & - & -  \\ 
3FHLJ0811.9+0237 & 0.1726  & 4   & 0. $\pm$10   &  1013 & 3.6 & X & - & -  \\ 
3FHLJ0905.5+1357 & 0.2239: & 29  & 0.3$\pm$0.1  &  3767 & 9.1 & X & - & -  \\ 
3FHLJ1549.9-0659 & 0.418   & 17  & 0.2$\pm$0.1  &  824 & 3.5 & X & - & -  \\ 
3FHLJ1800.5+7827 & 0.683   & 284 & 0.2$\pm$0.1  &  643 & 5.1 & X & - & -  \\ 
3FHLJ1904.1+3627 & 0.08977 & 32  & 0.6$\pm$0.2  &  3298 & 5.7 & X & - & -  \\ 
3FHLJ1911.5-1908 & 0.138   & 24  & 0.4$\pm$0.2  &  3375 & 5.0 & X & - & -  \\ 
\hline
\end{tabular}
\end{center}
%\tablenotetext{}{
\raggedright
\footnotesize %\\
\texttt{Col.1}: Object name reported in the 3FHL catalog; \texttt{Col.2}: Redshift of the source \citep[(as reported in ][]{paiano2020} ; \texttt{Col.3-4}: Variability Index and Fractional Index from 4FGL catalog; \texttt{Col.5-6-7}: Possible detection in the VHE band by CTA (North or South), MAGIC and ASTRI mini array. 
\end{table*}

\setcounter{table}{1}
\begin{table*}
\begin{center}
\caption{3FHL objects of spectroscopic redshift lower limit from intervening absorption lines in their optical spectra.}\label{tab:results}
\begin{tabular}{llccccc} 
\hline
OBJECT           & z & Variability & Fractional  & CTA  & MAGIC & ASTRI \\
                 &   & Index       & variability &      &       &       \\
\hline                                                     
3FGLJ0141.4-0929 & $\geq$~0.501 - 0.873 & 257 & 0.5$\pm$0.1 & X & - & -  \\ 
3FHLJ0612.8+4122 & $\geq$1.107  - 1.805 & 82  & 0.2$\pm$0.1 & X & - & -   \\  
3FHLJ0706.5+3744 & $\geq$0.1042 - 0.199 & 8   & 0.$\pm$10   & X & X & -  \\ 
3FHLJ1253.1+5300 & $\geq$0.6638 - 1.13  & 116 & 0.2$\pm$0.1 & X & - & -   \\
3FHLJ1447.9+3608 & $\geq$0.738  - 1.245 & 25  & 0.2$\pm$0.1 & X & - & - \\ 
\hline
\end{tabular}
\end{center}
%\tablenotetext{}{
\raggedright
\footnotesize %\\
\texttt{Col.1}: Object name reported in the 3FHL catalog; \texttt{Col.2}: Redshift interval between the spectroscopic redshift lower limit of the source, derived by intervening absorption lines \citep[as reported in ][]{paiano2020}, and the redshift corresponding to a distance a factor 2 larger; \texttt{Col.3-4}: Variability Index and Fractional Index from 4FGL catalog ;\texttt{Col.5-6-7}: Possible detection in the VHE band by CTA (North or South), MAGIC and ASTRI mini array. 
\end{table*}

\setcounter{table}{2}
\begin{table*}
\begin{center}
\caption{3FHL objects of unknown redshift. }\label{tab:results}
\begin{tabular}{llccccc} 
\hline
OBJECT           & z & Variability & Fractional  & CTA  & MAGIC & ASTRI \\
                 &   & Index       & variability &      &       &       \\
\hline             
3FHLJ0009.4+5030 & ($>$~0.60 - 1.025) & 149 & 0.4$\pm$0.1 & X & - & - \\ 
3FHLJ0131.1+6120 & ($>$~0.10 - 0.192)  & 74 & 0.5$\pm$0.1 & X & X & X \\ 
3FHLJ0134.4+2638 & ($>$~0.15 - 0.283)  & 11 & 0.1$\pm$0.1 & X & - & - \\
3FHLJ0322.0+2336 & ($>$~0.25 - 0.458)  & 18 & 0.2$\pm$0.1 & X & - & - \\
3FHLJ0433.1+3227 & ($>$~0.45 - 0.790)  & 13 & 0.3$\pm$0.3 & - & - & - \\
3FHLJ0434.7+0921 & ($>$~0.10 - 0.192)  & 34 & 0.5$\pm$0.2 & X & - & - \\
3FHLJ0515.8+1528 & ($>$~0.20 - 0.371)  & 41 & 0.4$\pm$0.1 & X & - & - \\
3FHLJ0540.5+5823 & ($>$~0.10 - 0.192)  & 16 & 0.2$\pm$0.1 & X & - & - \\
3FHLJ0607.4+4739 & ($>$~0.10 - 0.192) & 102 & 0.3$\pm$0.1 & X & - & - \\
3FHLJ0702.6-1950 & ($>$~0.10 - 0.192) & 163 & 0.6$\pm$0.2 & X & - & - \\
3FHLJ0816.4-1311 & ($>$~0.40 - 0.709)  & 85 & 0.4$\pm$0.1 & X & X & - \\
3FHLJ0910.5+3329 & ($>$~0.15 - 0.283)  & 21 & 0.2$\pm$0.1 & X & - & - \\
3FHLJ0953.0-0840 & ($>$~0.15 - 0.283)  & 40 & 0.2$\pm$0.1 & X & X & -  \\
3FHLJ1037.6+5711 & ($>$~0.25 - 0.458) & 153 & 0.3$\pm$0.1 & X & X & - \\
3FHLJ1055.6-0125 & ($>$~0.55 - 0.951)  & 10 & 0.2$\pm$0.3 & - & - & - \\
3FHLJ1059.1-1134 & ($>$~0.10 - 0.192) & 120 & 0.3$\pm$0.1 & X & - & -\\
3FHLJ1150.5+4154 & ($>$~0.25 - 0.458)  & 64 & 0.3$\pm$0.1 & X & X & - \\
3FHLJ1233.7-0145 & ($>$~0.10 - 0.192)  & 29 & 0.3$\pm$0.1 & X & - & - \\
3FHLJ1418.4-0233 & ($>$~0.12 - 0.228)  & 86 & 0.3$\pm$0.1 & X & X & X \\
3FHLJ1445.0-0326 & ($>$~0.45 - 0.790)  & 10 & 0.04$\pm$0.4& X & - & - \\
3FHLJ1454.5+5124 & ($>$~0.40 - 0.709) & 399 & 0.5$\pm$0.1 & X & - & - \\
3FHLJ1503.7-1541 & ($>$~0.10 - 0.192)  & 14 & 0.2$\pm$0.1 & X & X & X \\
3FHLJ1748.6+7006 & ($>$~0.30 - 0.543) & 541 & 0.4$\pm$0.1 & X & X & - \\
3FHLJ1841.3+2909 & ($>$~0.10 - 0.192)  & 8  & 0.$\pm$10   & X & X & -  \\
3FHLJ1921.8-1607 & ($>$~0.12 - 0.228)  & 42 & 0.4$\pm$0.1 & X & X & -  \\
\hline
\end{tabular}
\end{center}
%\tablenotetext{}{
\raggedright
\footnotesize %\\
\texttt{Col.1}: Object name reported in the 3FHL catalog; \texttt{Col.2}: Redshift interval between the redshift lower limit of the source estimated by the lack of detection of absorption lines due to the stellar population of the blazar host galaxy \citep[as reported in ][]{paiano2020} and the redshift corresponding to a distance a factor 2 larger; \texttt{Col.3-4}: Variability Index and Fractional Index from 4FGL catalog ;\texttt{Col.5-6-7}: Possible detection in the VHE band by CTA (North or South), MAGIC and ASTRI mini array.     %}
\end{table*}

\newpage
\setcounter{figure}{0}
\begin{figure*}%[htbp]
\includegraphics[width=0.35\textwidth,angle=-90]{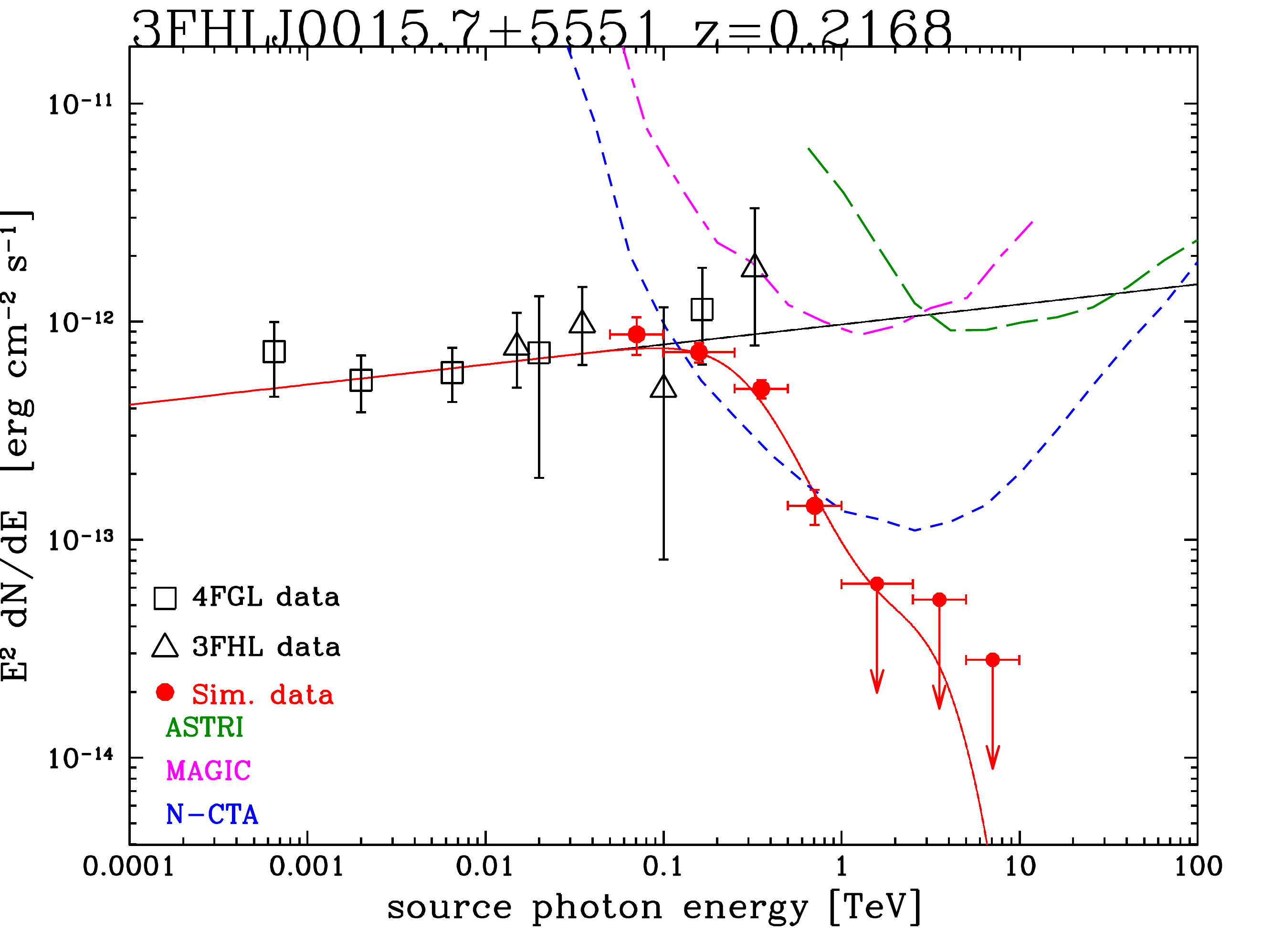}
\includegraphics[width=0.35\textwidth,angle=-90]{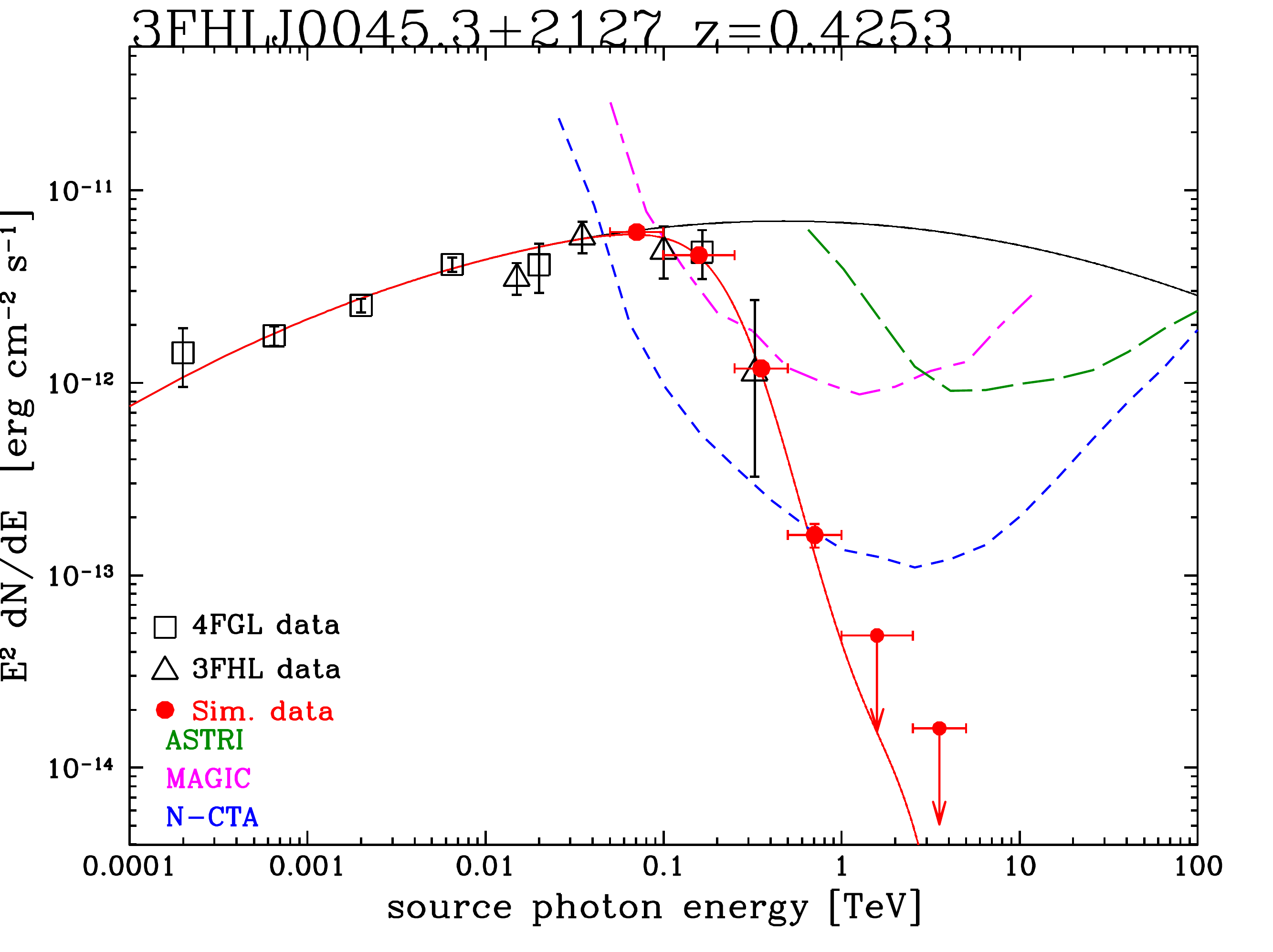}
\includegraphics[width=0.35\textwidth,angle=-90]{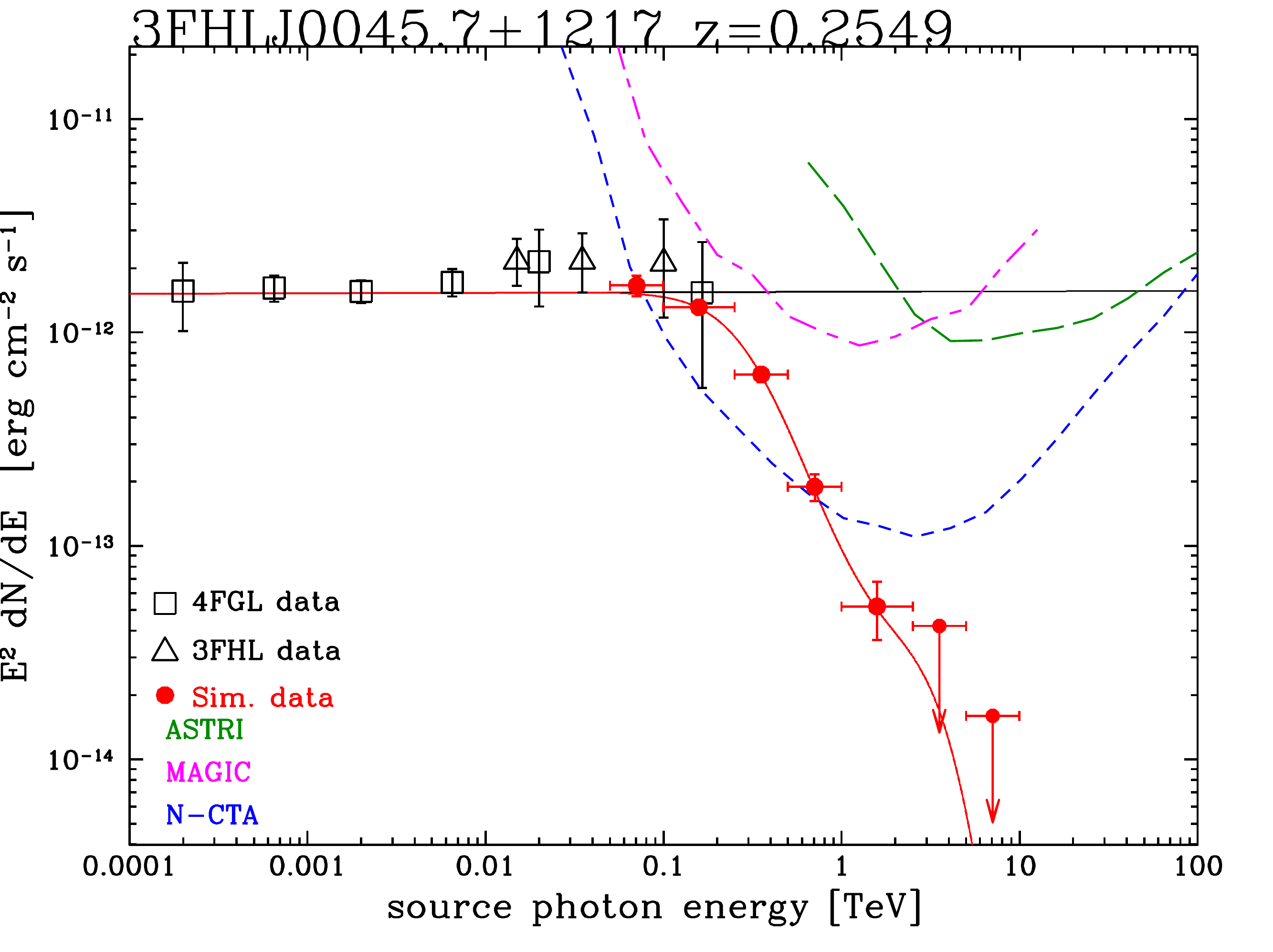}
\includegraphics[width=0.35\textwidth,angle=-90]{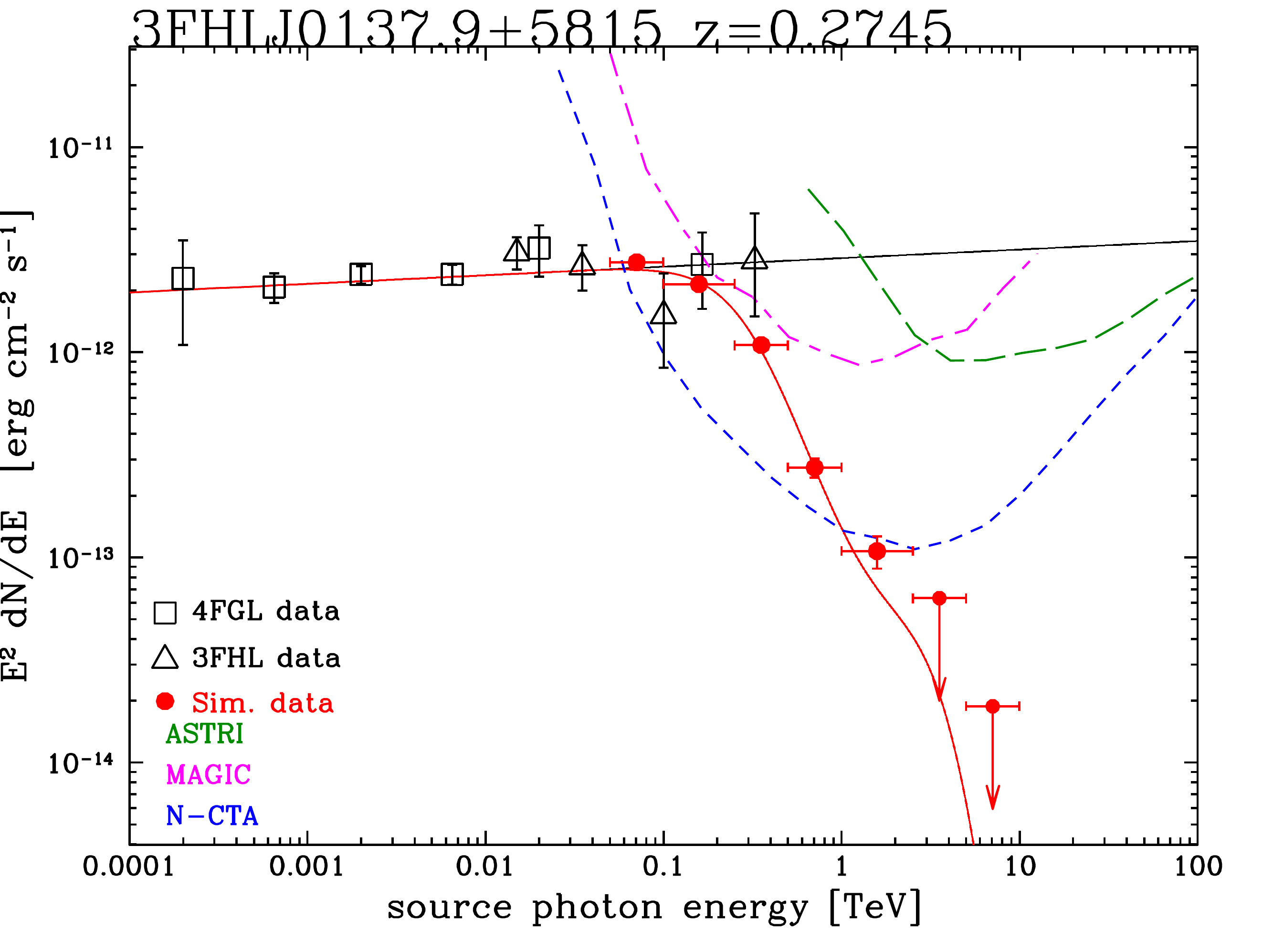}
\includegraphics[width=0.35\textwidth,angle=-90]{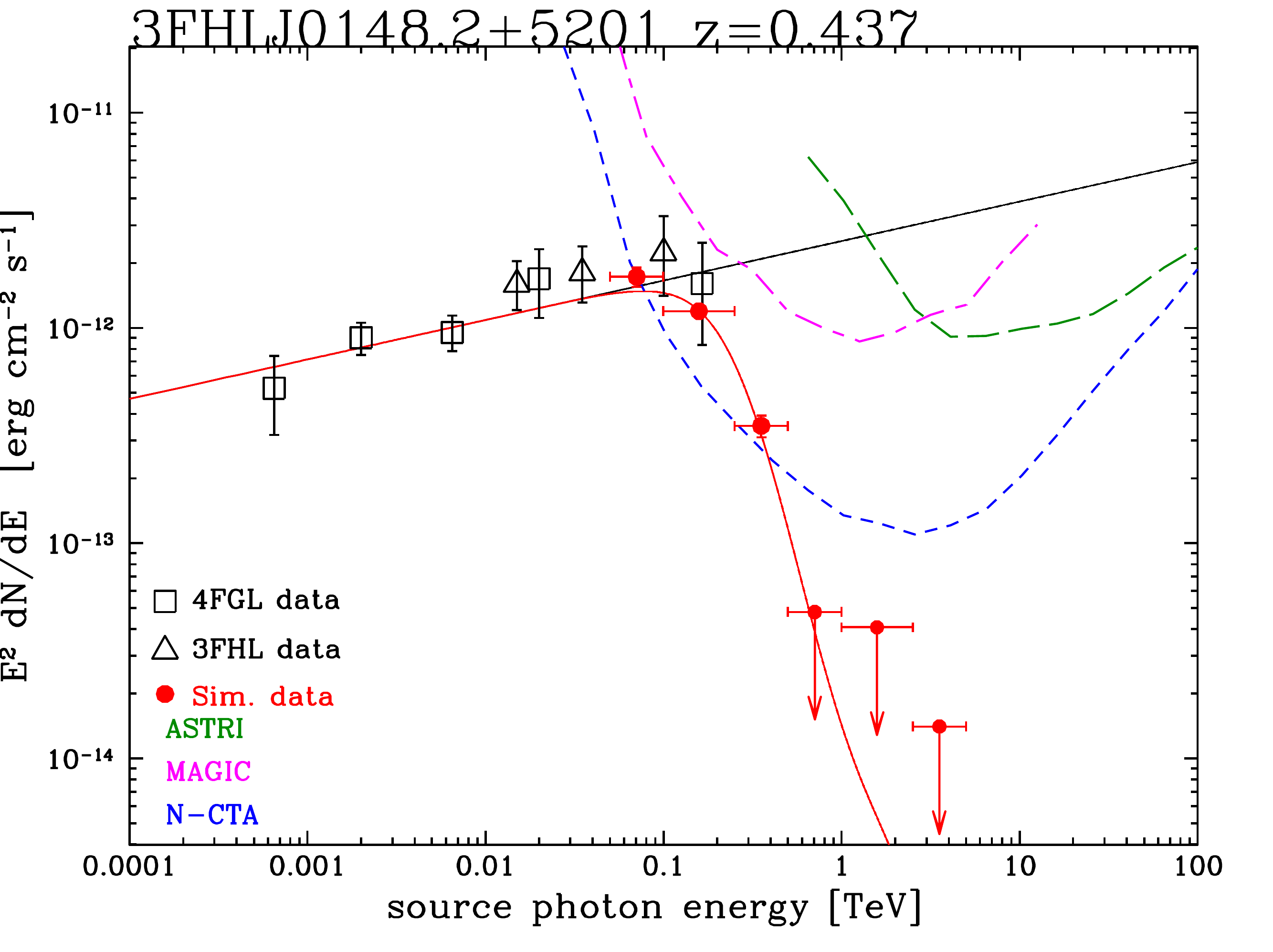}
\includegraphics[width=0.35\textwidth,angle=-90]{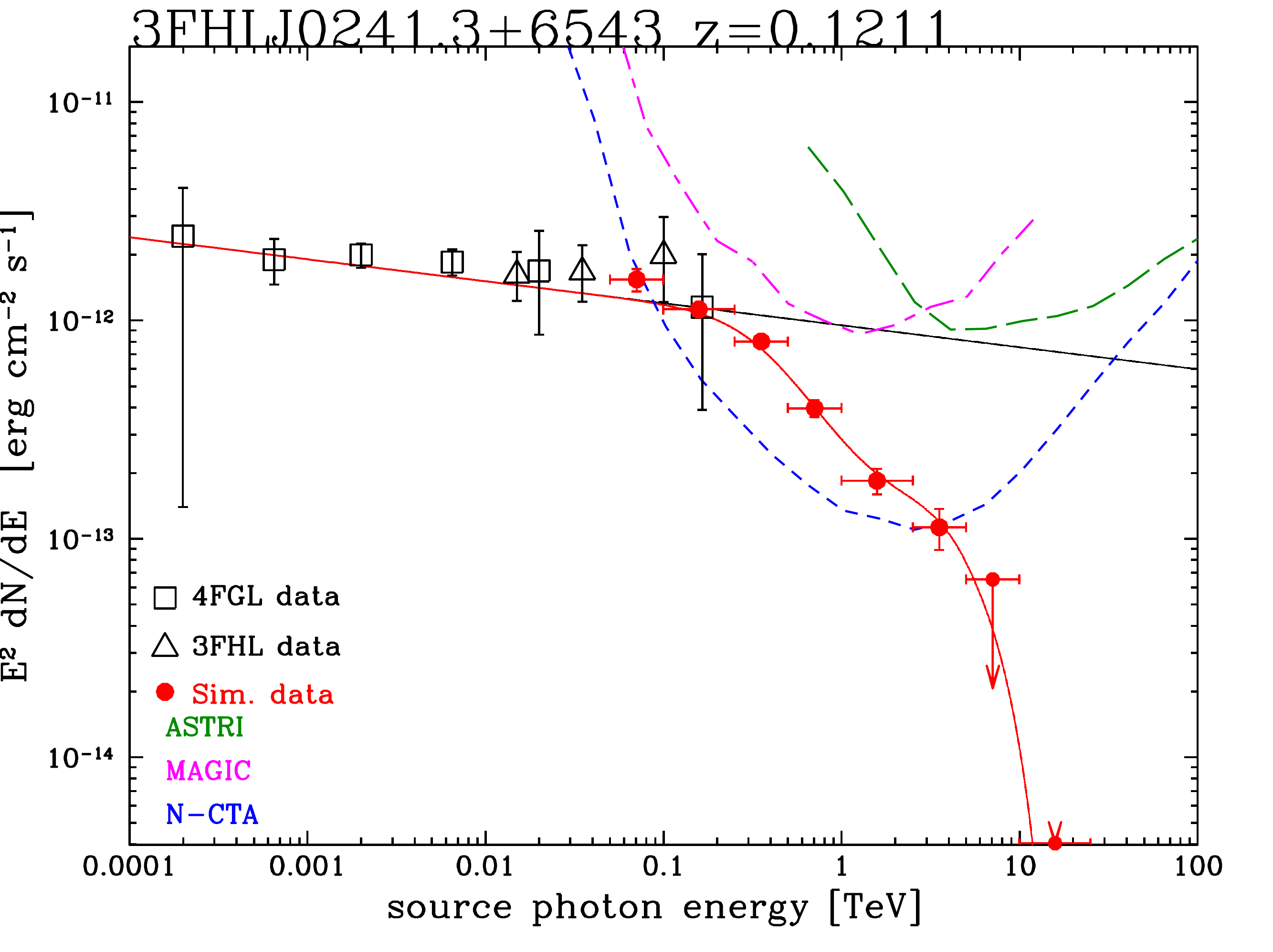}
\caption{ VHE extrapolation before (black solid curve) and after (red solid curve) the EBL model correction (see text in Sect. 2 and 3 for details) for the 25 sources of well-known redshift of the BLL sample of Paper~I. We overplot the sensitivity curves at 5$\sigma$ for 50 hour and zenith angle $<$~20 degrees for CTA-north or CTA-south (blue short dashed curve),  for MAGIC at 5$\sigma$ for 50 hour and zenith angle $<$~35 degrees (magenta dashed-dotted curve) and for ASTRI mini array (green long dashed curve). We superpose the \textit{Fermi} flux data from the 4FGL catalogue (black opened squared), the 3FHL catalogue (black opened triangles) and the simulated CTA spectra (red points). For the latter, data points and upper limits (red arrows) are all at $3\sigma$ significance.}
\label{fig:fig1}
\end{figure*}%[htbp]

\newpage

\setcounter{figure}{0}
\begin{figure*}%[htbp]
\includegraphics[width=0.35\textwidth,angle=-90]{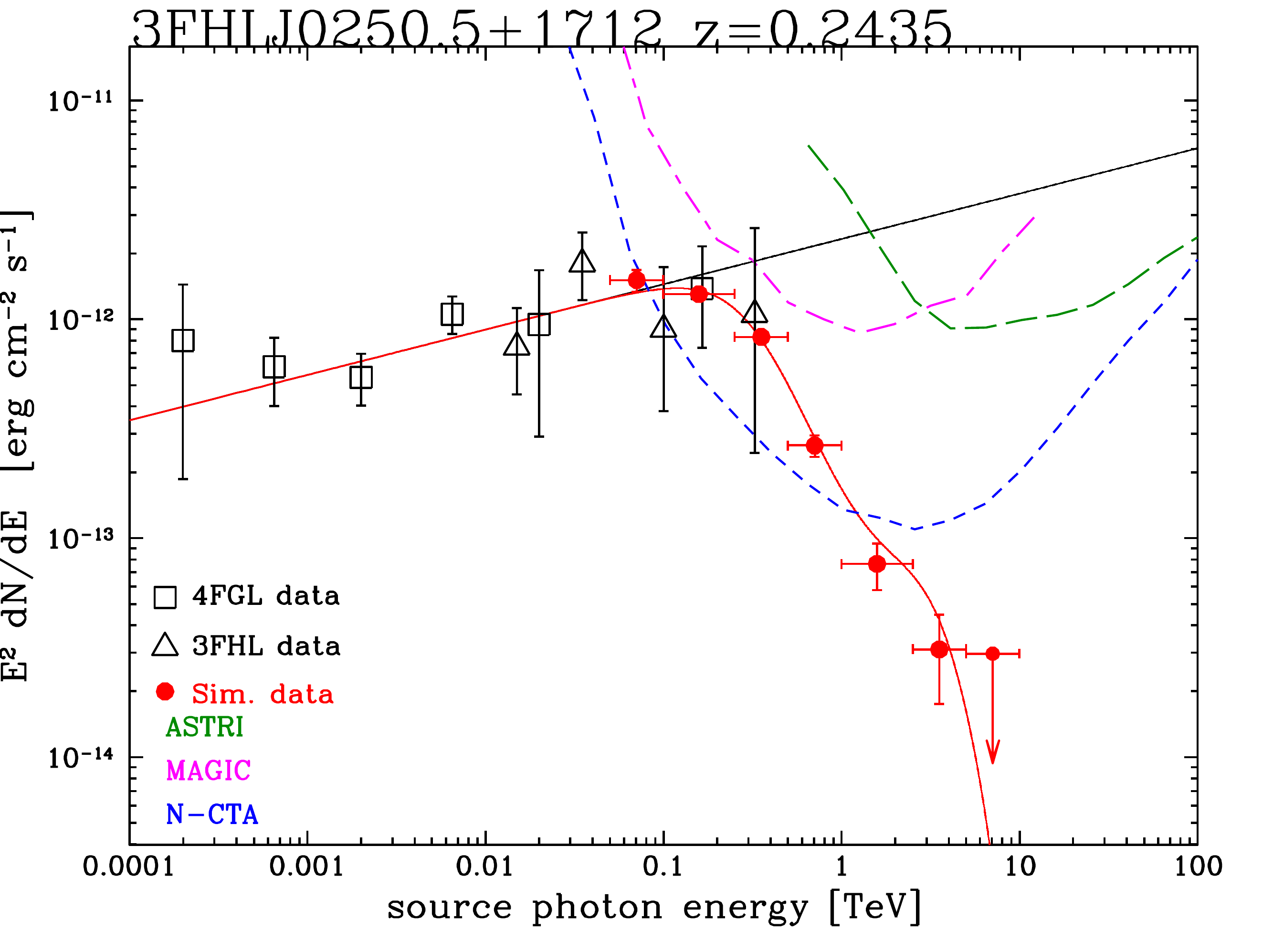}
\includegraphics[width=0.35\textwidth,angle=-90]{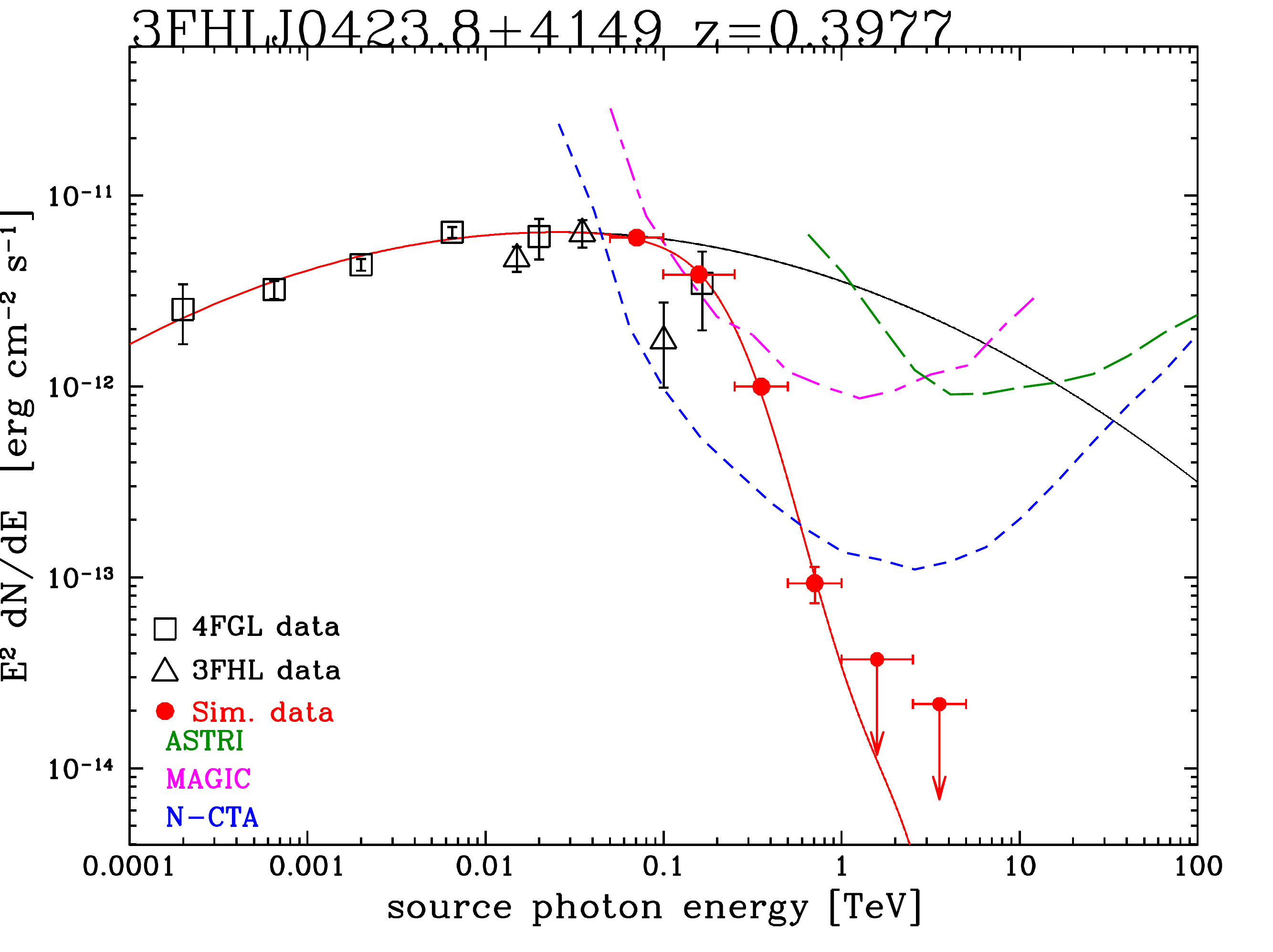}
\includegraphics[width=0.35\textwidth,angle=-90]{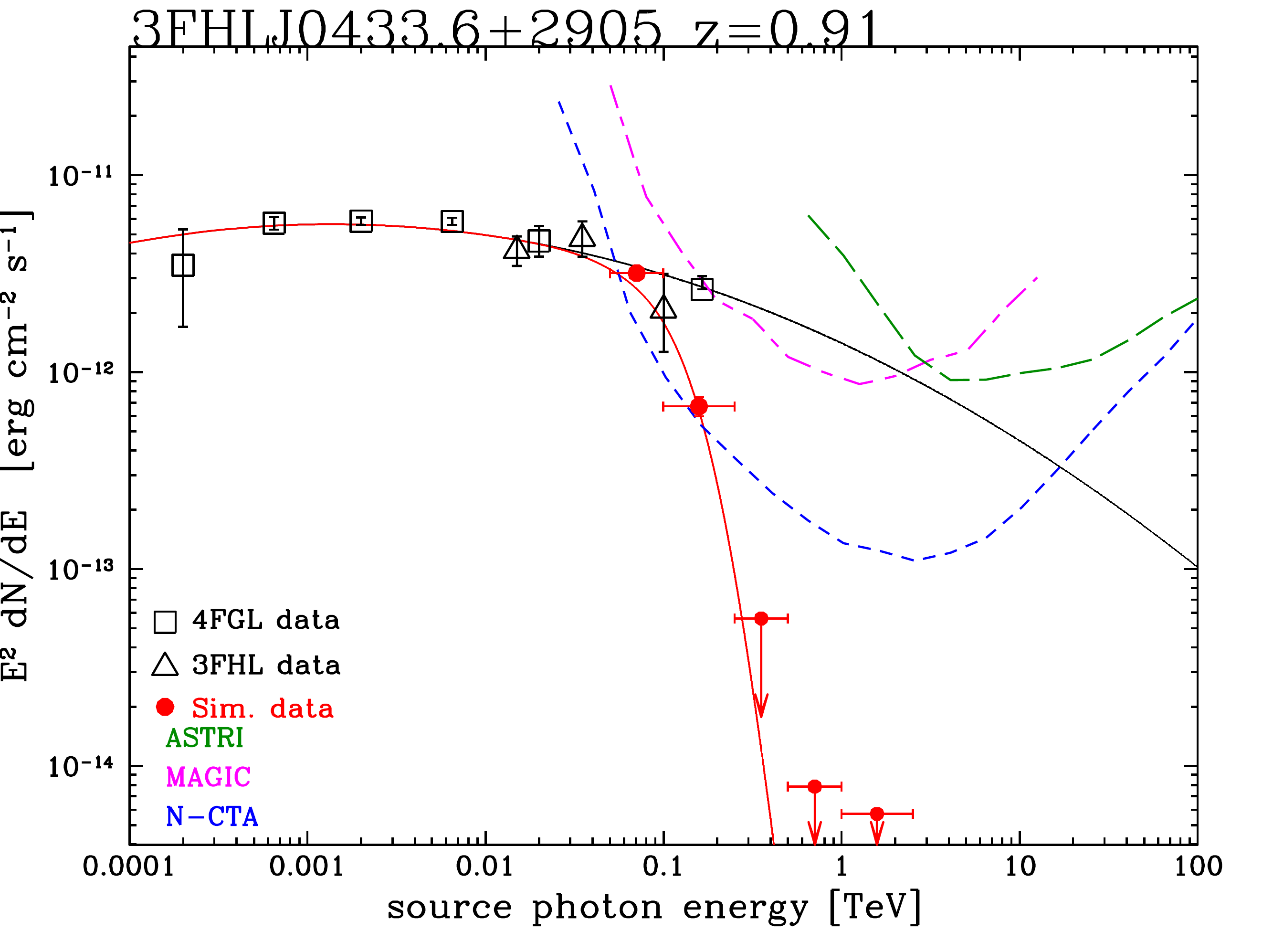}
\includegraphics[width=0.35\textwidth,angle=-90]{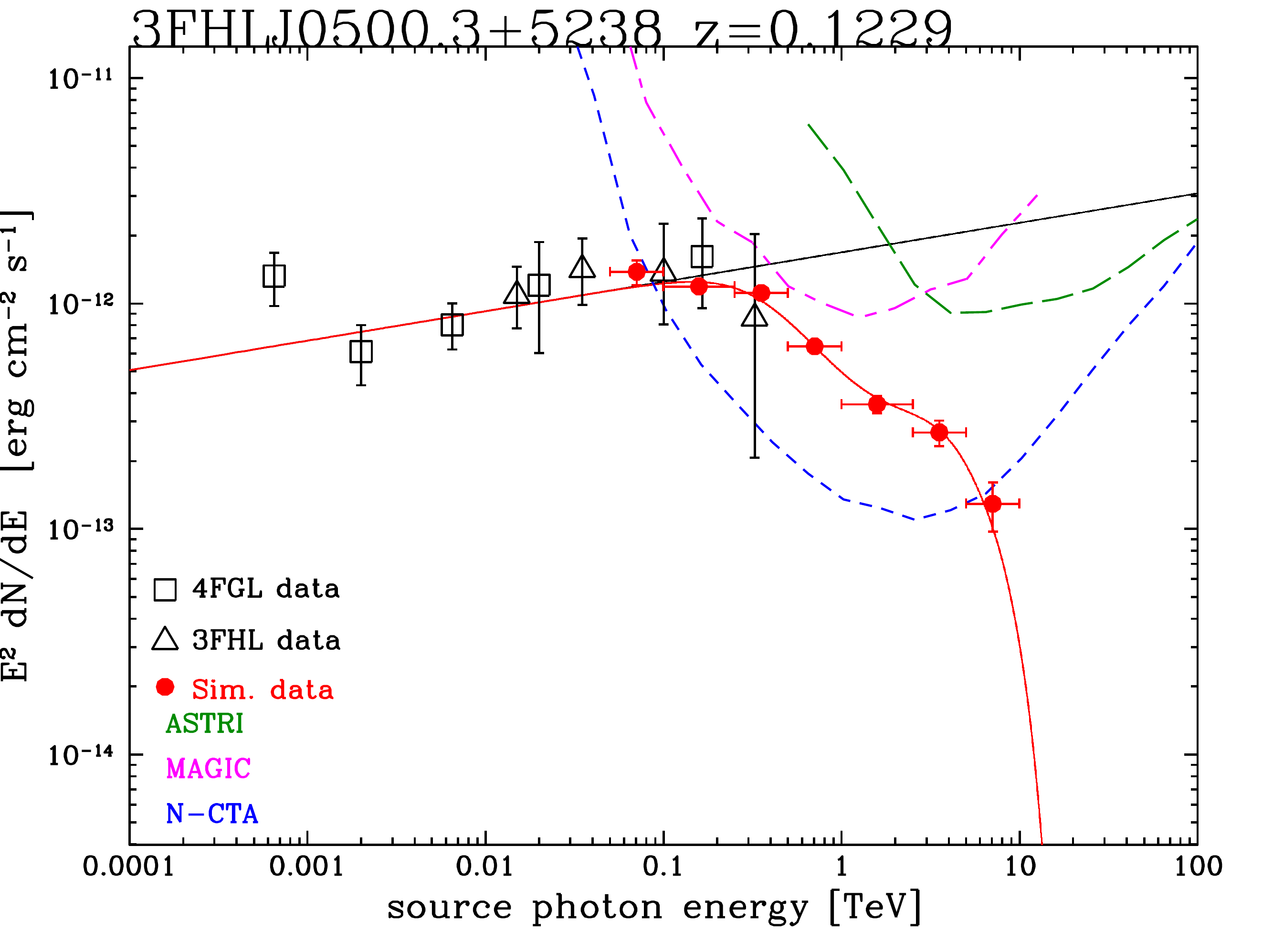}
\includegraphics[width=0.35\textwidth,angle=-90]{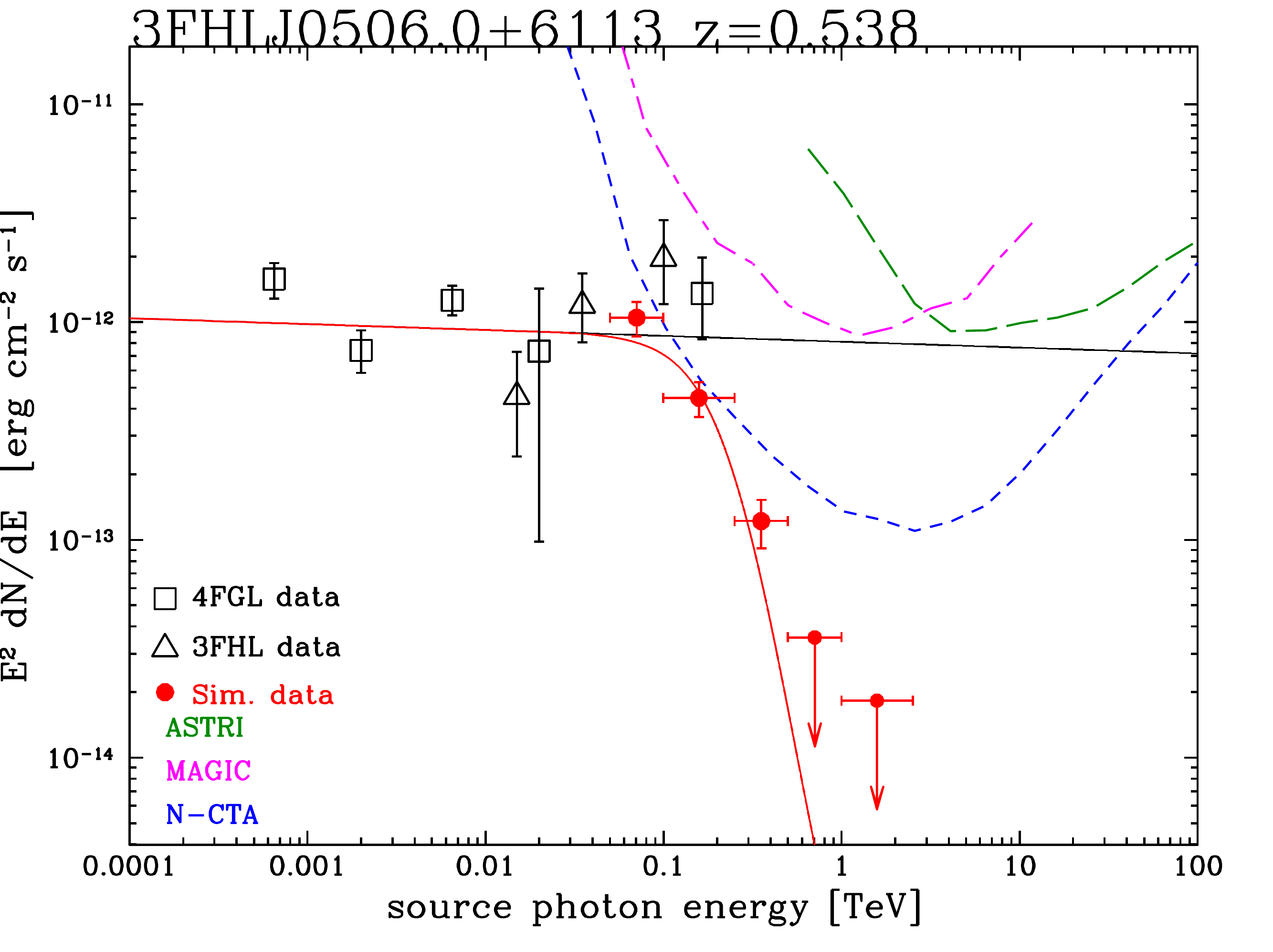}
\includegraphics[width=0.35\textwidth,angle=-90]{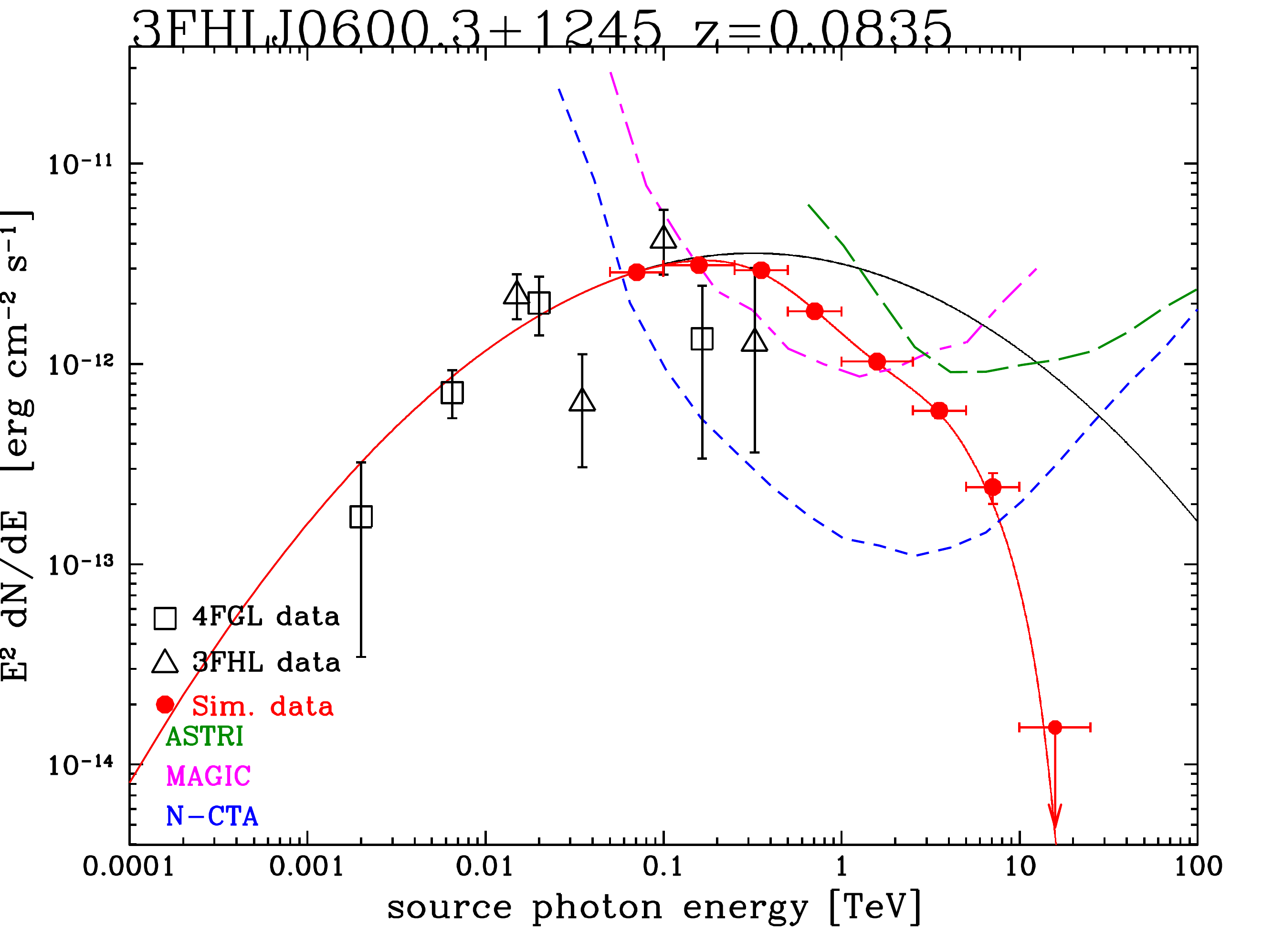}
\caption{Continued from Fig.~1.}
\label{fig:fig1}
\end{figure*}%[htbp]

\setcounter{figure}{0}
\begin{figure*}%[htbp]
\includegraphics[width=0.35\textwidth,angle=-90]{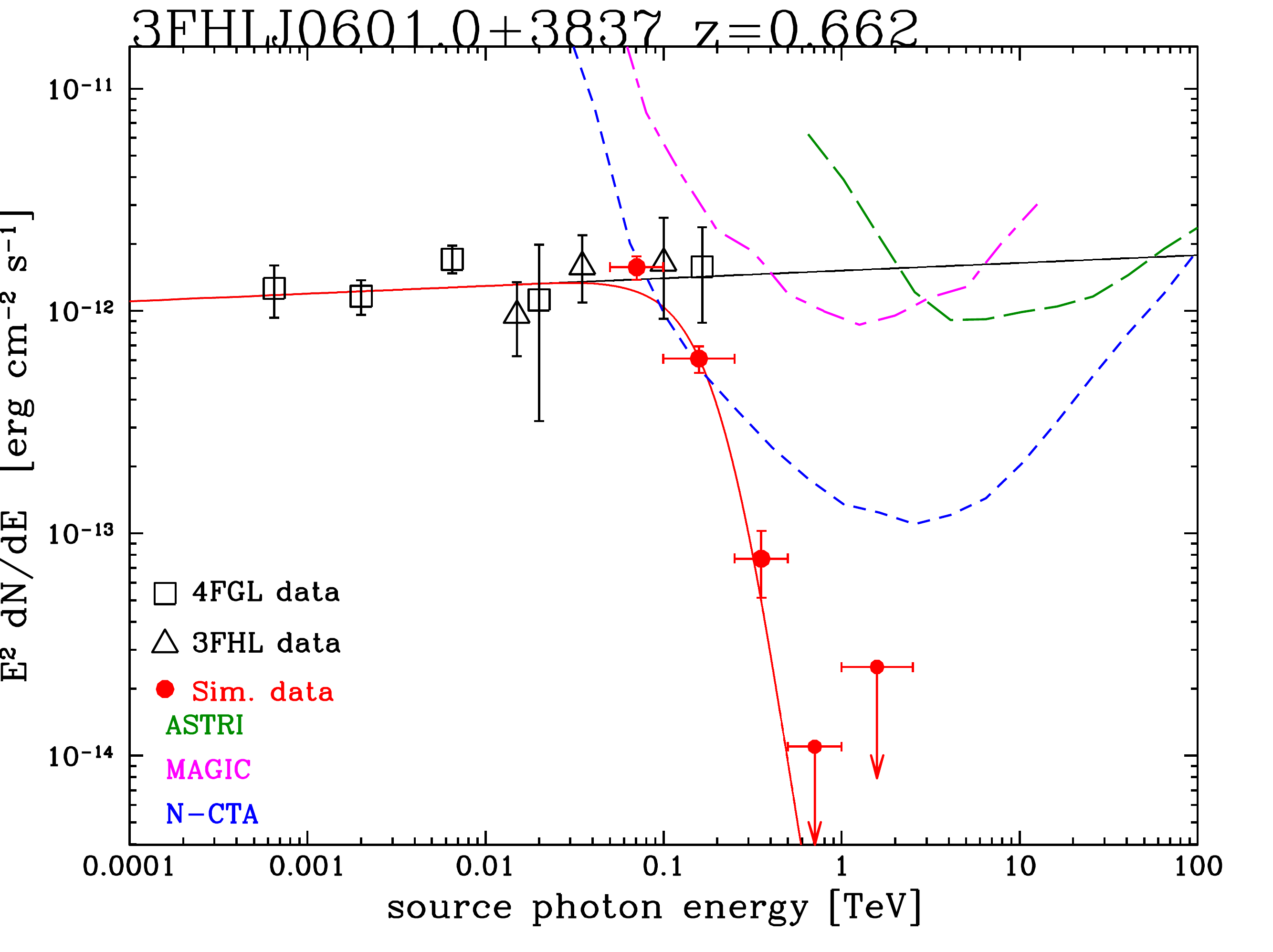}
\includegraphics[width=0.35\textwidth,angle=-90]{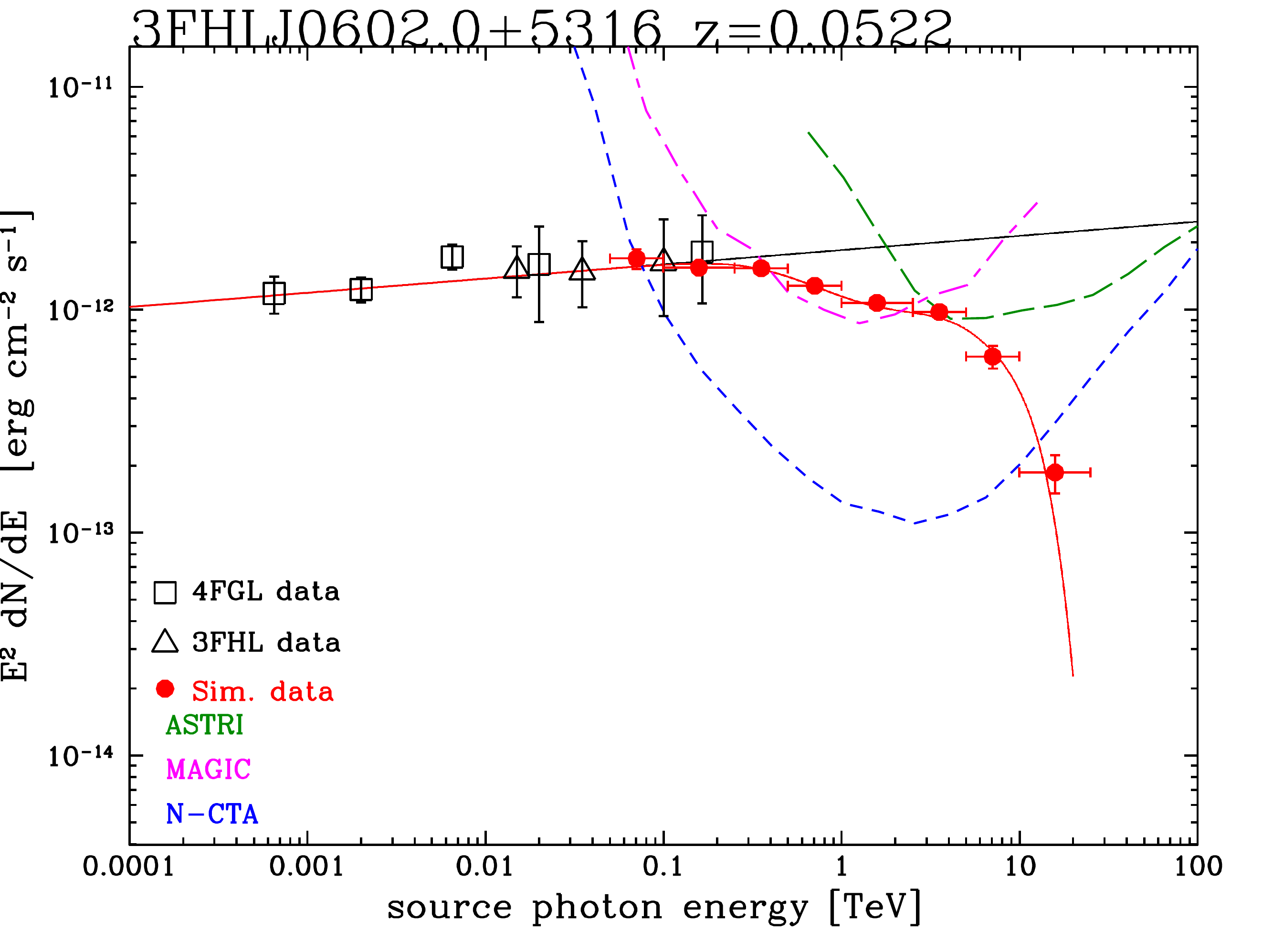}
\includegraphics[width=0.35\textwidth,angle=-90]{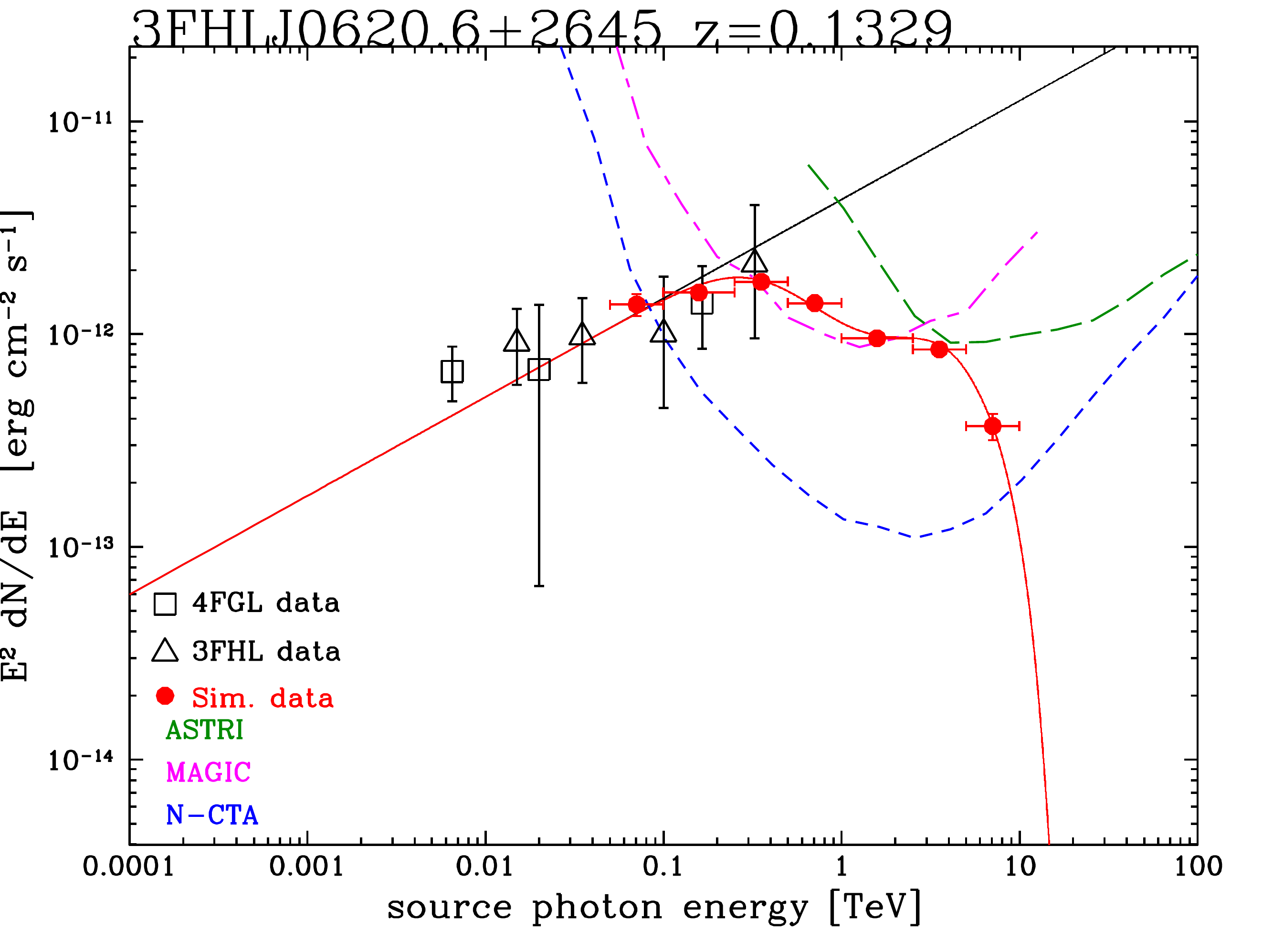}
\includegraphics[width=0.35\textwidth,angle=-90]{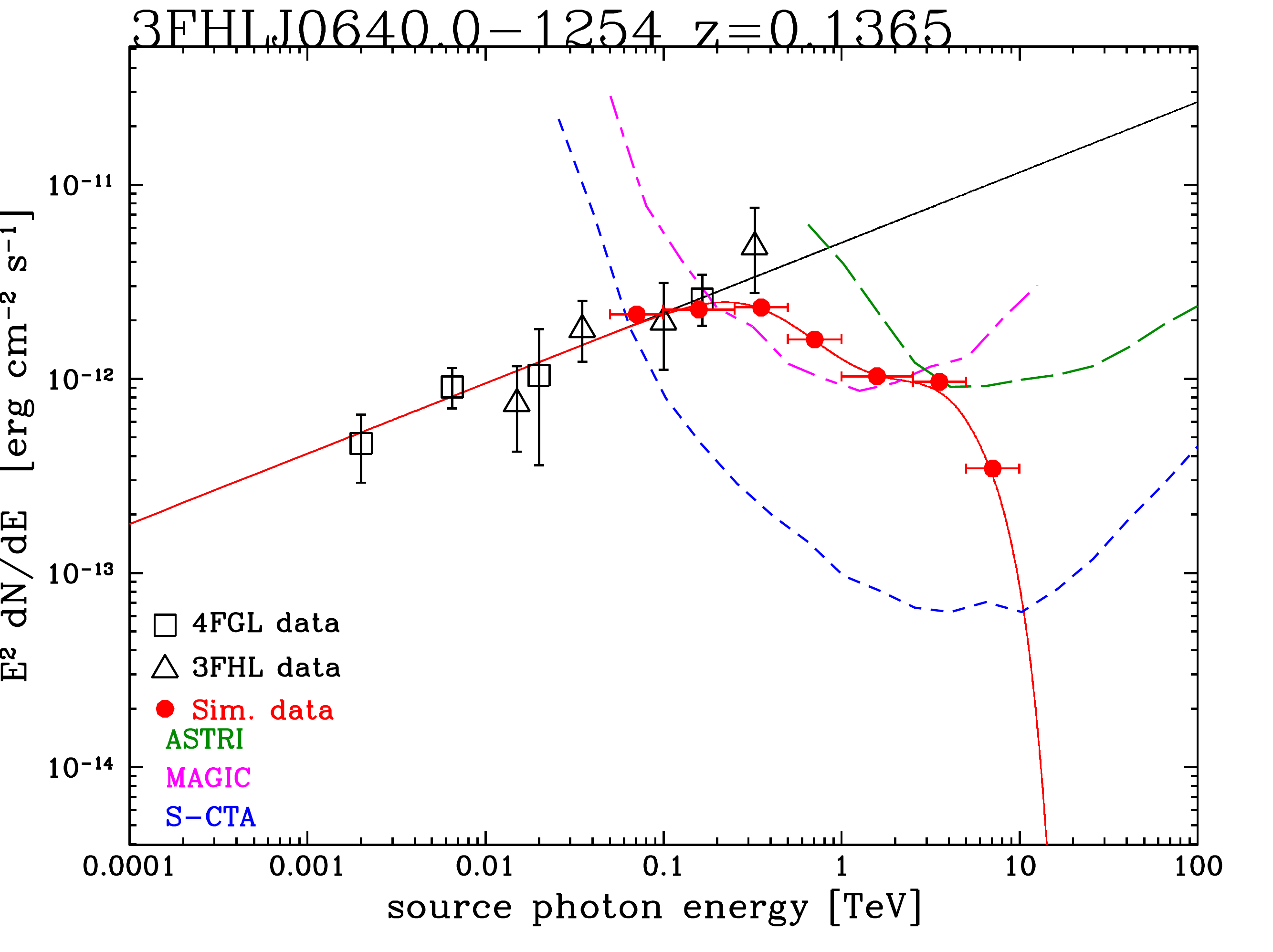}
\includegraphics[width=0.35\textwidth,angle=-90]{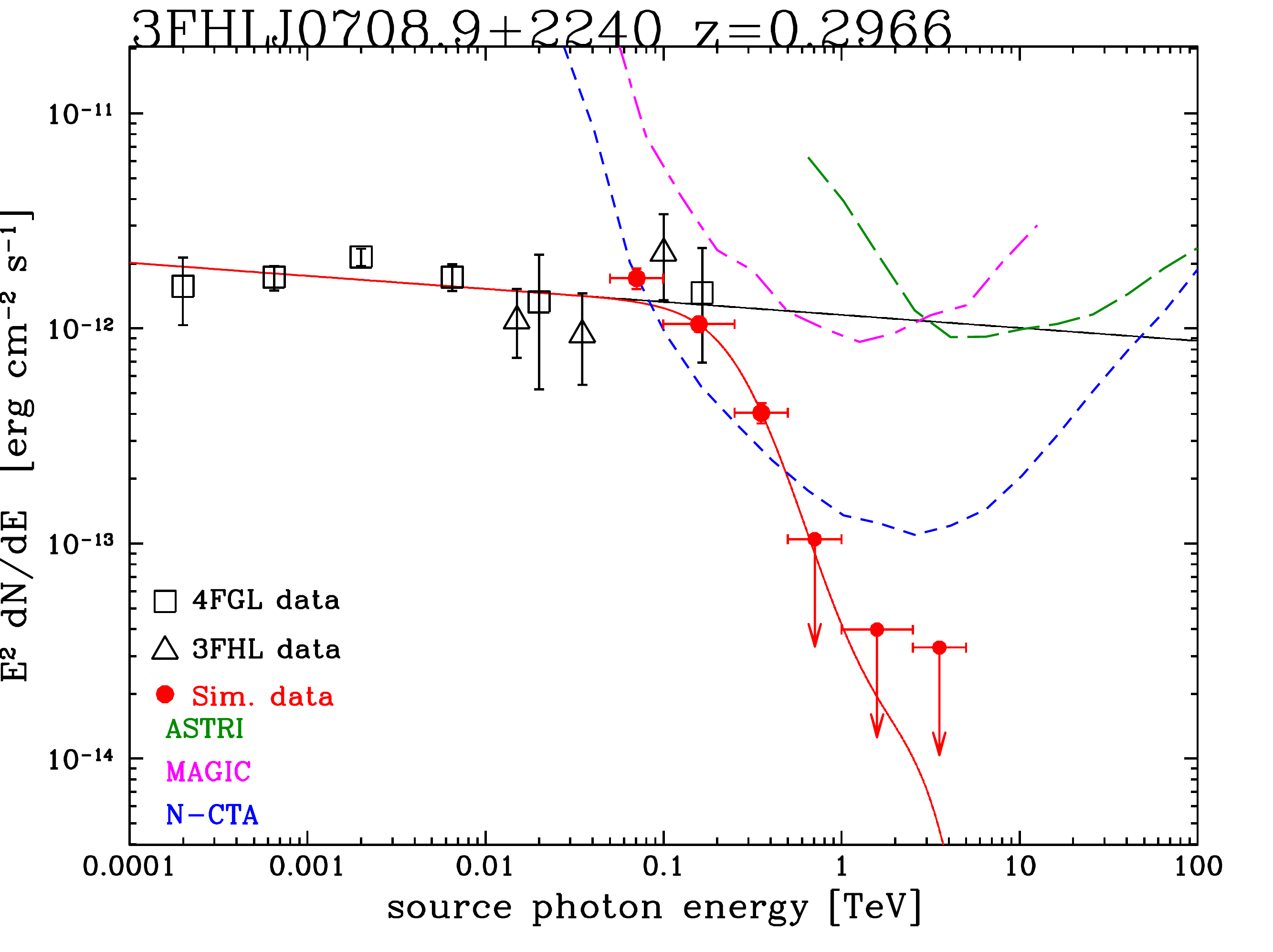}
\includegraphics[width=0.35\textwidth,angle=-90]{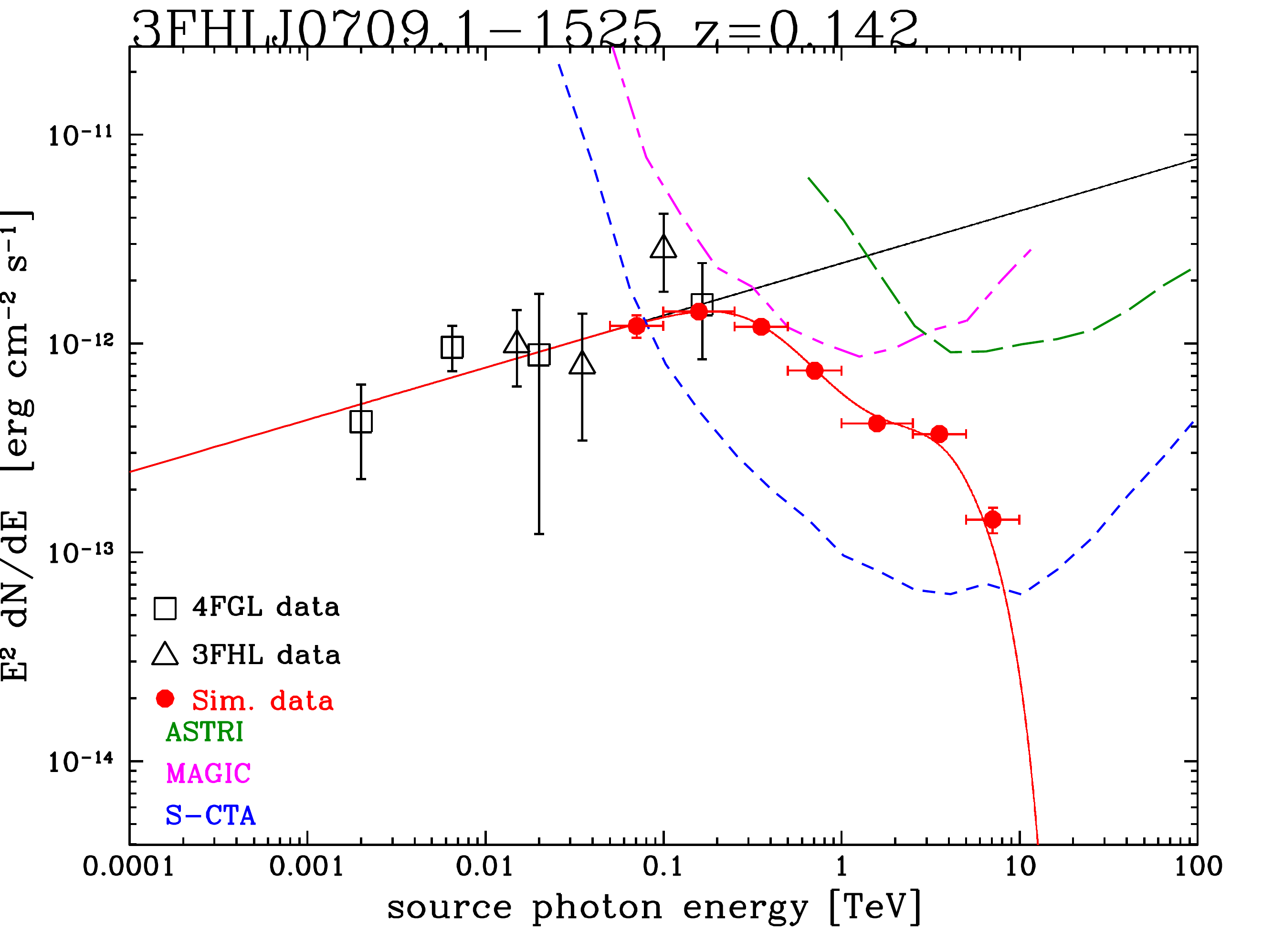}
\caption{Continued from Fig.~1.}
\label{fig:fig1}
\end{figure*}%[htbp]

\setcounter{figure}{0}
\begin{figure*}%[htbp]
\includegraphics[width=0.35\textwidth,angle=-90]{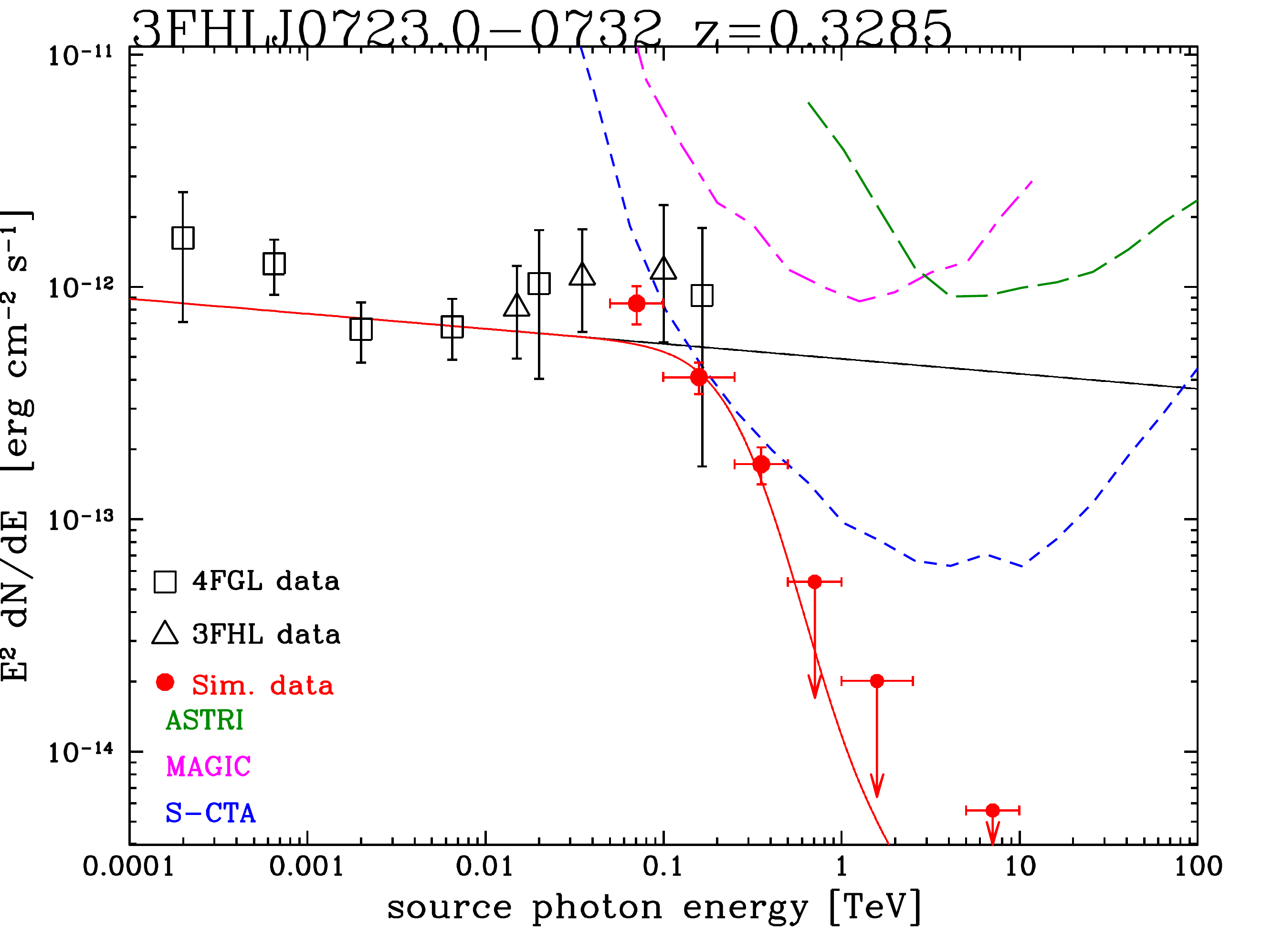}
\includegraphics[width=0.35\textwidth,angle=-90]{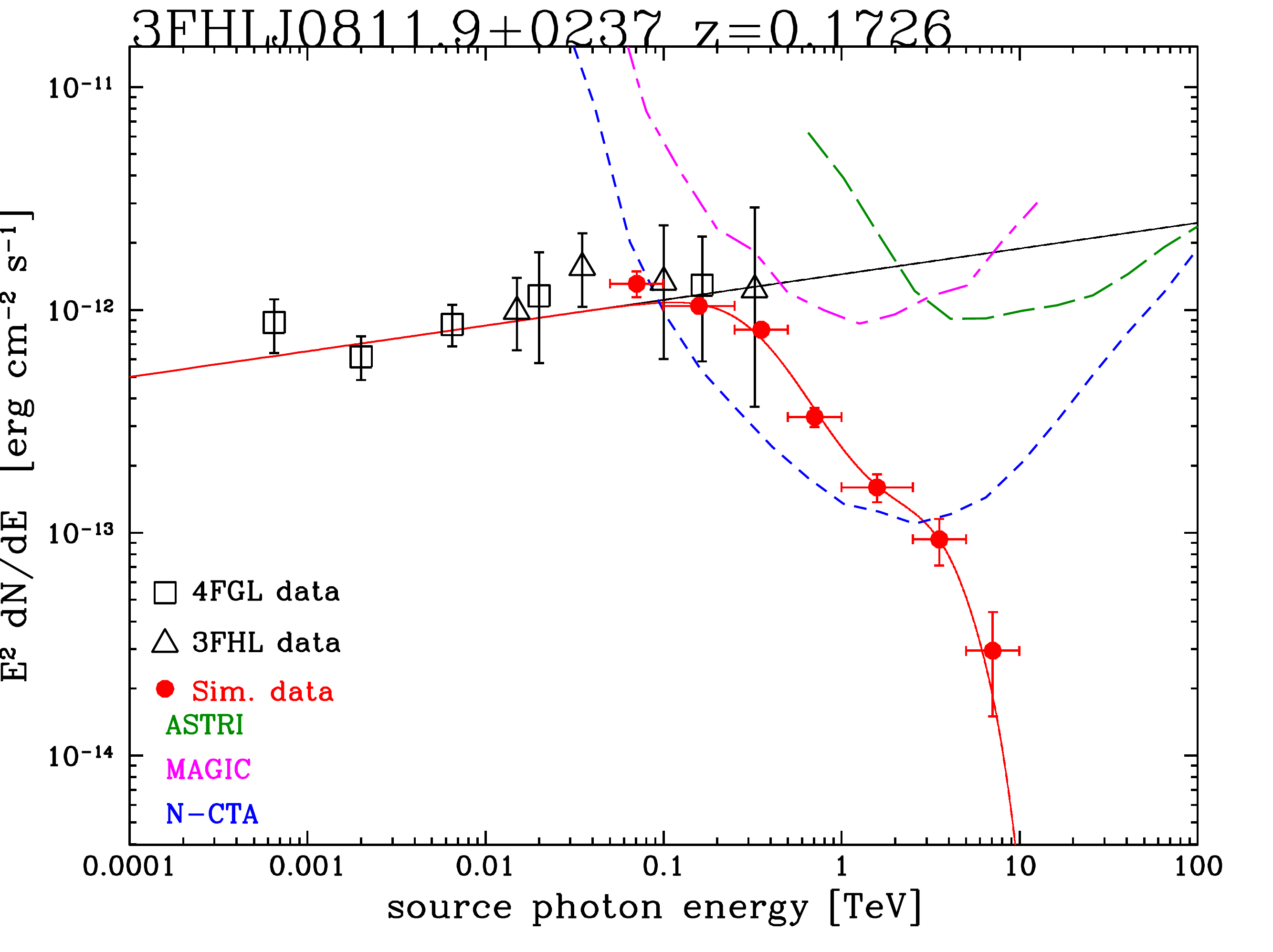}
\includegraphics[width=0.35\textwidth,angle=-90]{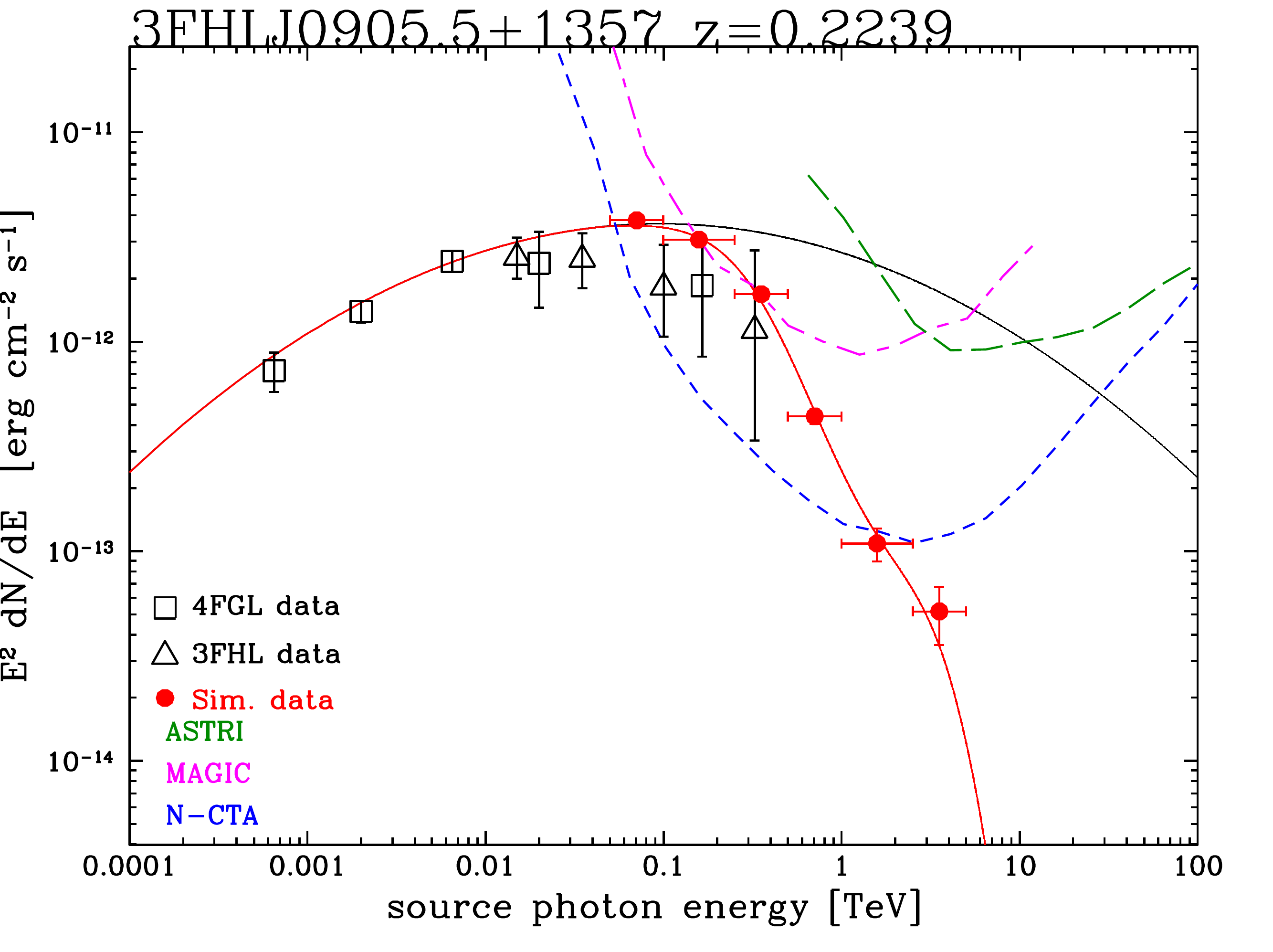}
\includegraphics[width=0.35\textwidth,angle=-90]{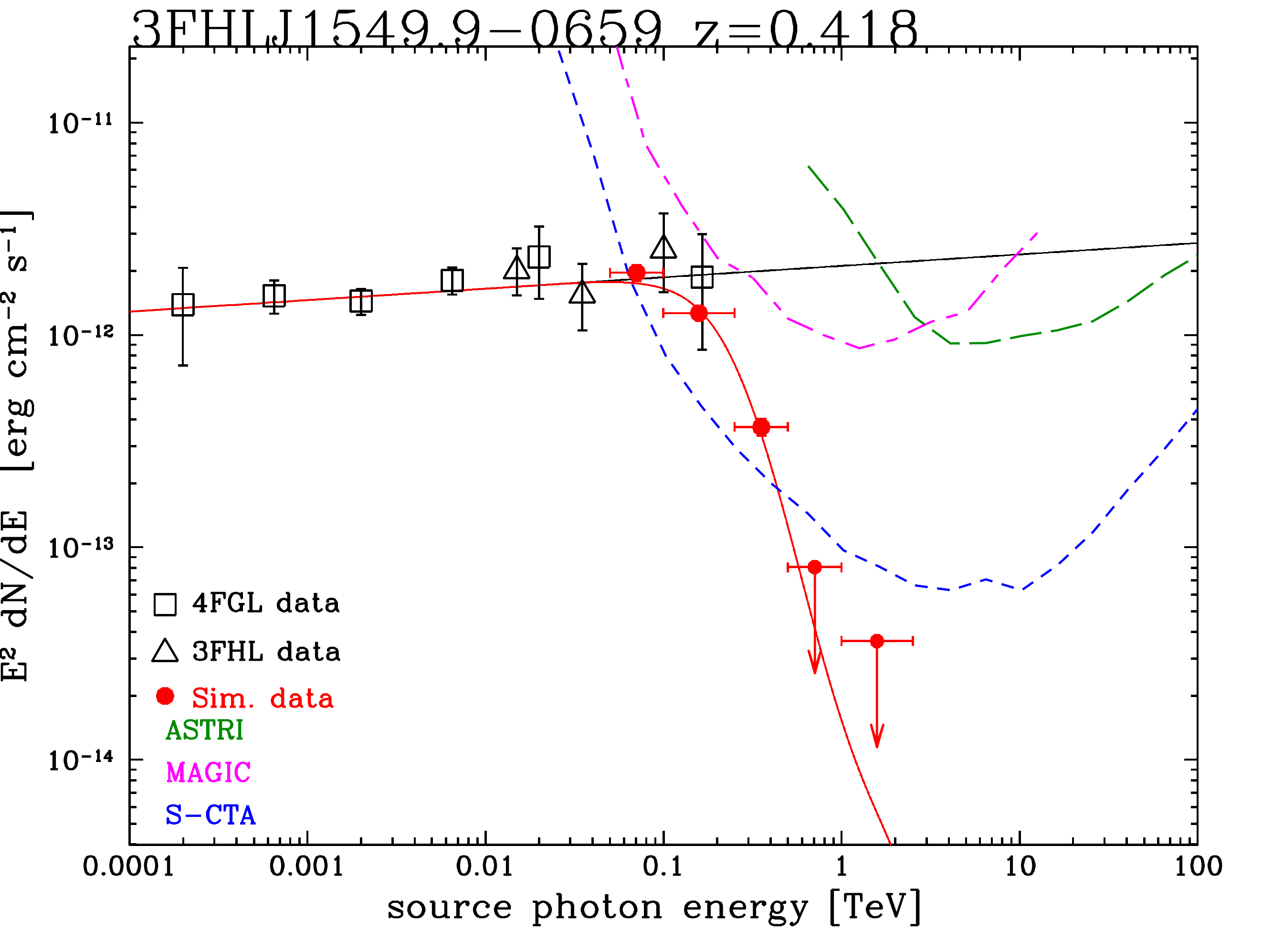}
\includegraphics[width=0.35\textwidth,angle=-90]{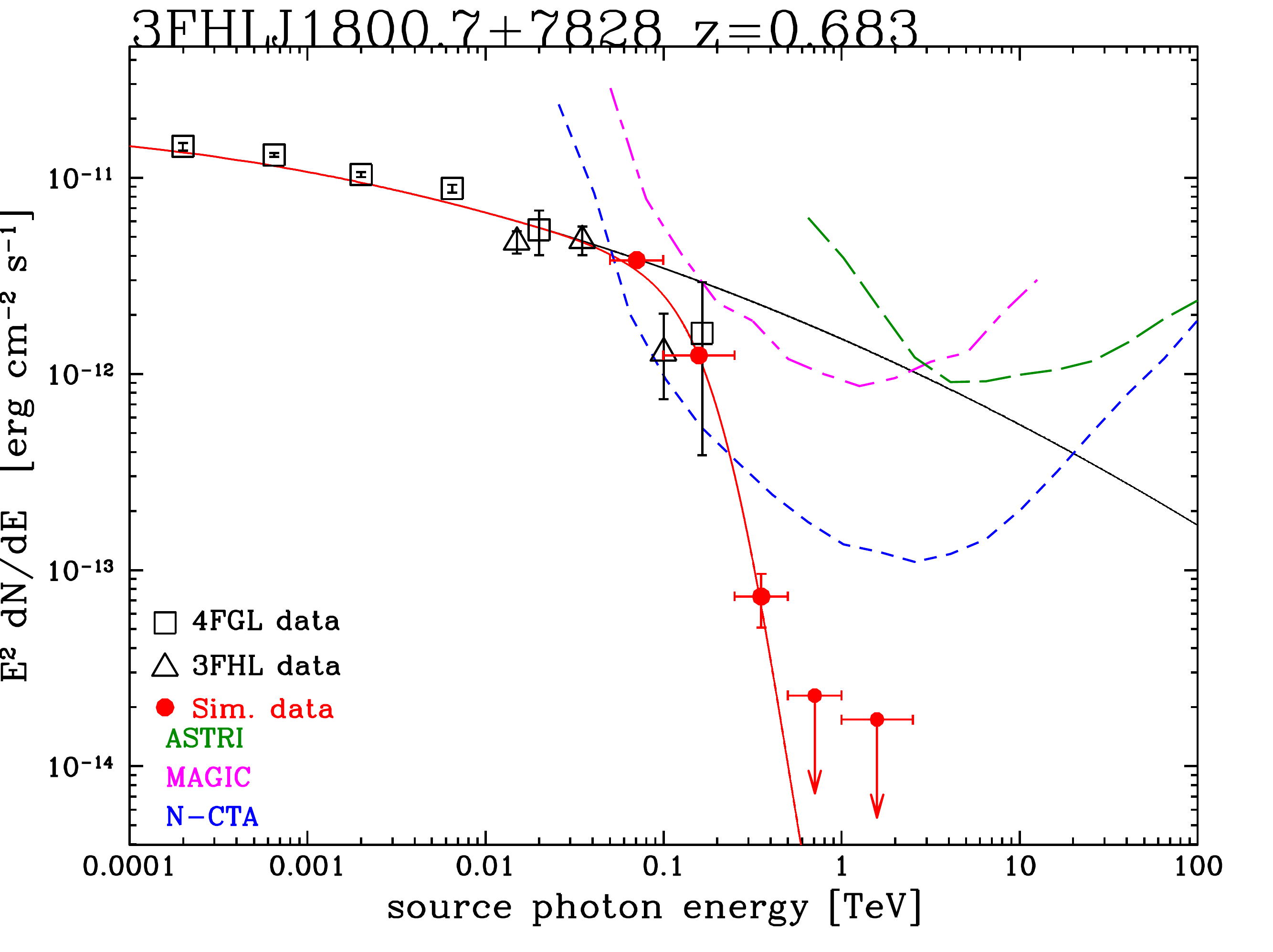}
\includegraphics[width=0.35\textwidth,angle=-90]{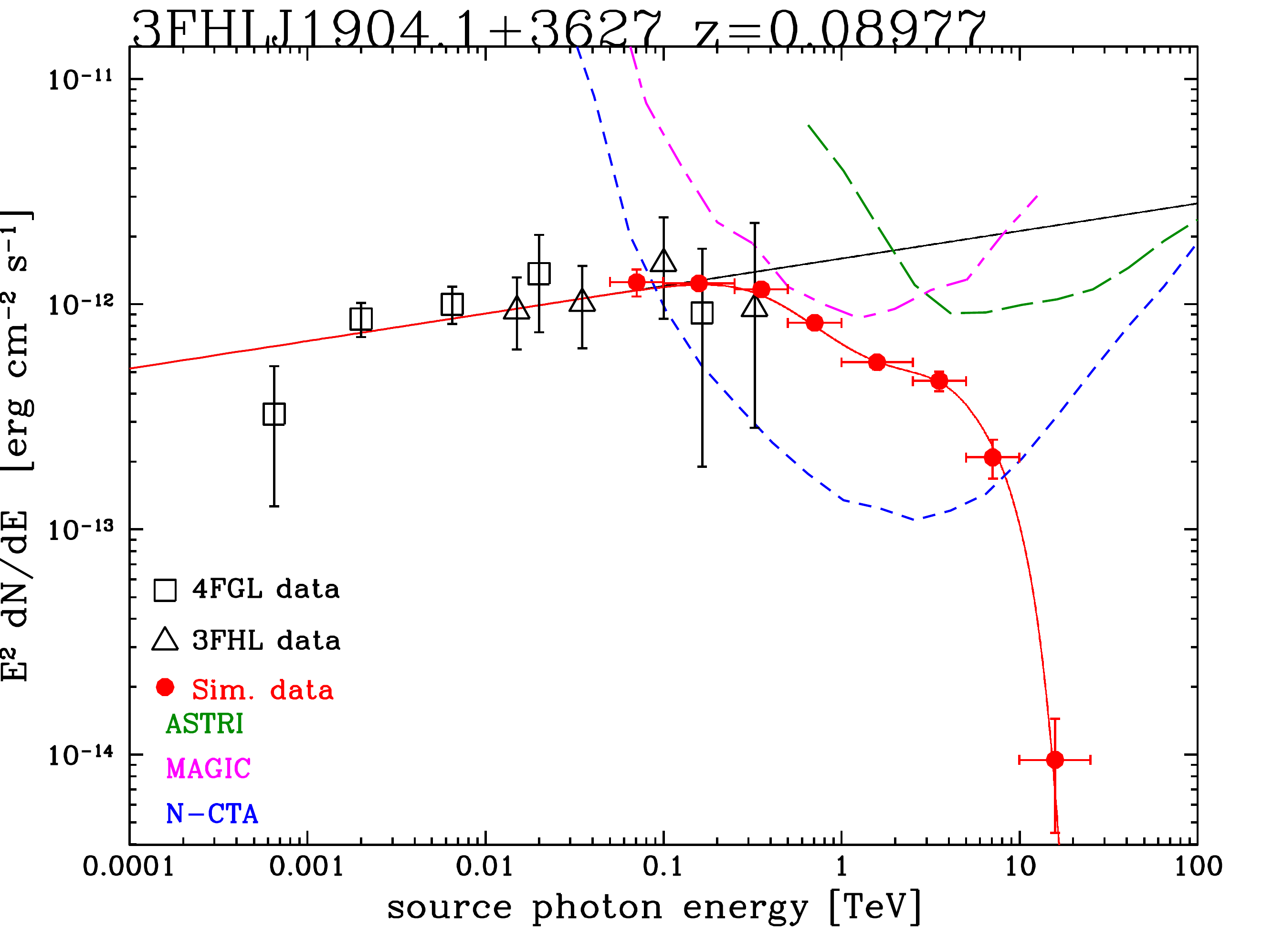}
\includegraphics[width=0.35\textwidth,angle=-90]{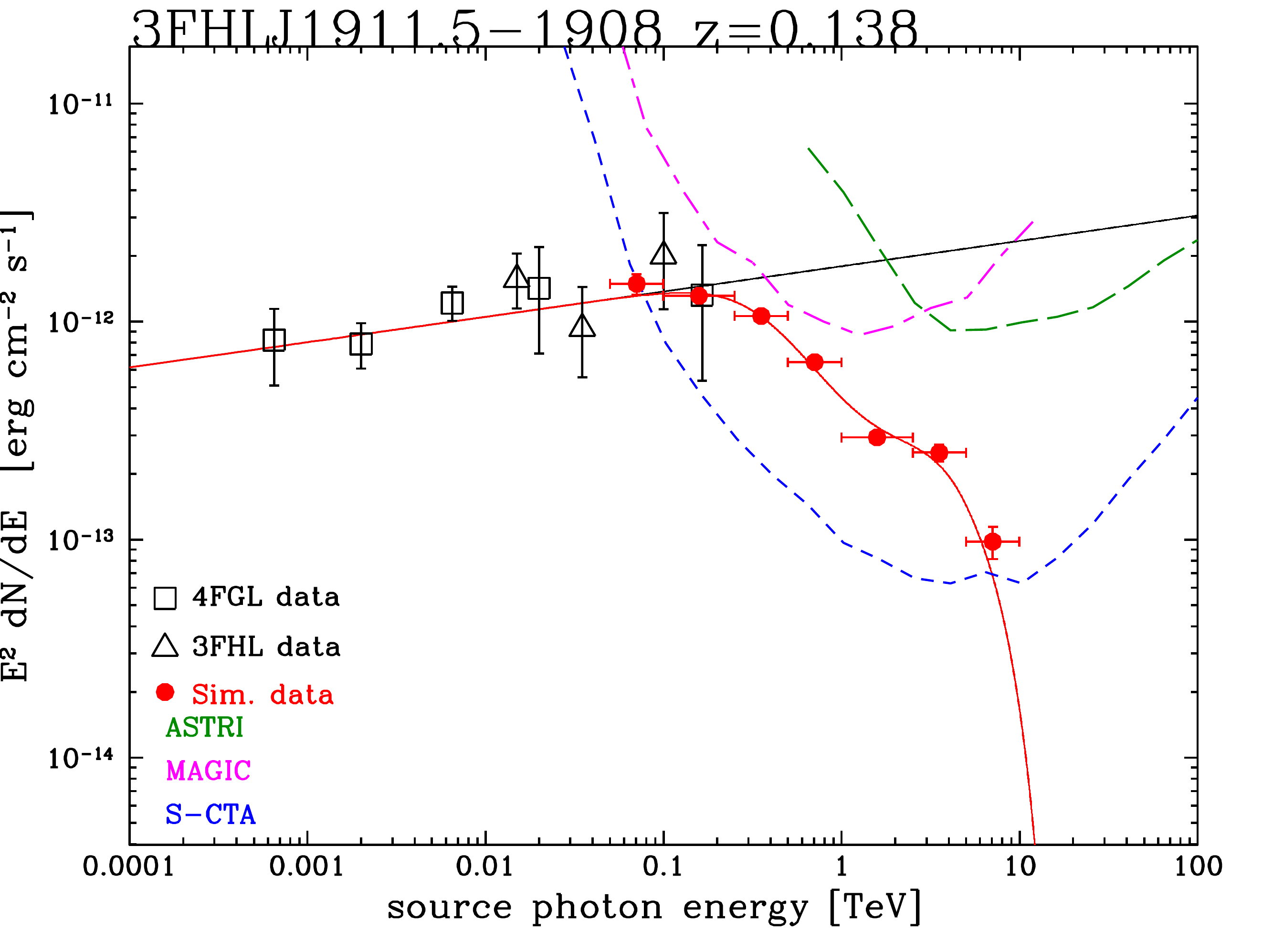}
\caption{Continued from Fig.~1.}
\label{fig:fig1}
\end{figure*}%[htbp]

\newpage
\setcounter{figure}{1}
\begin{figure*}%[htbp]
\includegraphics[width=0.35\textwidth,angle=-90]{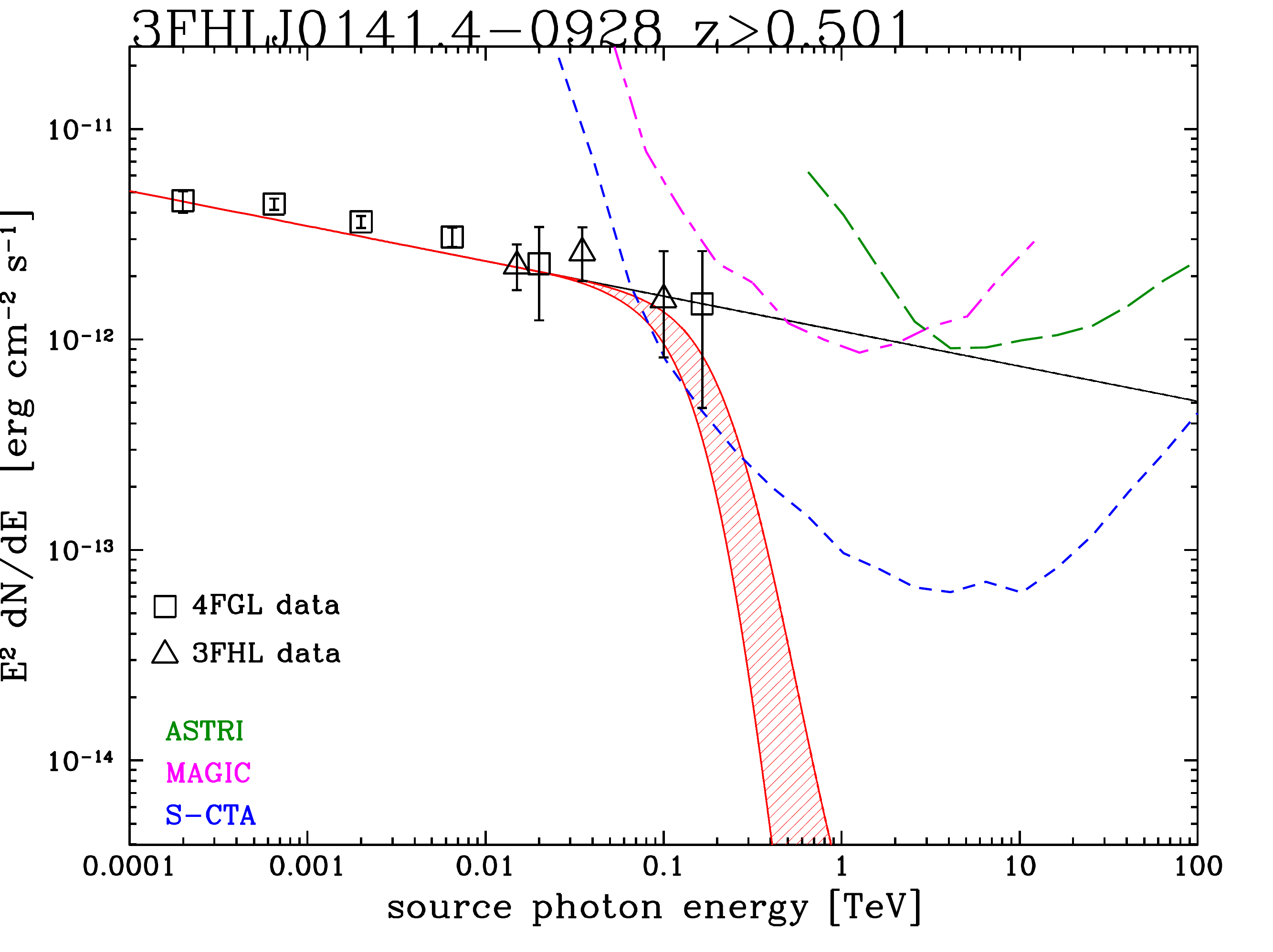}
\includegraphics[width=0.35\textwidth,angle=-90]{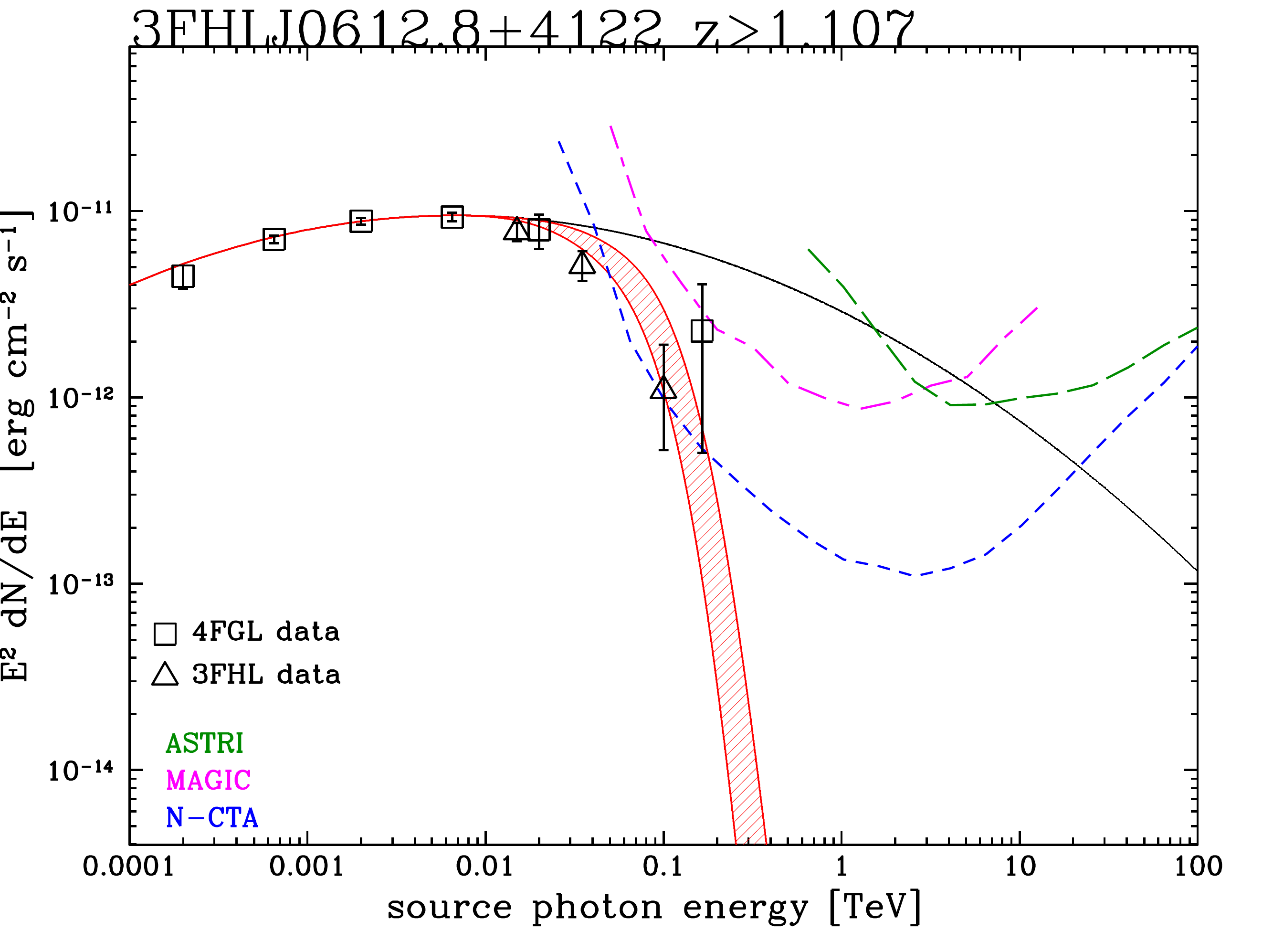}
\includegraphics[width=0.35\textwidth,angle=-90]{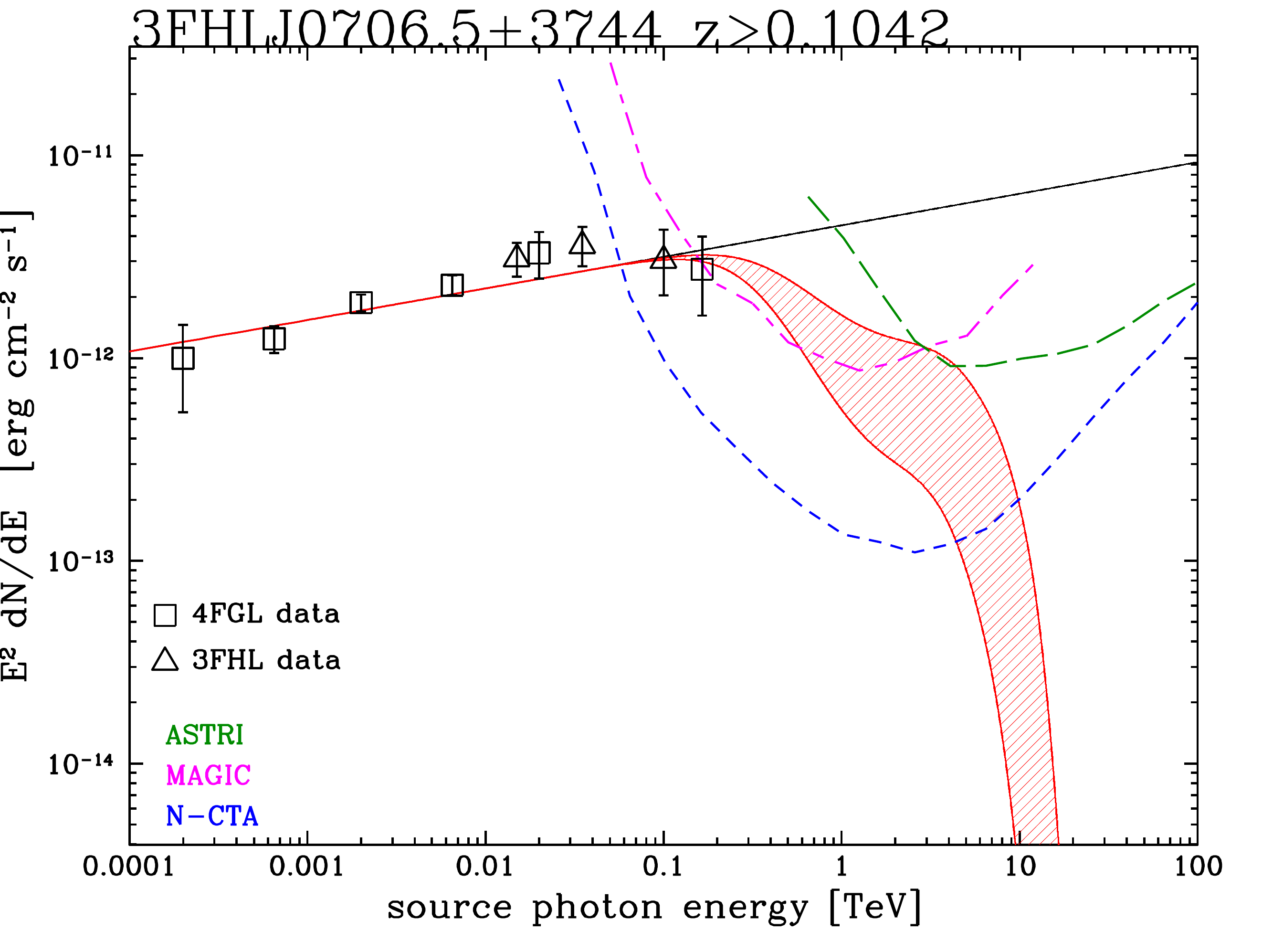}
\includegraphics[width=0.35\textwidth,angle=-90]{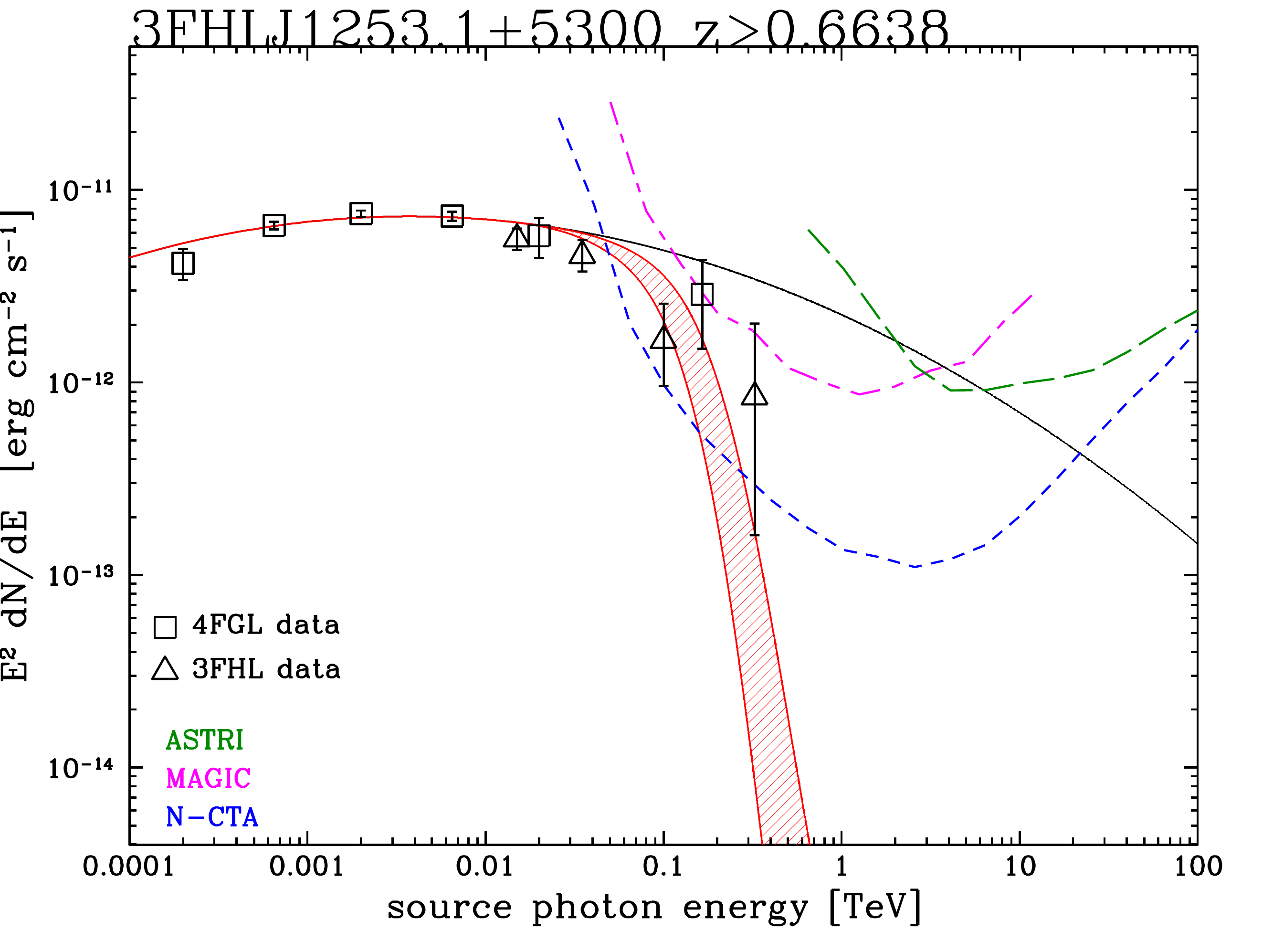}
\includegraphics[width=0.35\textwidth,angle=-90]{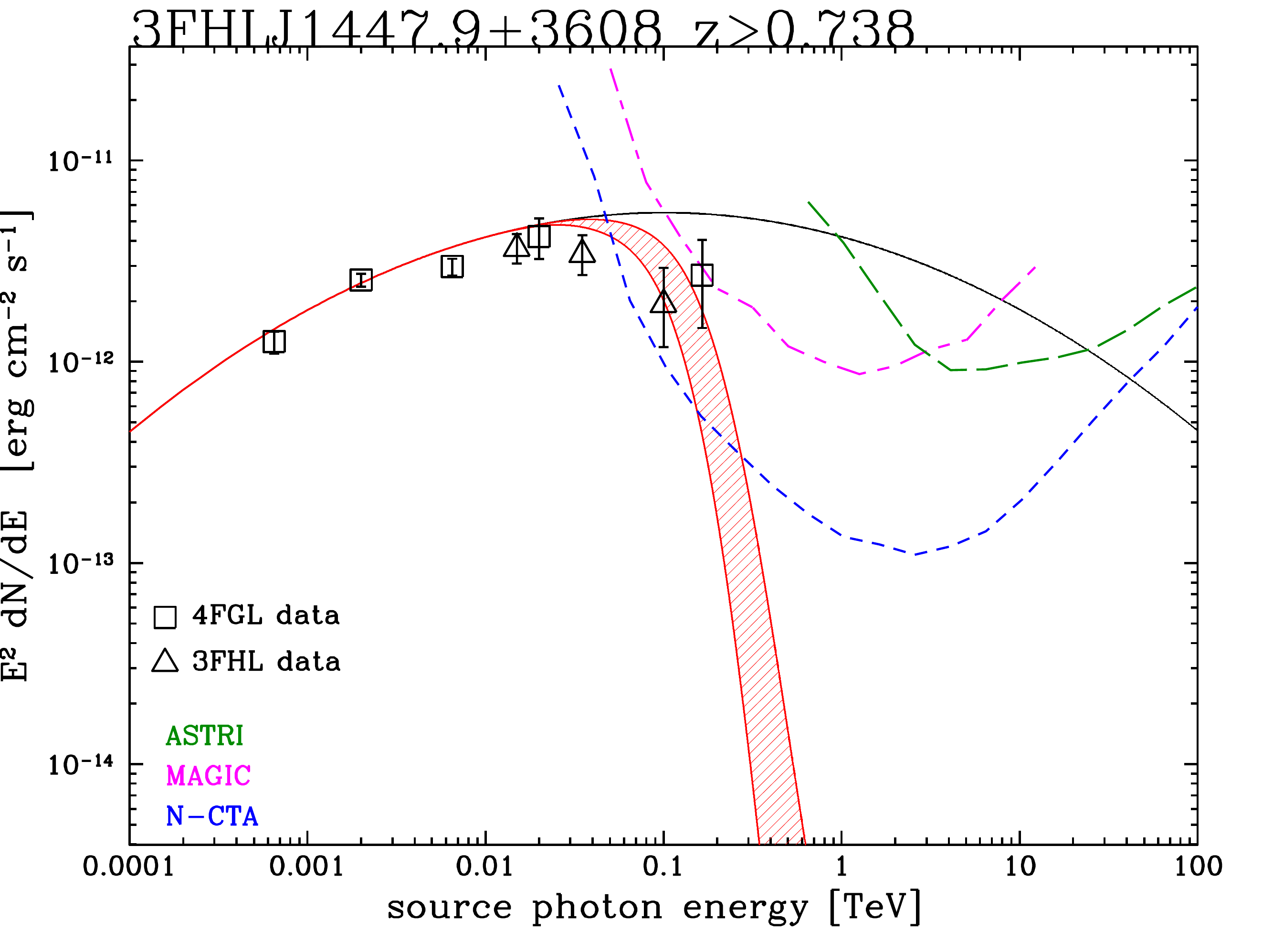}
\caption{VHE extrapolation before (black solid curve) and after (as a red-filled band considering the redshifts between the measured lower limit and one corresponding to a distance a factor 2 larger) the EBL model  correction (see text in Sections 2 and 3 for details) for the 5 sources of the BLL sample of Paper~I for which we provide a robust spectroscopic lower limit on \textit{z} derived by intervening absorbers. We overplot the sensitivity curves at 5$\sigma$ for 50 hour and zenith angle $<$~20 degrees for CTA-north or CTA-south (blue short dashed curve), for MAGIC at 5$\sigma$ for 50 hour and zenith angle $<$~35 degrees (magenta dashed-dotted curve) and for ASTRI mini array (green long dashed curve). We superpose the \textit{Fermi} flux data from the 4FGL catalogue (black opened squared) and the 3FHL catalogue (black opened triangles). }
\label{fig:fig2}
\end{figure*}%[htbp]

\setcounter{figure}{2}
\begin{figure*}%[htbp]
\includegraphics[width=0.35\textwidth,angle=-90]{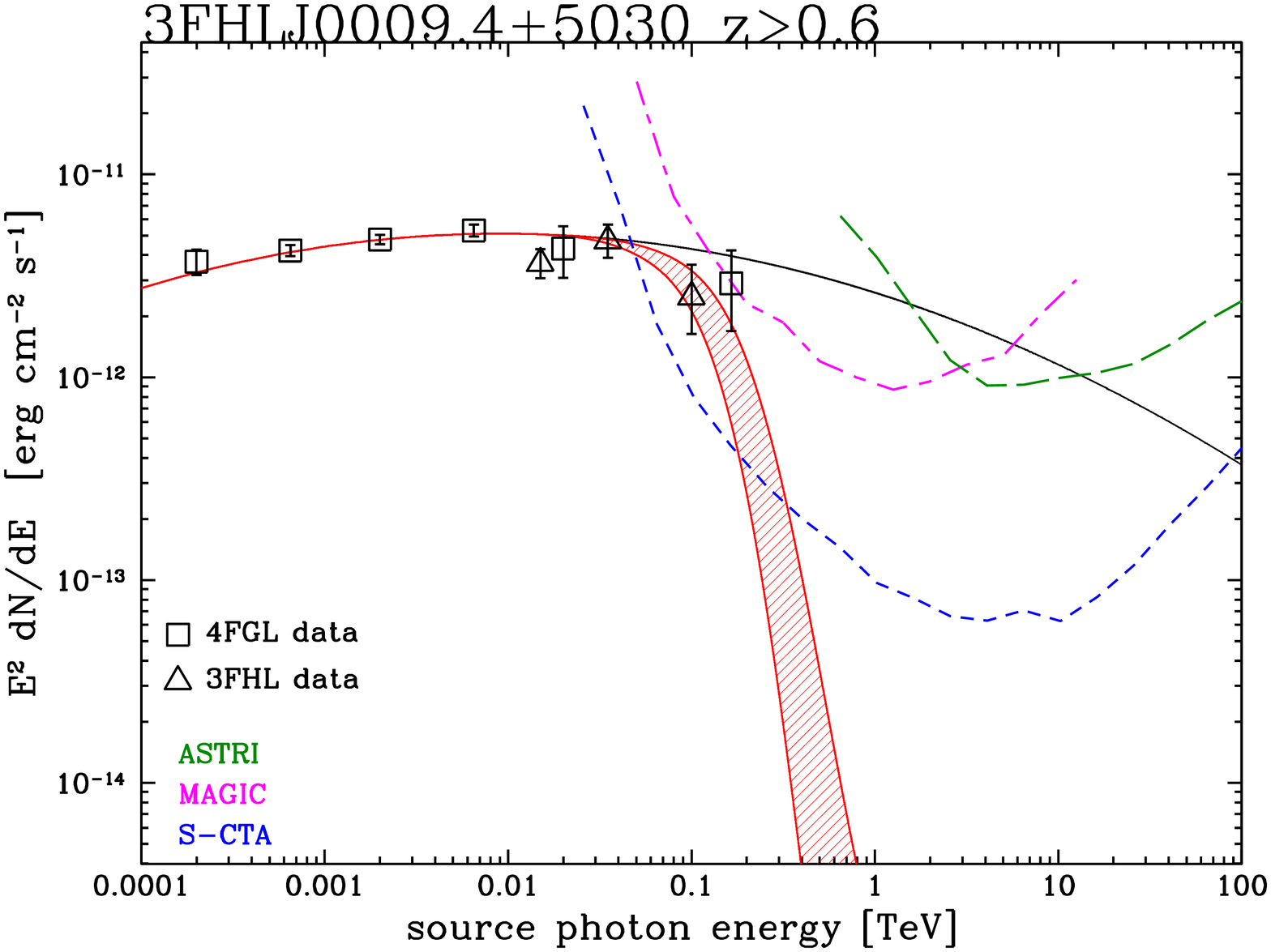}
\includegraphics[width=0.35\textwidth,angle=-90]{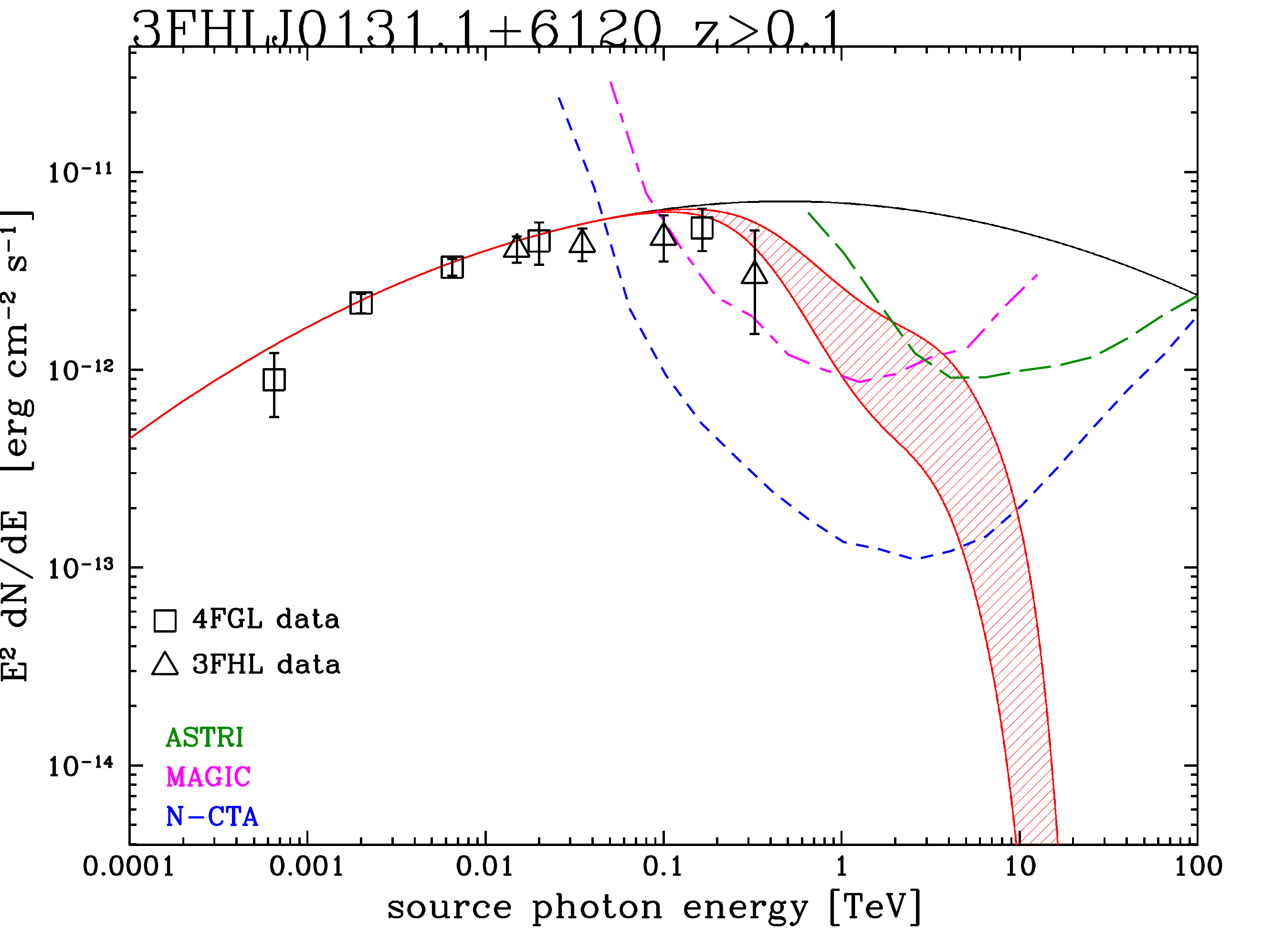}
\includegraphics[width=0.35\textwidth,angle=-90]{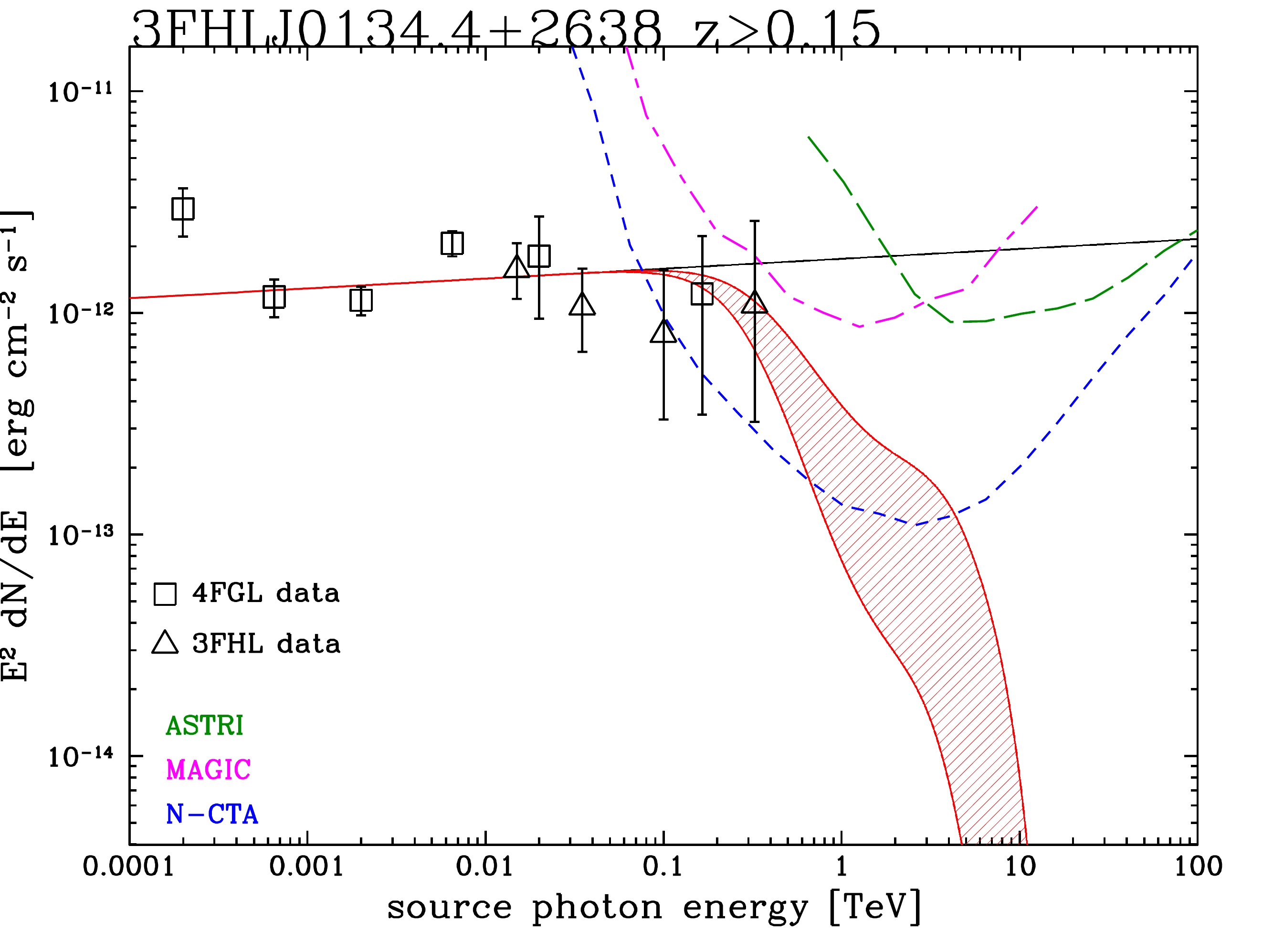}
\includegraphics[width=0.35\textwidth,angle=-90]{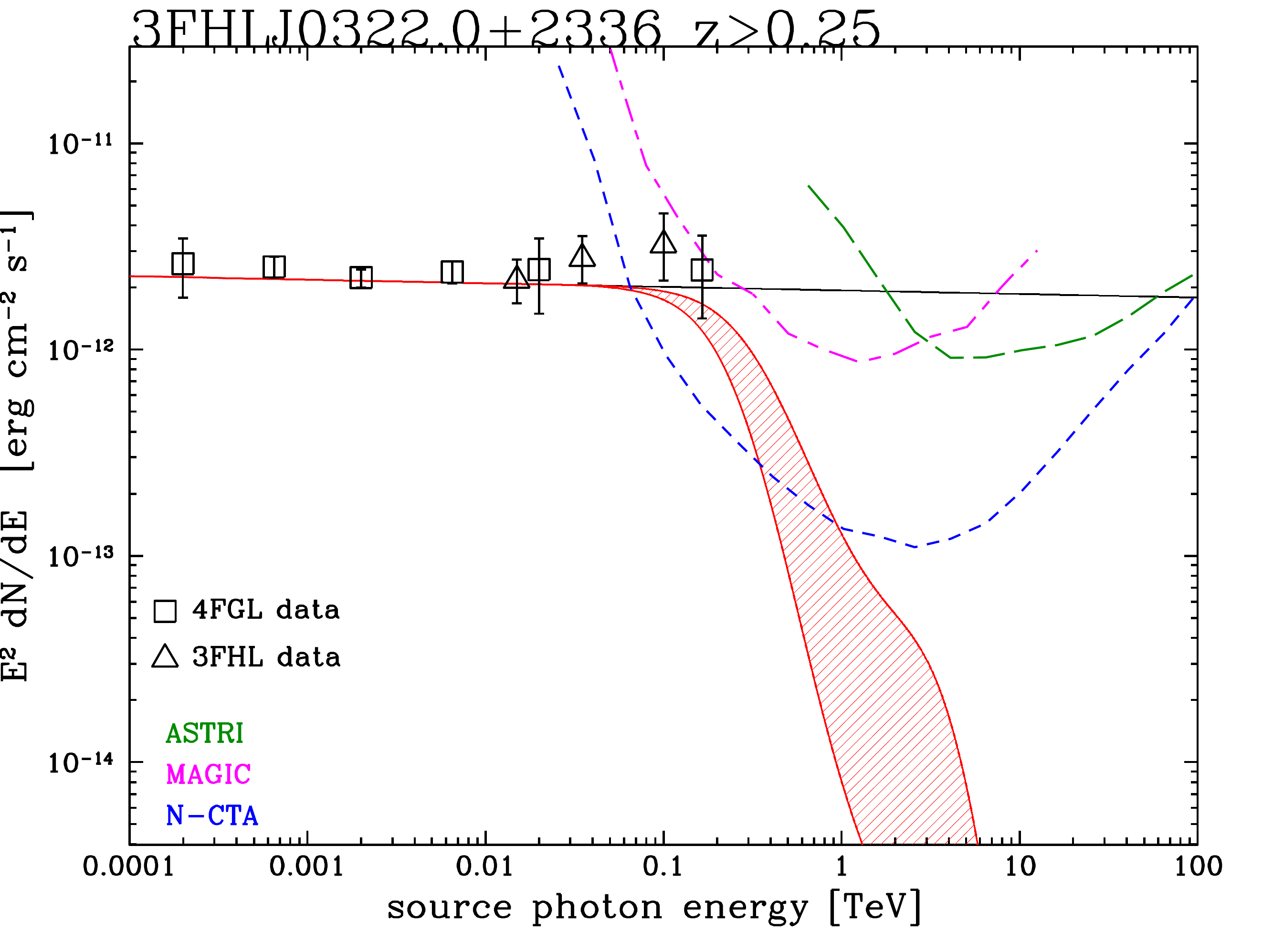}
\includegraphics[width=0.35\textwidth,angle=-90]{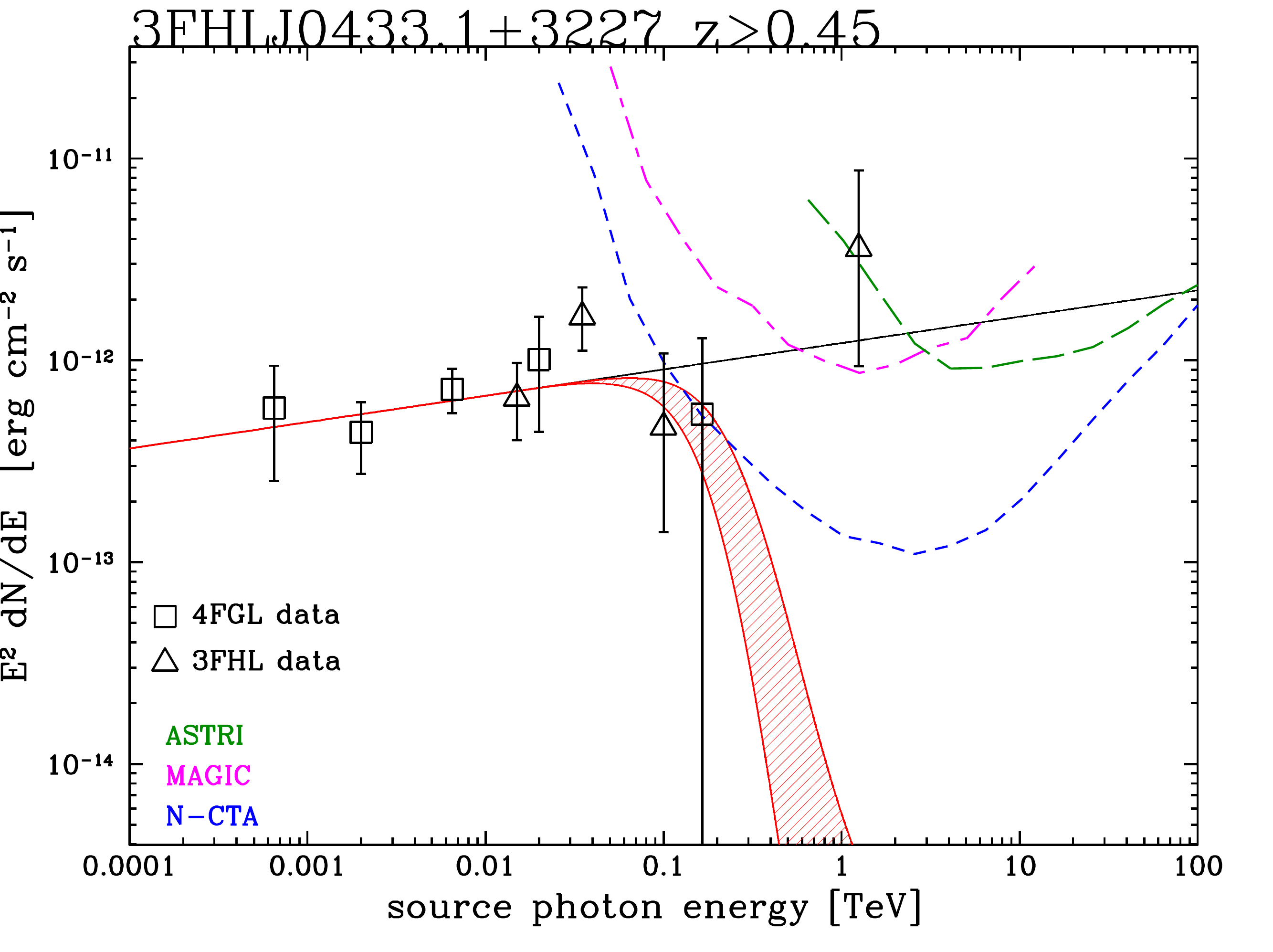}
\includegraphics[width=0.35\textwidth,angle=-90]{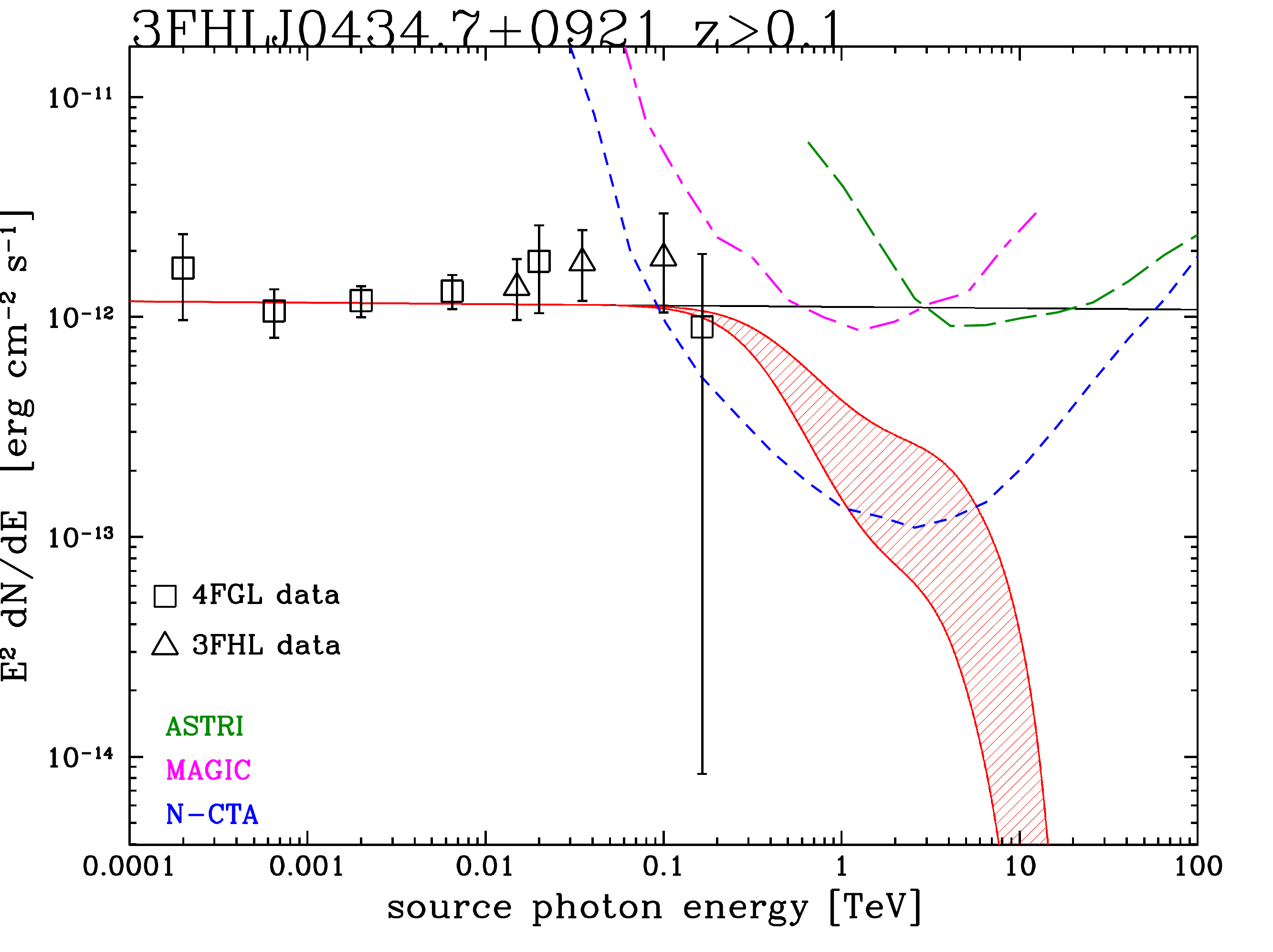}
\caption{VHE extrapolation before (black solid curve) and after (as a red-filled band considering the redshifts between the measured lower limit and one corresponding to a distance a factor 2 larger) the EBL model  correction (see text in Sect. 2 and 3 for details) for the 25 sources of the BLL sample of Paper~I for which we provide a redshift lower limit derived from the absence of detected lines from the BLL host galaxy. We overplot the sensitivity curves at 5$\sigma$ for 50 hour and zenith angle $<$~20 degrees for CTA-north or CTA-south (blue short dashed curve), for MAGIC at 5$\sigma$ for 50 hour and zenith angle $<$~35 degrees (magenta dashed-dotted curve) and for ASTRI mini array (green long dashed curve). We superpose the \textit{Fermi} flux data from the 4FGL catalogue (black opened squared) and the 3FHL catalogue (black opened triangles).}
\label{fig:fig3}
\end{figure*}%[htbp]

\setcounter{figure}{2}
\begin{figure*}%[htbp]
\includegraphics[width=0.35\textwidth,angle=-90]{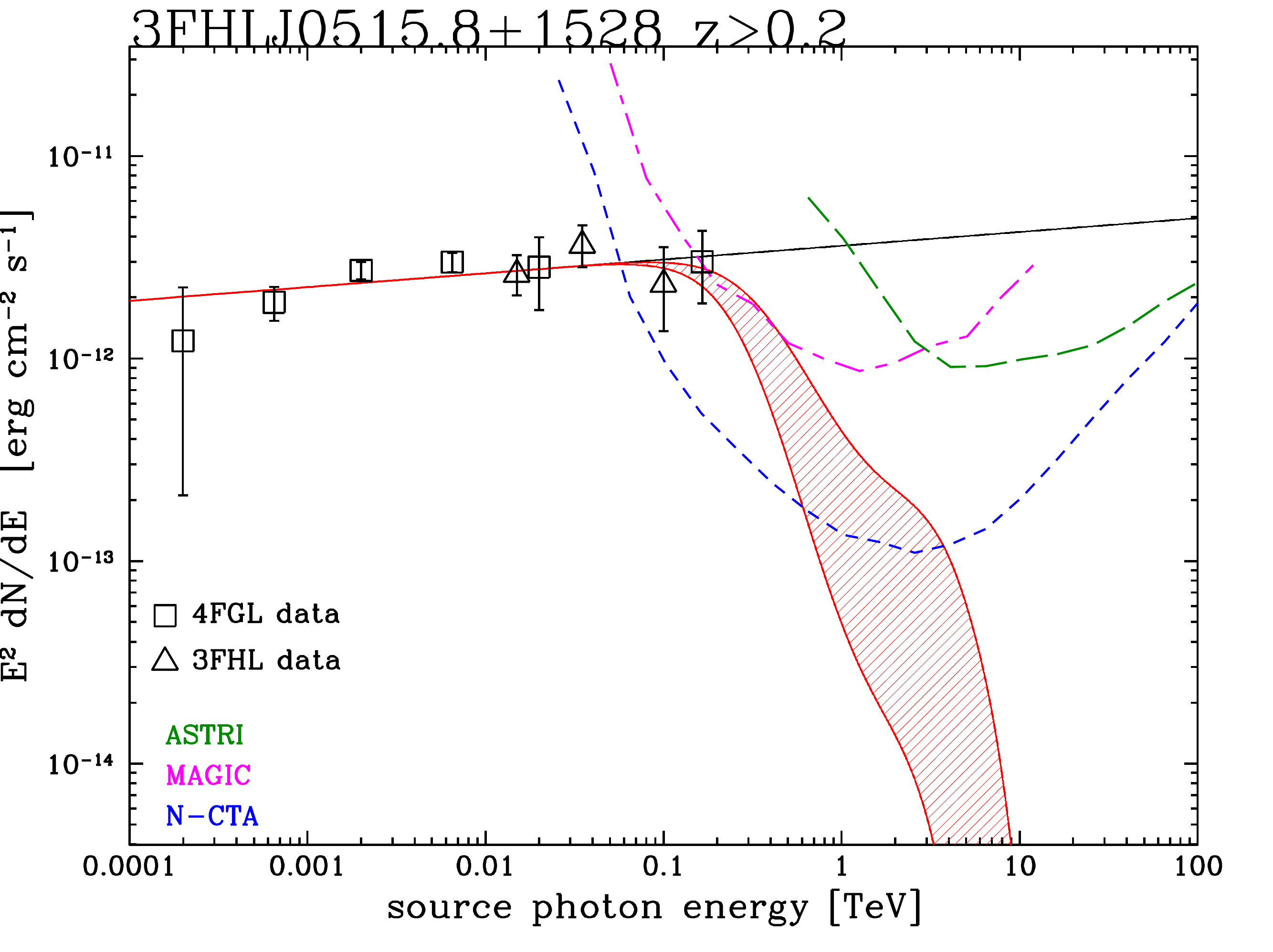}
\includegraphics[width=0.35\textwidth,angle=-90]{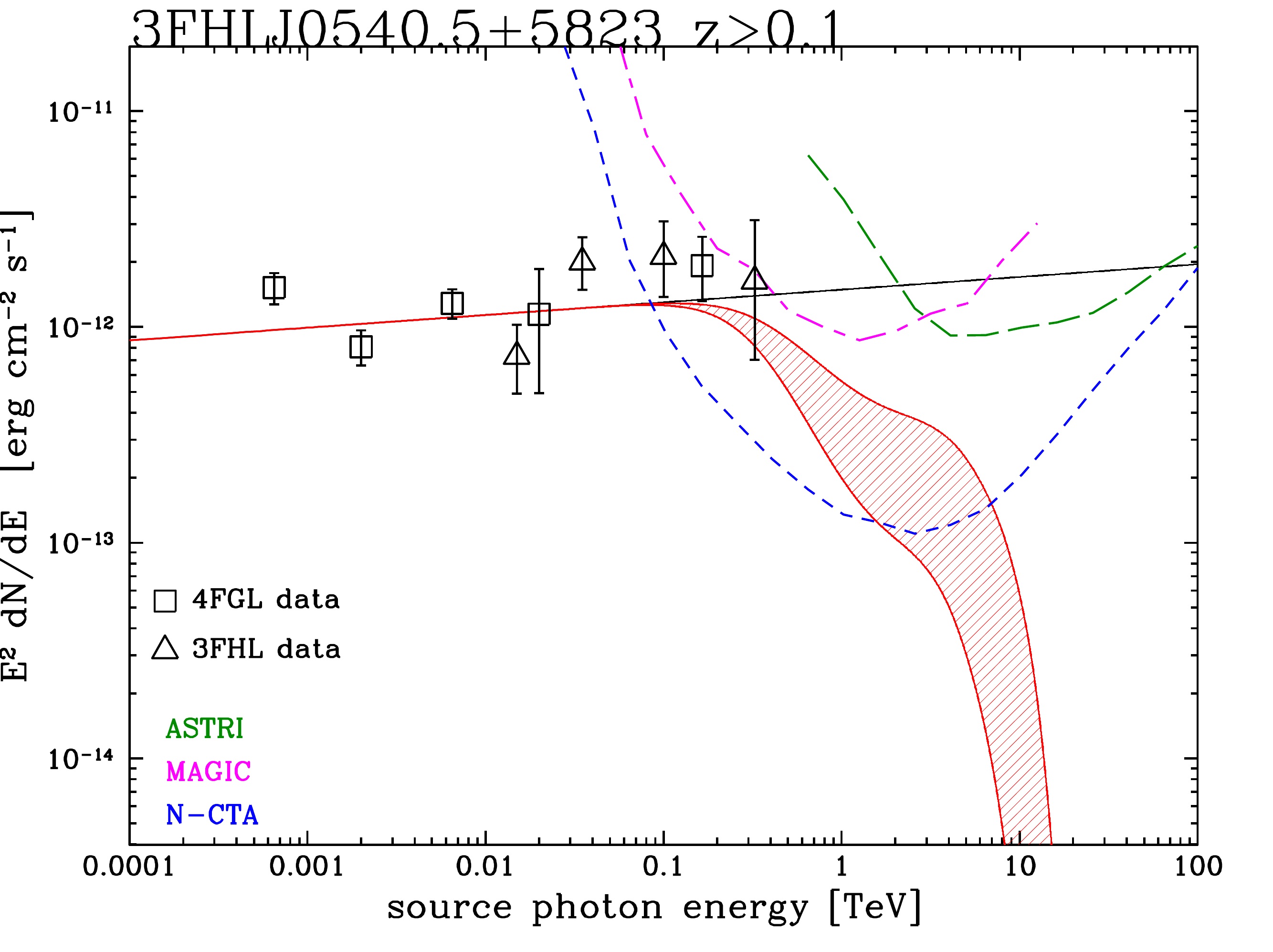}
\includegraphics[width=0.35\textwidth,angle=-90]{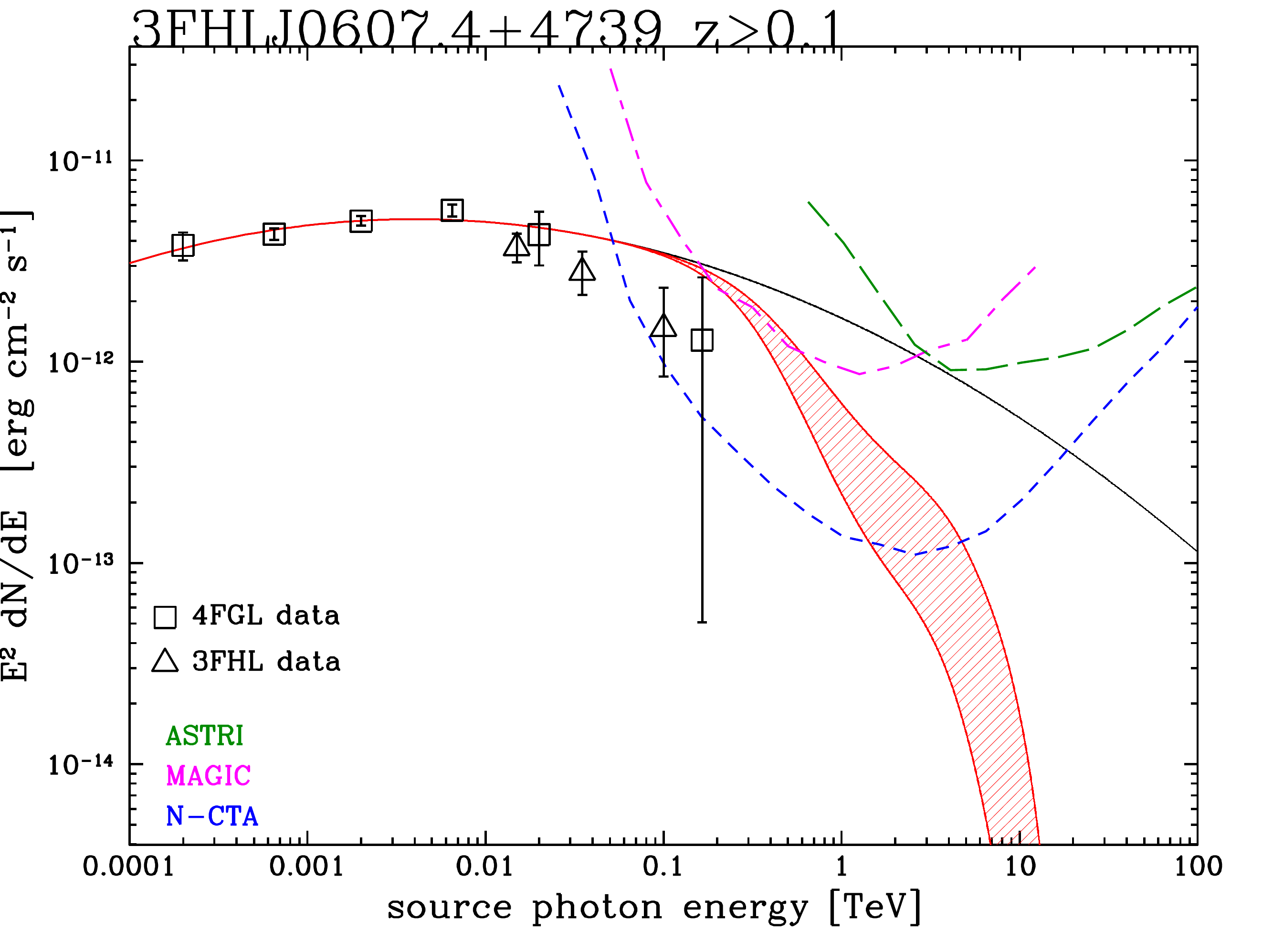}
\includegraphics[width=0.35\textwidth,angle=-90]{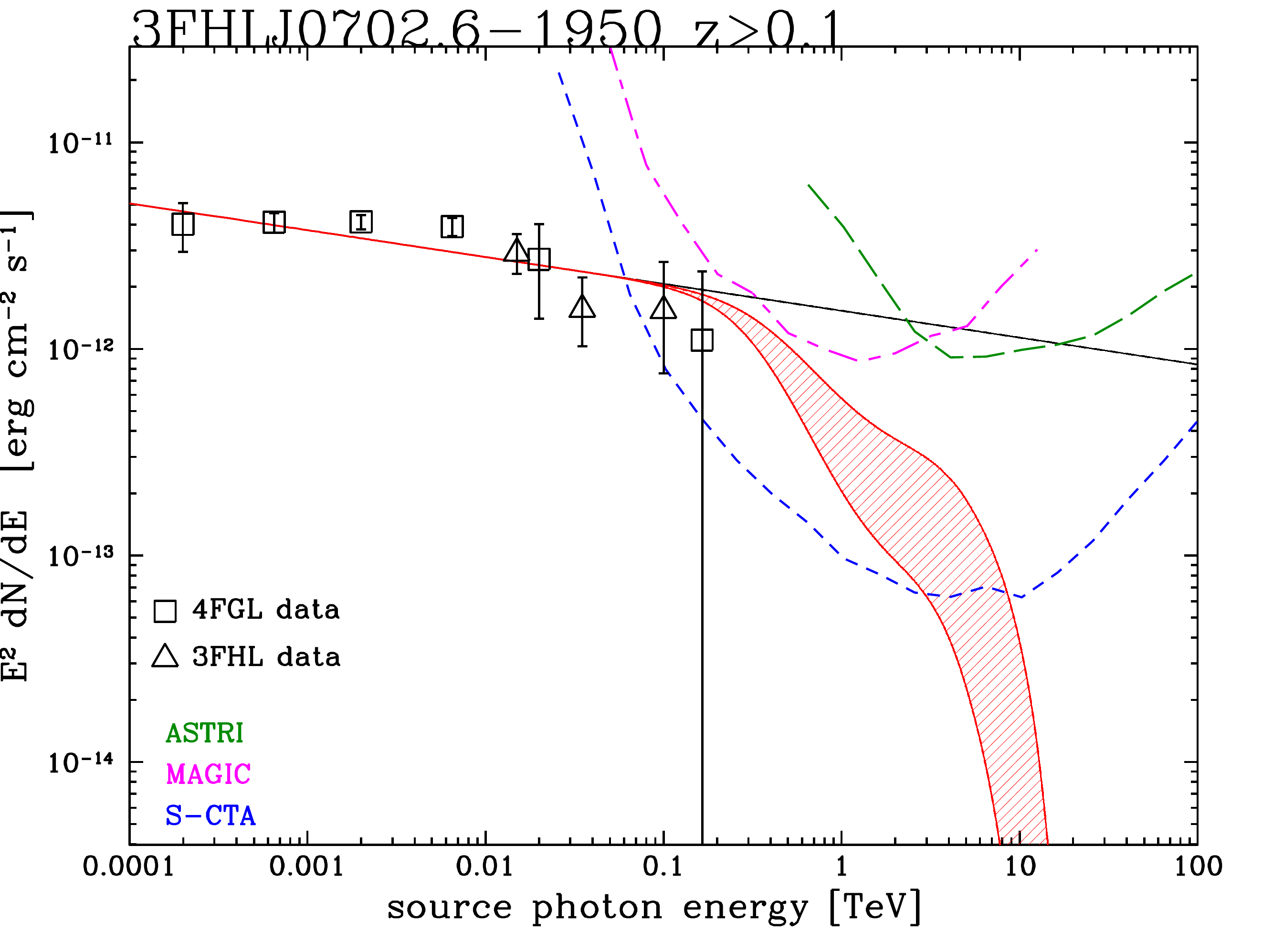}
\includegraphics[width=0.35\textwidth,angle=-90]{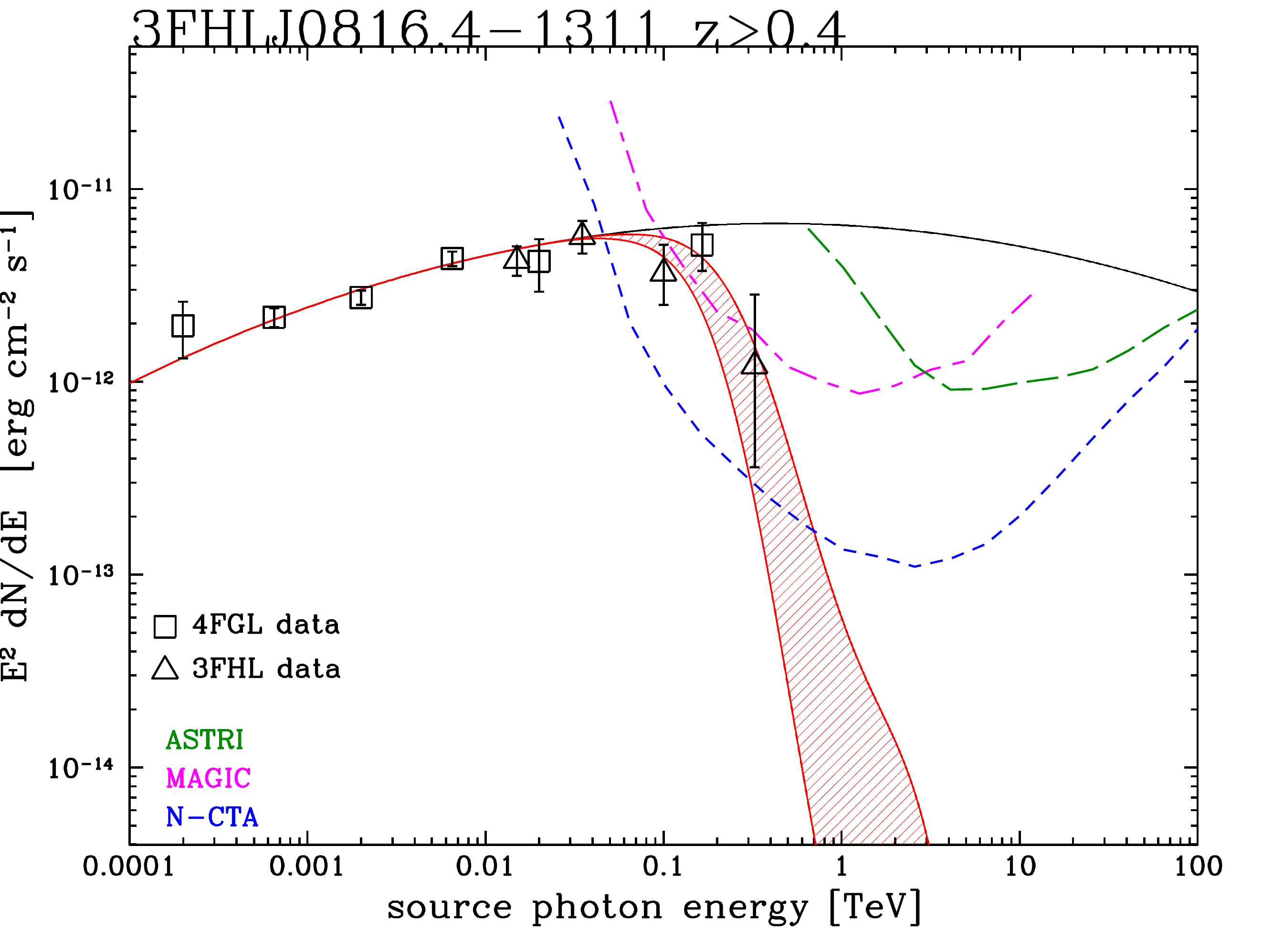}
\includegraphics[width=0.35\textwidth,angle=-90]{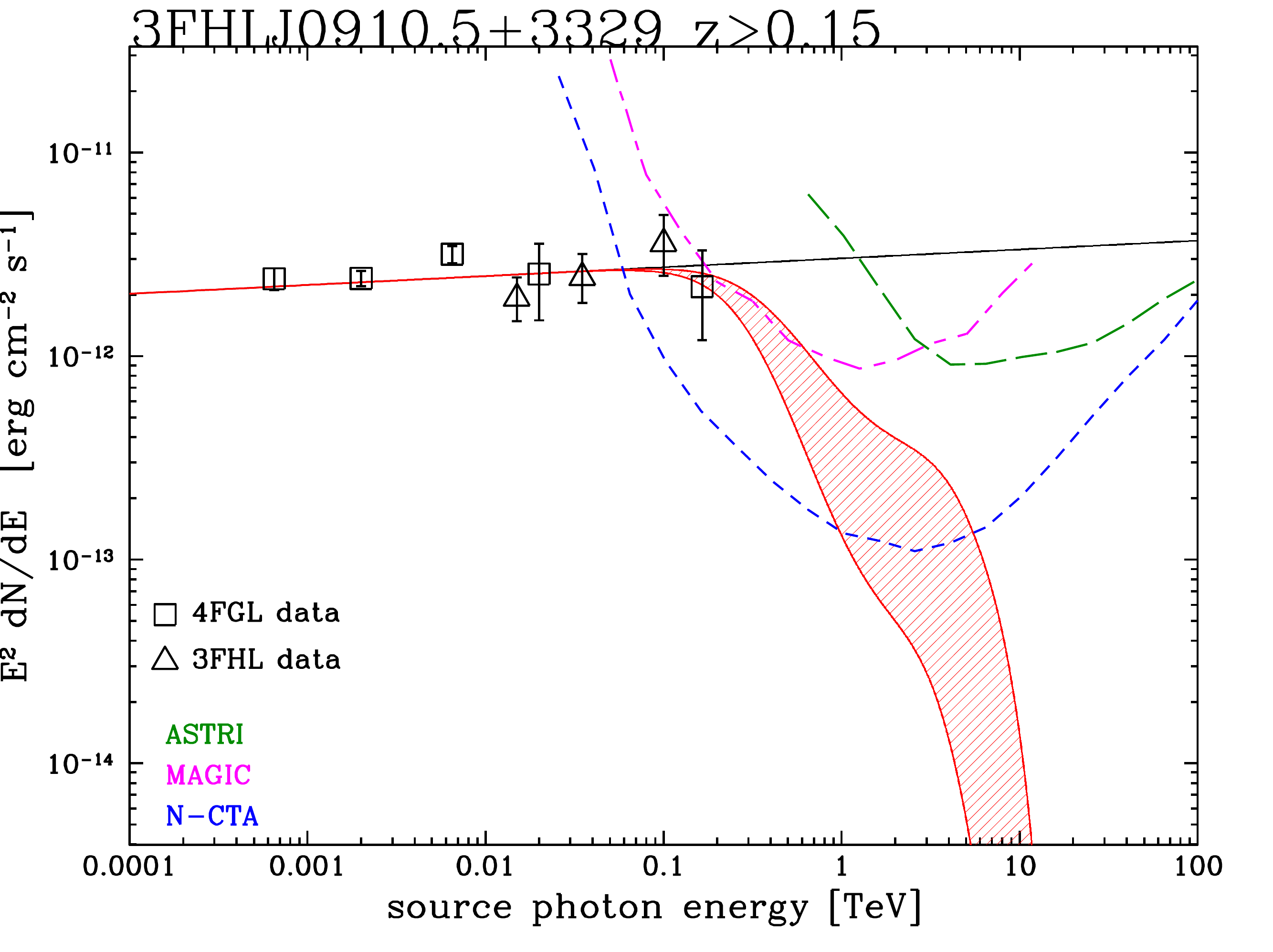}
\caption{Continued from Fig.~3.}
\label{fig:fig3}
\end{figure*}%[htbp]

\setcounter{figure}{2}
\begin{figure*}%[htbp]
\includegraphics[width=0.35\textwidth,angle=-90]{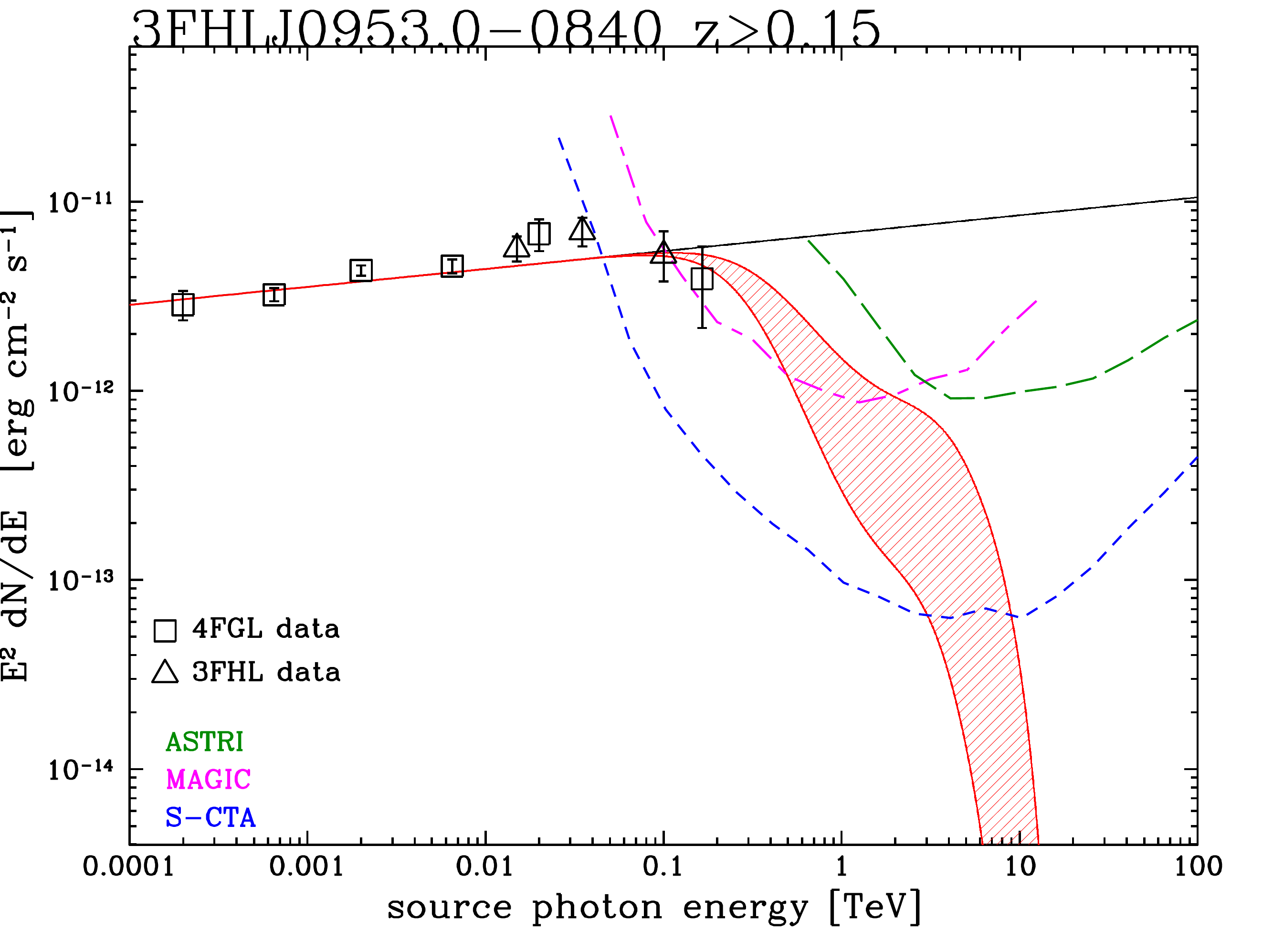}
\includegraphics[width=0.35\textwidth,angle=-90]{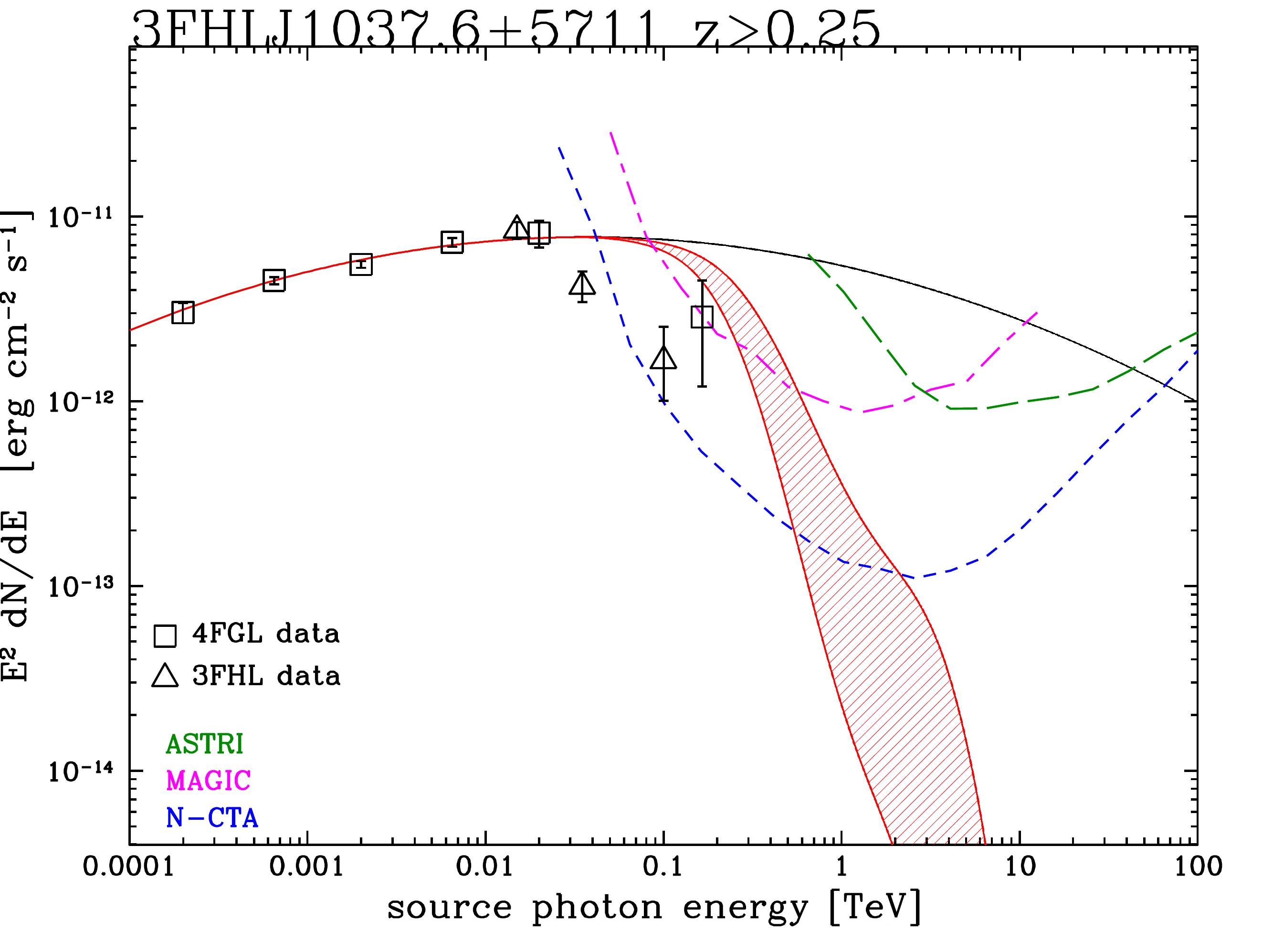}
\includegraphics[width=0.35\textwidth,angle=-90]{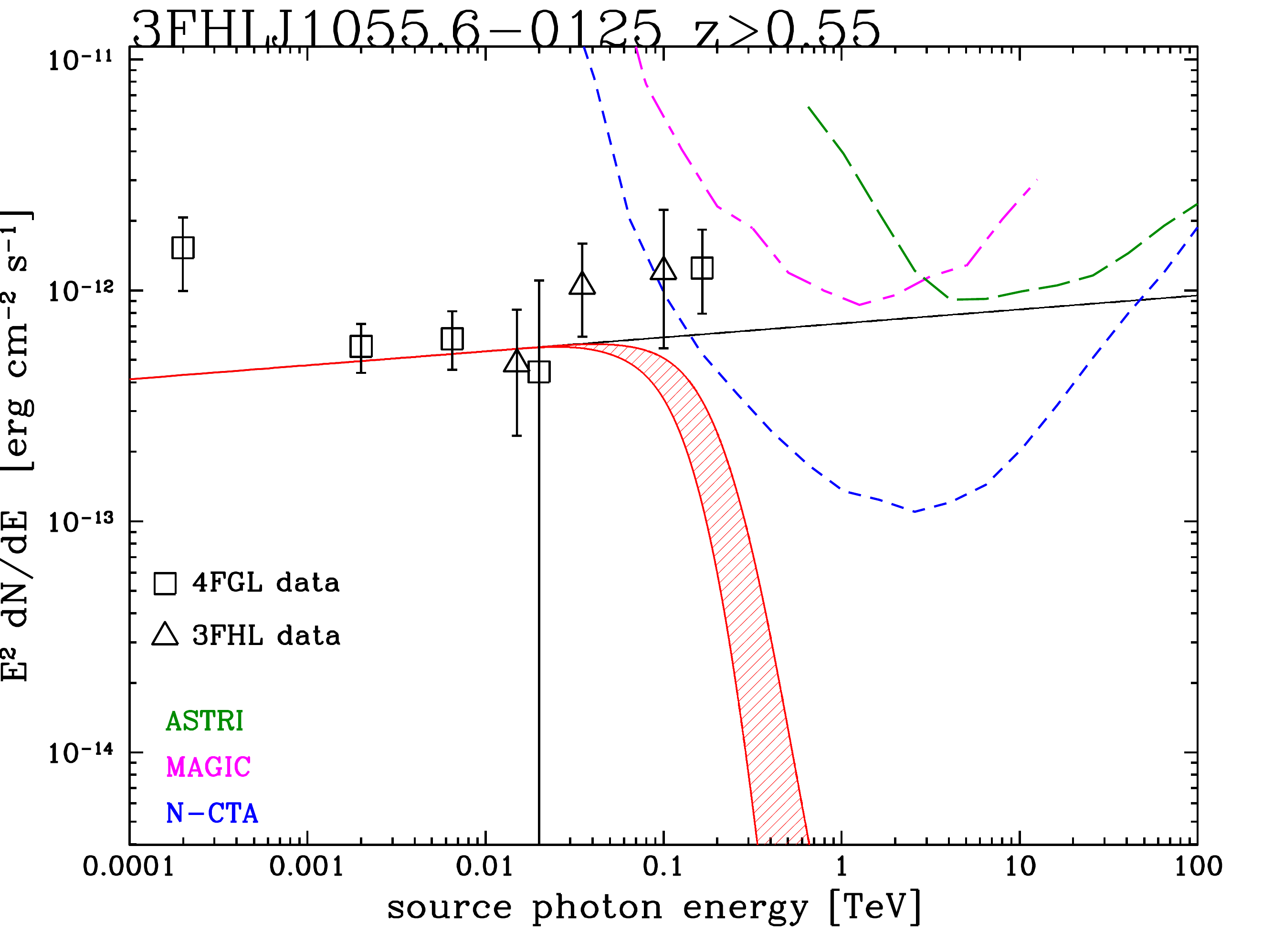}
\includegraphics[width=0.35\textwidth,angle=-90]{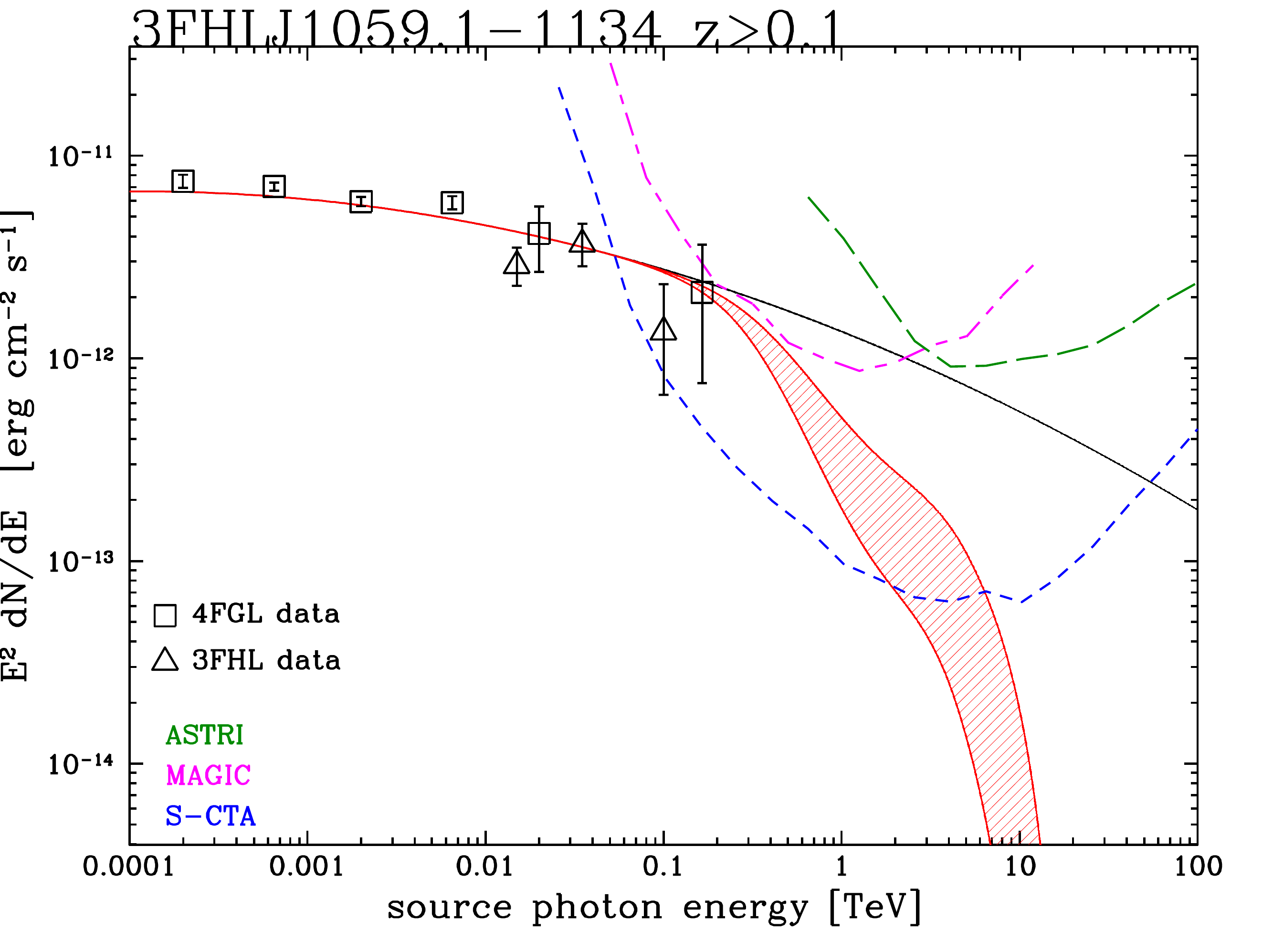}
\includegraphics[width=0.35\textwidth,angle=-90]{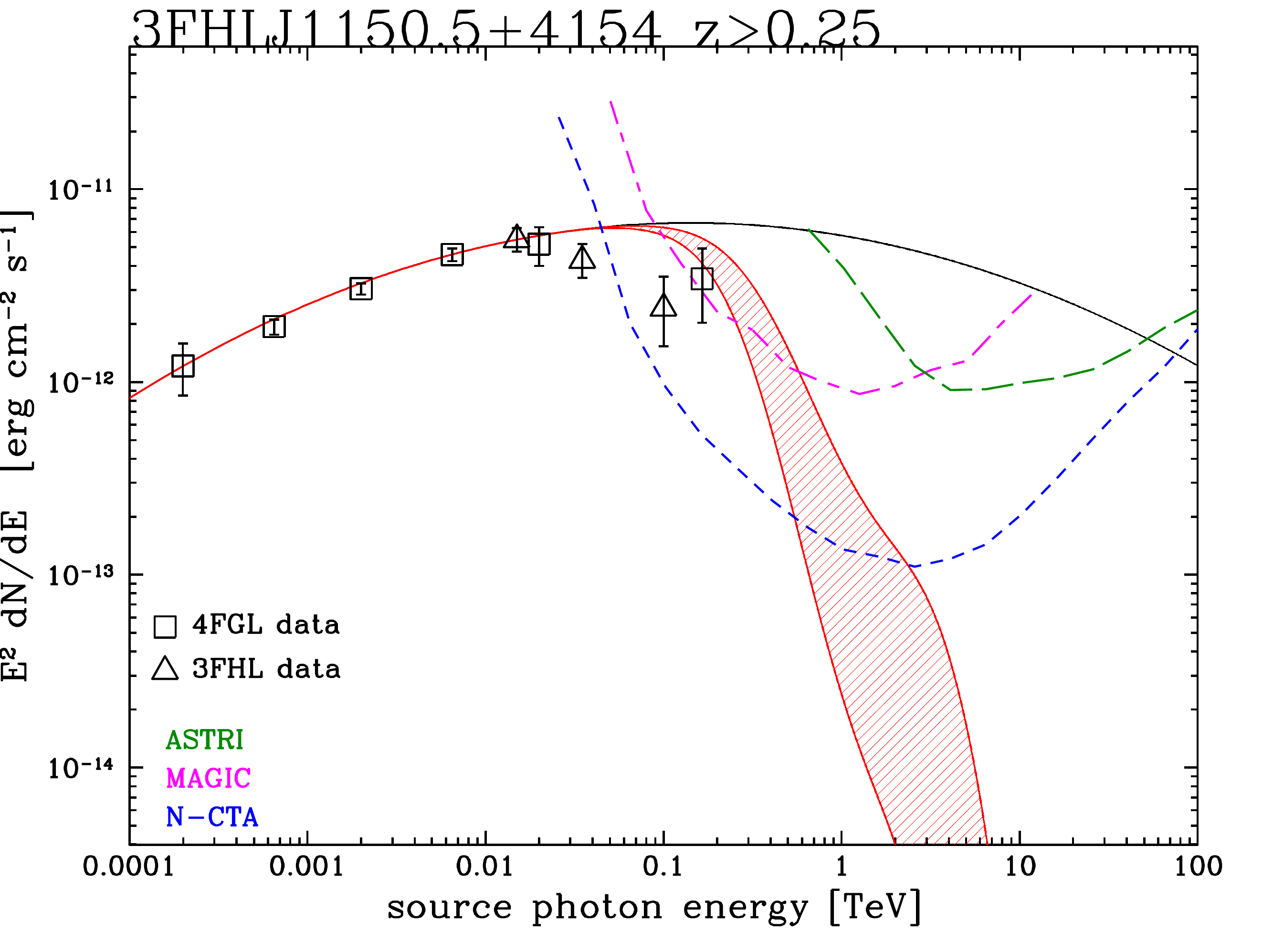}
\includegraphics[width=0.35\textwidth,angle=-90]{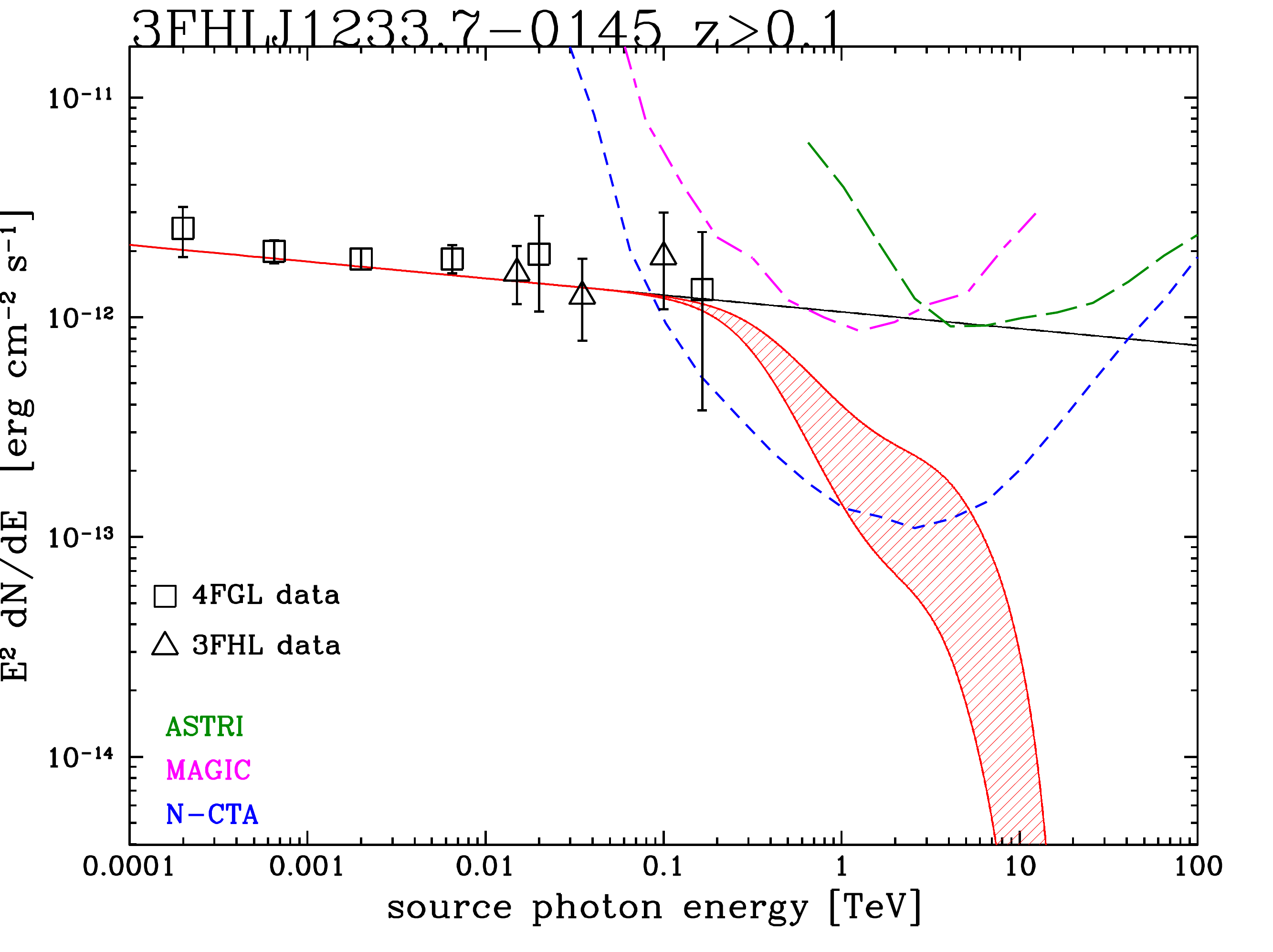}
\caption{Continued from Fig.~3.}
\label{fig:fig1}
\end{figure*}%[htbp]

\setcounter{figure}{2}
\begin{figure*}%[htbp]
\includegraphics[width=0.35\textwidth,angle=-90]{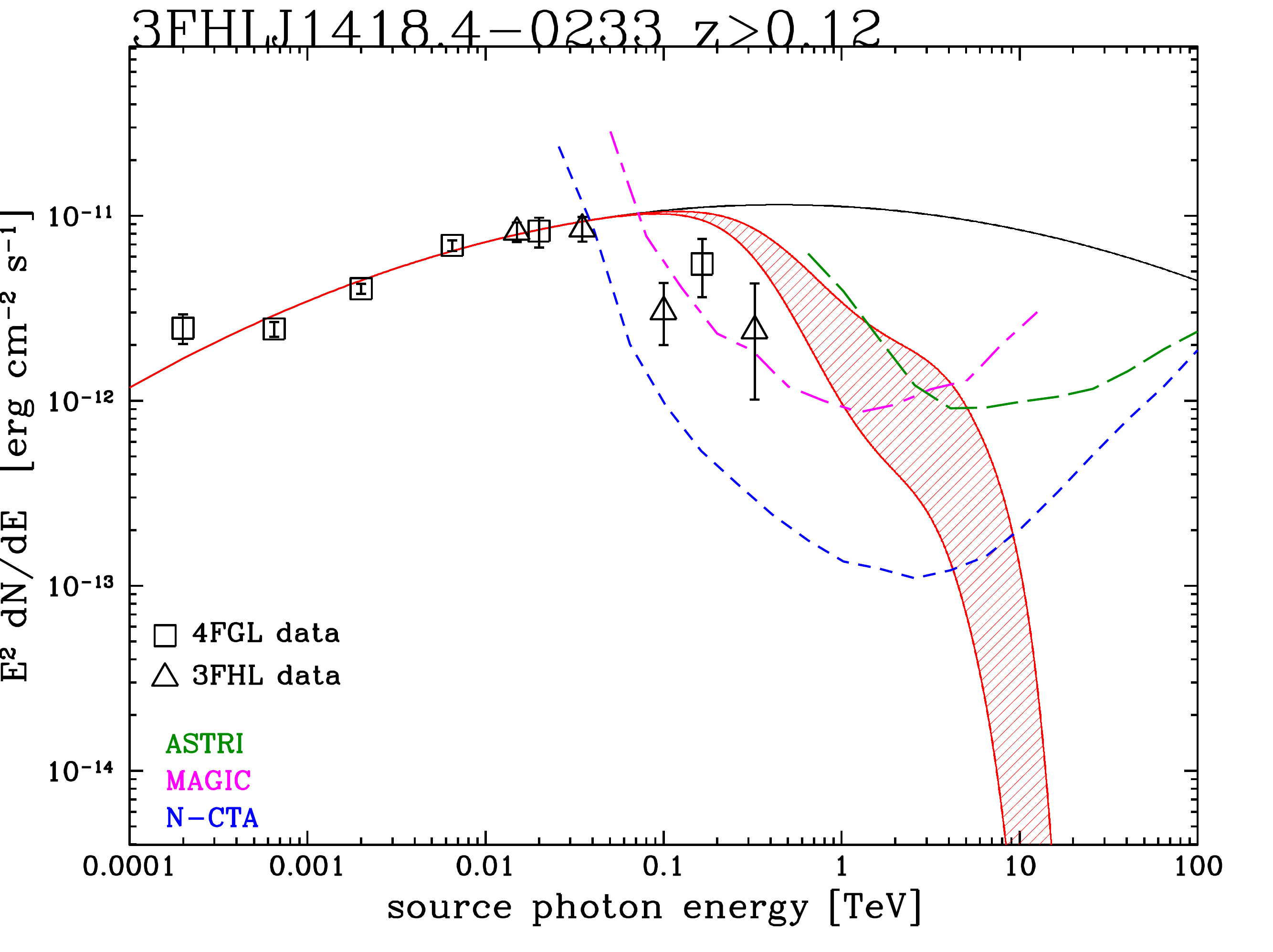}
\includegraphics[width=0.35\textwidth,angle=-90]{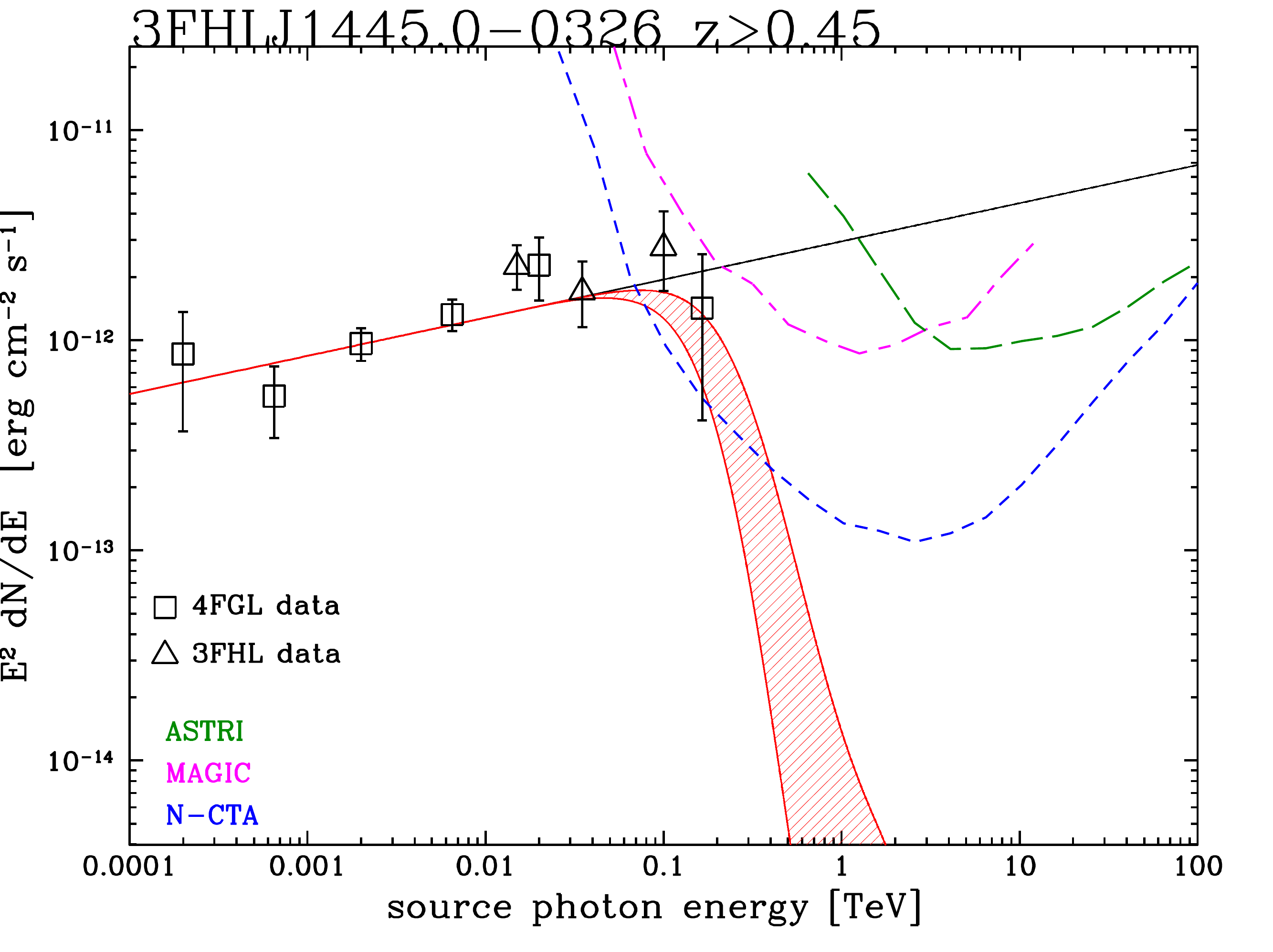}
\includegraphics[width=0.35\textwidth,angle=-90]{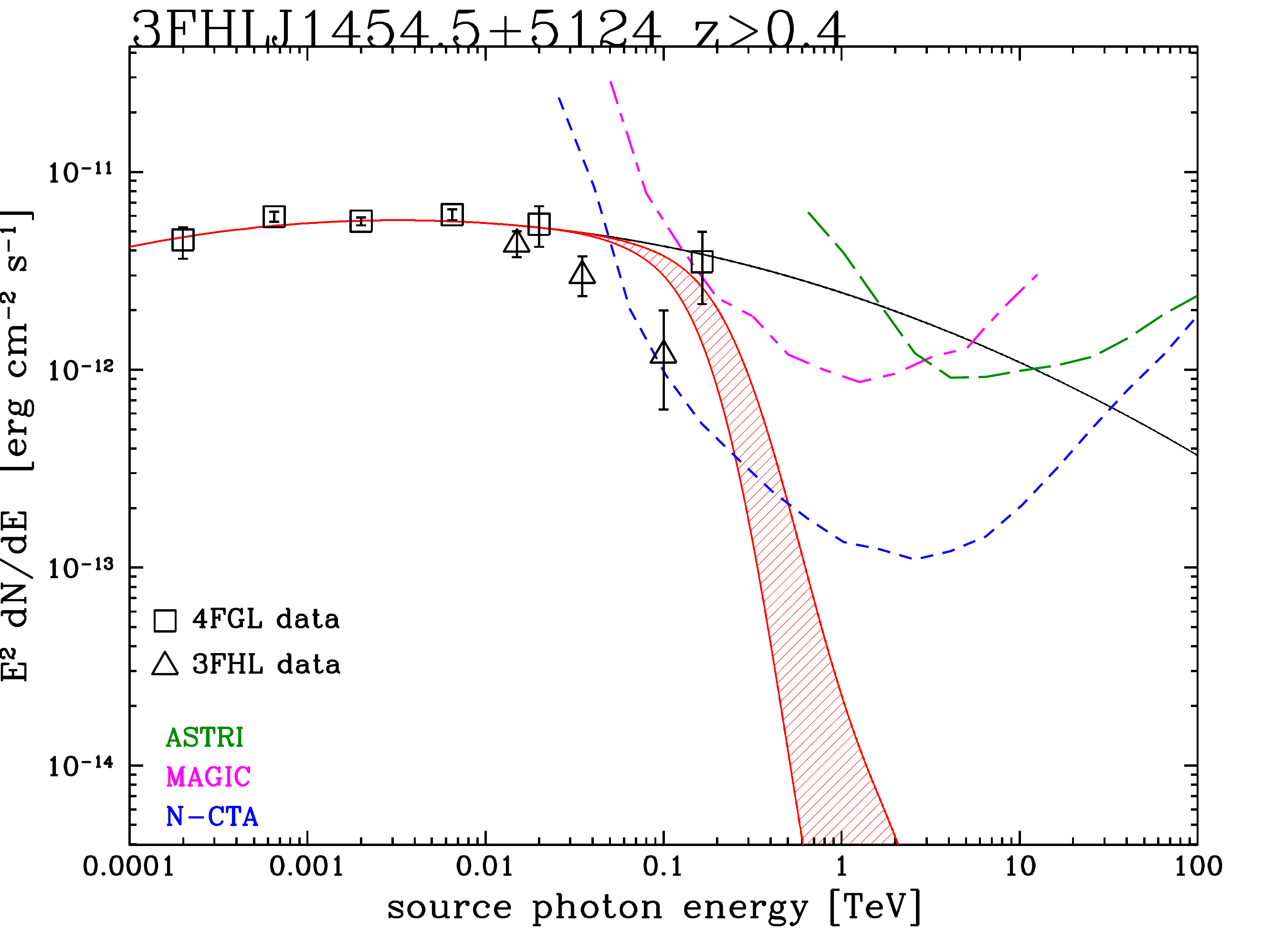}
\includegraphics[width=0.35\textwidth,angle=-90]{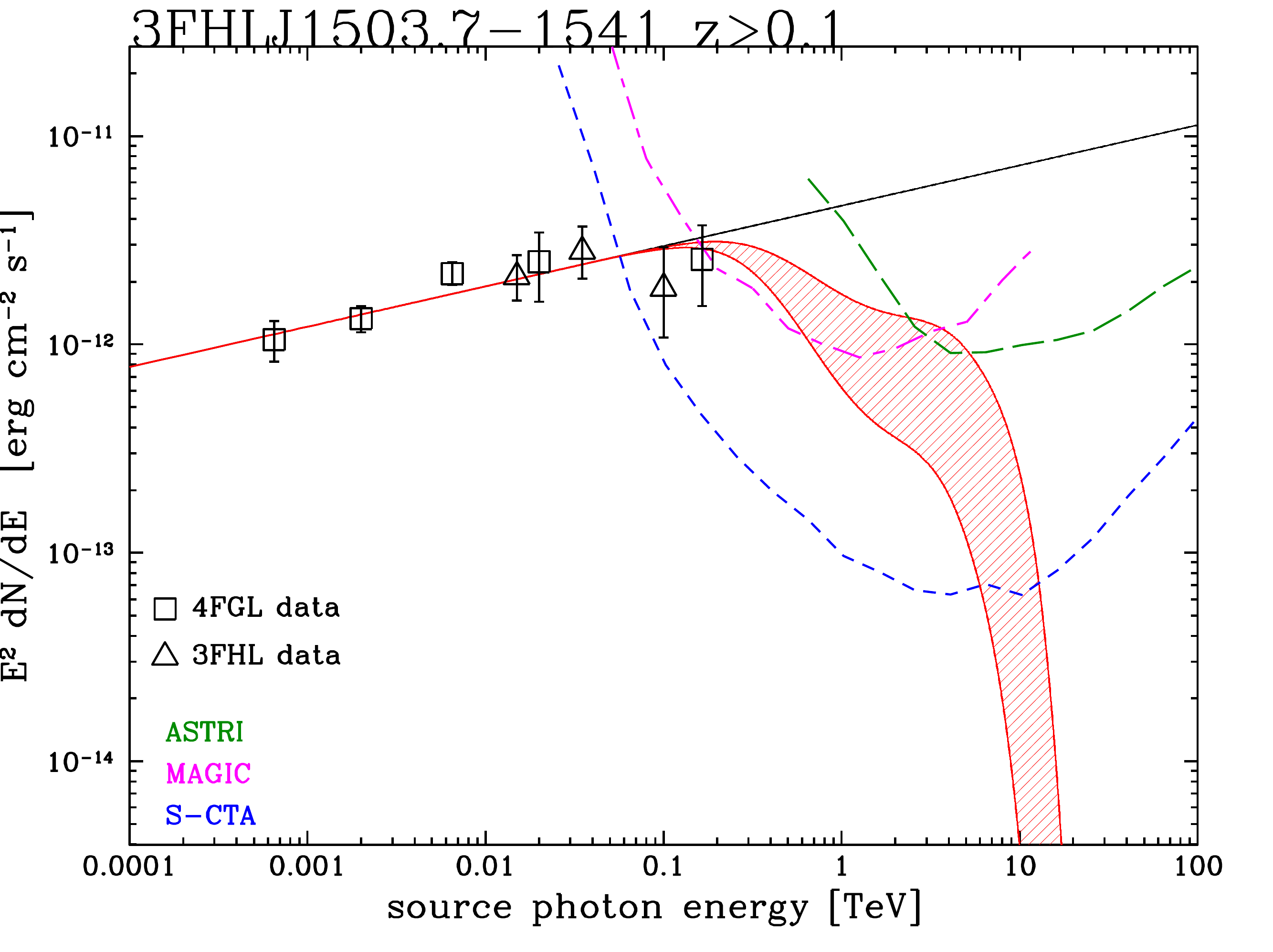}
\includegraphics[width=0.35\textwidth,angle=-90]{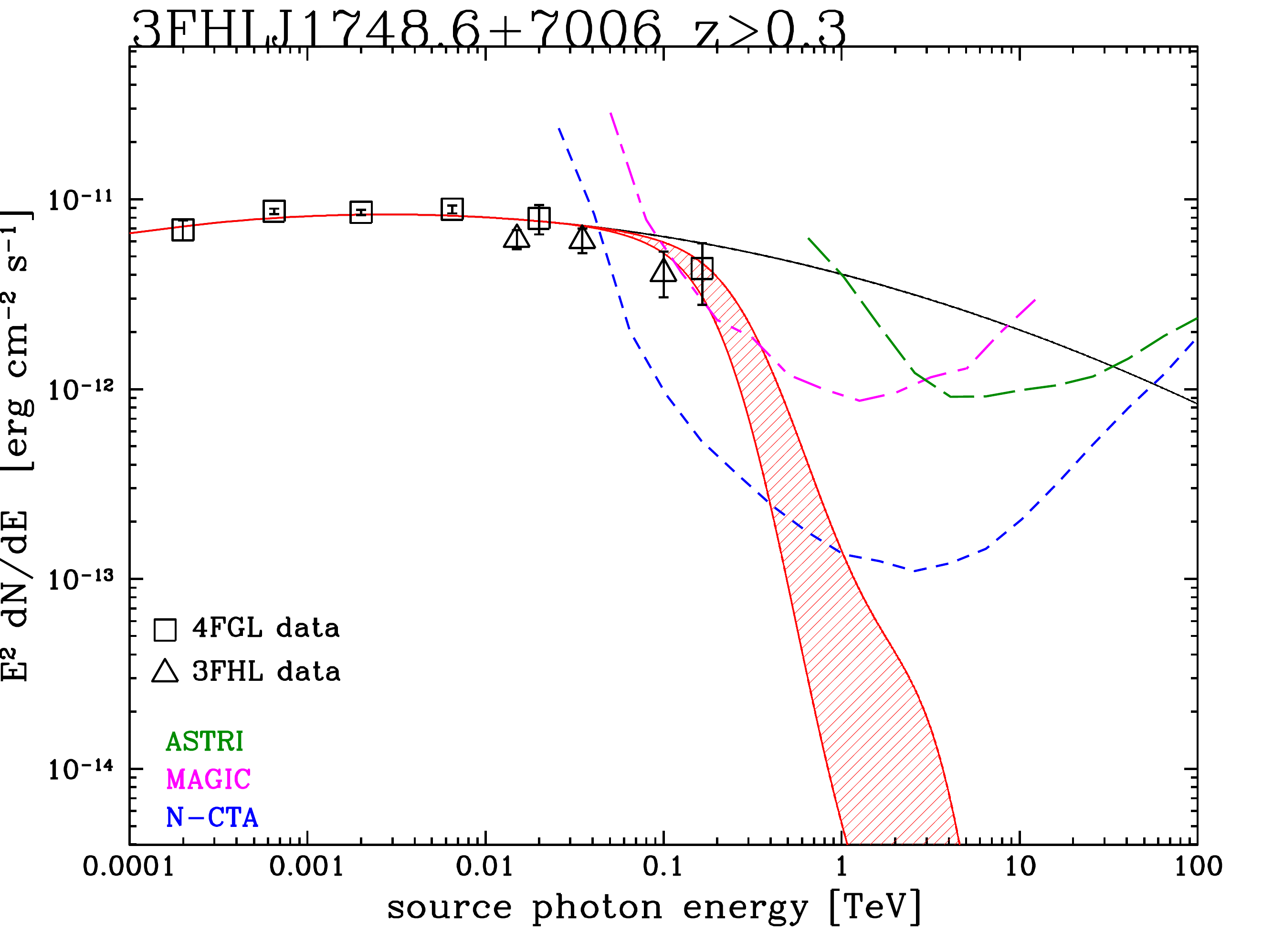}
\includegraphics[width=0.35\textwidth,angle=-90]{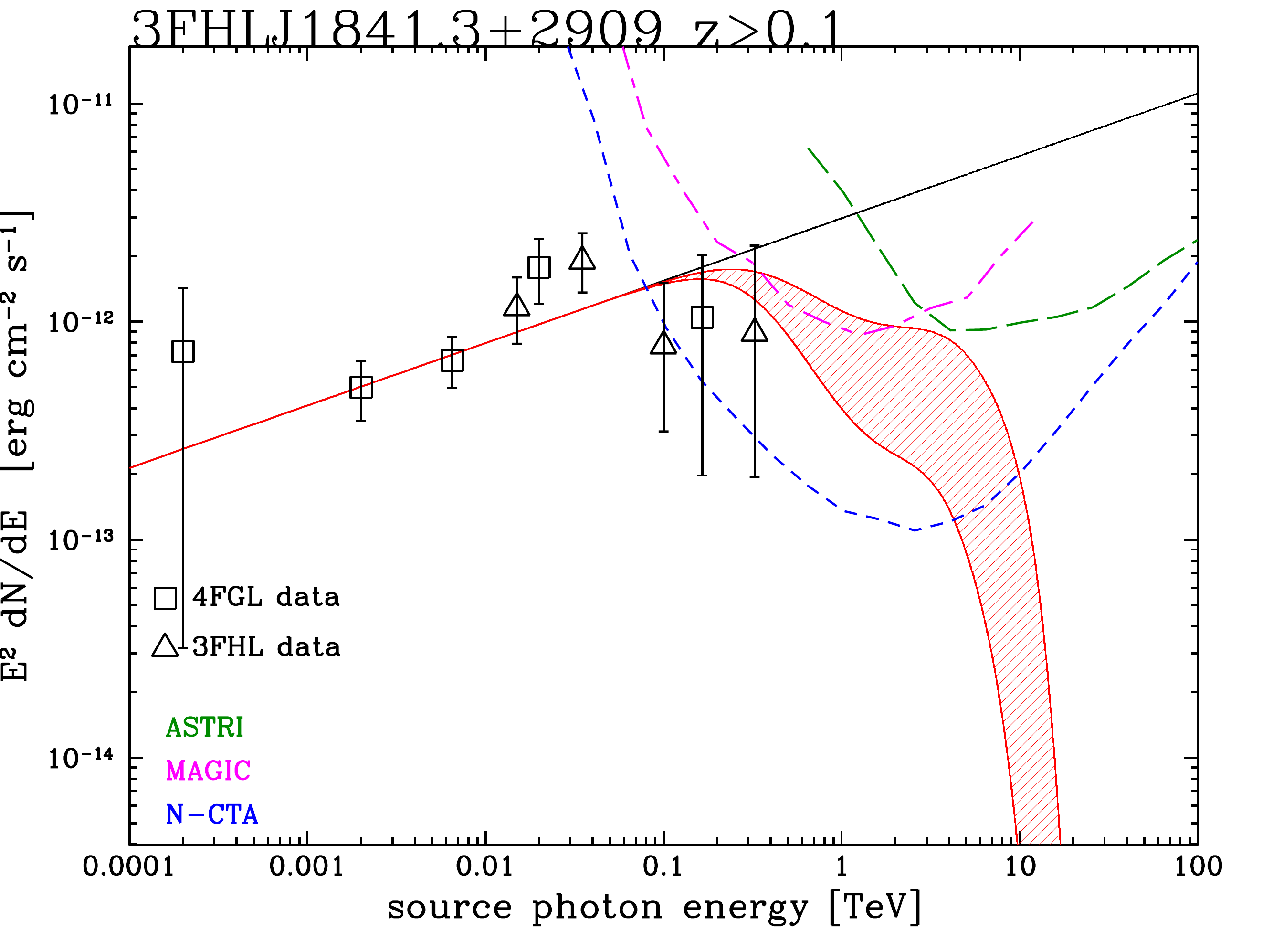}
\includegraphics[width=0.35\textwidth,angle=-90]{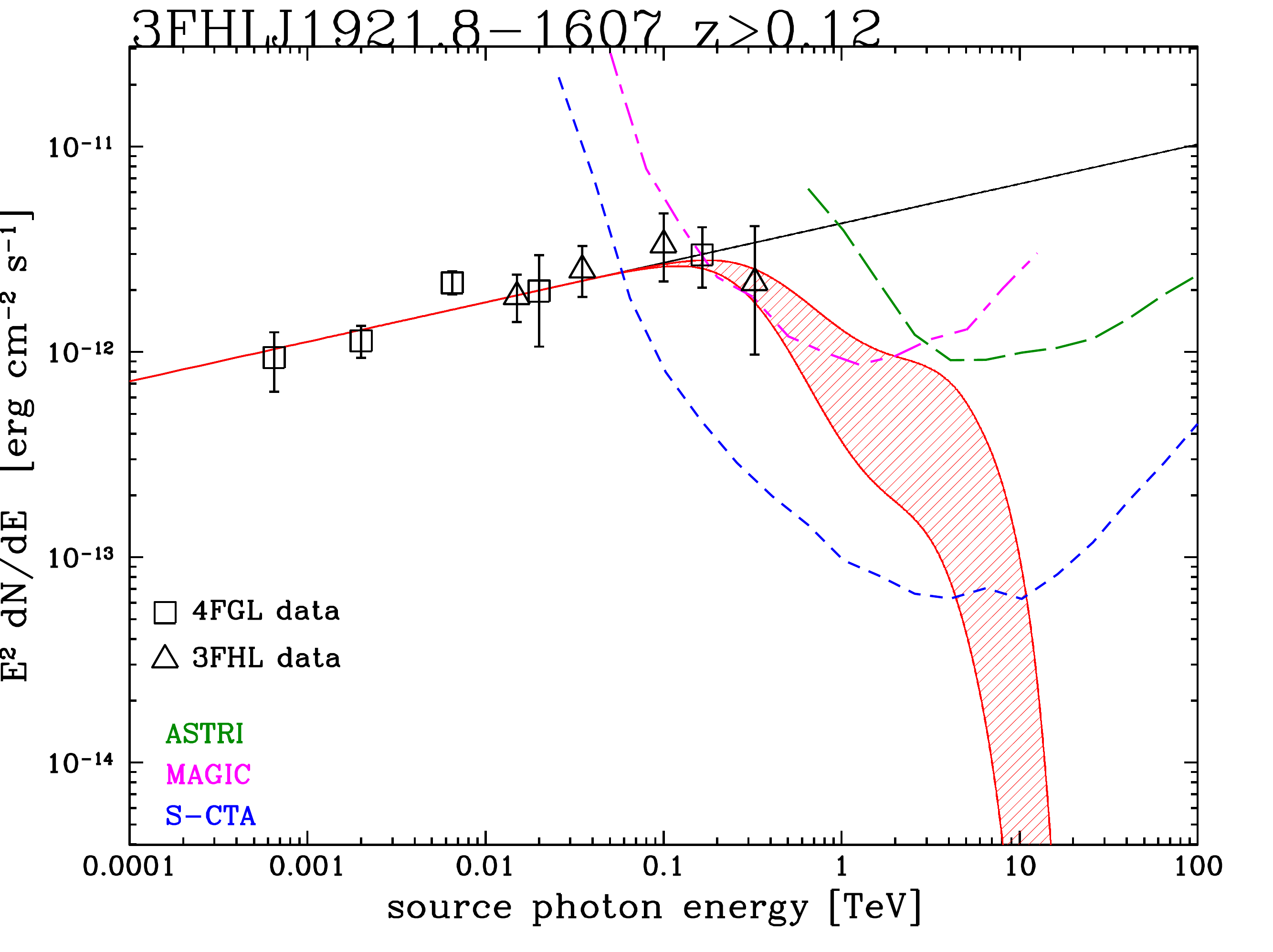}
\caption{Continued from Fig.~3.}
\label{fig:fig1}
\end{figure*}%[htbp]

\label{lastpage}
\end{document}